\documentclass[10pt,twocolumn,preprintnumbers,amsmath,amssymb,nofootinbib,superscriptaddress]{revtex4-1}
\usepackage{comment}
\usepackage{graphicx}
\usepackage{dcolumn}
\usepackage{amssymb,amsmath,bm}
\usepackage{color}
\usepackage[colorlinks,linkcolor=red,citecolor=blue,urlcolor=blue ]{hyperref}
\usepackage{multirow}
\usepackage{enumitem}
\usepackage{mathptmx} 
\usepackage[utf8]{inputenc}
\usepackage{tensor}
\usepackage{amssymb} 
\usepackage{lettrine}
\usepackage[normalem]{ulem}
\usepackage{empheq}

\usepackage{graphicx,rotating,hyperref,xcolor,colortbl,bm,amssymb,amsmath}
\usepackage{tabularx}
\usepackage[utf8]{inputenc} 
\usepackage[T1]{fontenc}
\usepackage{dutchcal}

\newcommand{\calh}{{\mathcal{h}}}

\def\beq{\begin{equation}}
\def\eeq{\end{equation}}

\def\bea{\begin{eqnarray}}
\def\eea{\end{eqnarray}}

\def\dd{{\rm d}}
\def\ii{{\rm i}}
\def\HH{\mathcal{H}}
\def\AmpGW{{\cal A}}
\def\AmpGWstar{\hat{\cal P}_\star}
\newcommand{\com}[1]{}
\newcommand{\sinc}{{\rm sinc}}

\newcommand{\be}{\begin{equation}}
\newcommand{\ee}{\end{equation}}

\newcommand{\gr}[1]{{\bm #1}}

\newcommand{\troisj}[6]{\left(\begin{array}{ccc}
      #1 & #2 & #3 \\
      #4 & #5 & #6\end{array}\right)}

\definecolor{rossos}{cmyk}{0,1,1,0.55}
\definecolor{blu}{cmyk}{1,1,0,0.3}
\definecolor{bluc}{cmyk}{1,1,0,0.1}
\definecolor{verde}{cmyk}{0.92,0,0.59,0.25}
\definecolor{verdec}{cmyk}{0.92,0,0.59,0.15}
\definecolor{verdes}{cmyk}{0.92,0,0.59,0.4}
\hypersetup{colorlinks, bookmarksopen, bookmarksnumbered,
citecolor=blu, linkcolor=bluc, pdfstartview=FitH, urlcolor=rossos}

\newcommand{\CP}[1]{{\color{violet} #1}}

\begin{document}

\title{Mitigating cosmic variance in the Hellings-Downs curve: a Cosmic Microwave Background analogy}
\author{Cyril Pitrou}
\email{pitrou@iap.fr}
\affiliation{Institut d'Astrophysique de Paris, UMR-7095 du CNRS et de Sorbonne Universit\'e, Paris, France}
\author{Giulia Cusin}
\email{cusin@iap.fr}
\affiliation{Institut d'Astrophysique de Paris, UMR-7095 du CNRS et de Sorbonne Universit\'e, Paris, France}
\affiliation{Département de Physique Théorique and Center for Astroparticle Physics, Université de Genève, Quai E. Ansermet 24, CH-1211 Genève 4, Switzerland}

\date{\today}

\begin{abstract}
\noindent
The Hellings-Downs (HD) correlation, which characterizes the signature of a stochastic gravitational wave background measured via Pulsar Timing Arrays (PTA), is derived using a harmonic formalism. This approach closely follows the theoretical framework traditionally employed to compute correlations of temperature fluctuations in the Cosmic Microwave Background (CMB). This parallel enables a direct comparison between the correlations observed in PTA and those in CMB experiments. After providing analytic estimates of the transmission functions due to the model-fitting procedure, we show that the covariance matrix in frequency space becomes very nondiagonal. We then build formally the quadratic estimator for the HD correlation in multipolar space, for both a perfect experiment, and for a realistic pulsar noise model. For a perfect experiment, we show that the signal to noise ratio grows with the observation time and the number of frequency bins, in turn determined by the cadence of observation. For an imperfect experiment, the behaviour is similar, with an effective multipole-dependent number of frequency, obtained after weighting with noise.  We predict that with $\sim 200$ pulsars monitored for $25$ years, multipoles of the HD correlation up to $\ell=4$ can be measured. Our findings clarify that what has been referred to as \emph{cosmic variance} in some of the previous literature is not an intrinsic limitation for PTA measurements.  Instead, when an optimal estimator is used, this variance can be mitigated by accumulating more observation time or improving the cadence of pulsar monitoring. Therefore, unlike CMB angular correlations, where cosmic variance represents an irreducible constraint, it can be reduced in PTA measurements and will continue to diminish in future experiments. Finally, we show that if the primordial power spectrum of tensor fluctuations was very blue with $n_T>4$, the CMB angular correlation due to these tensor modes would also exhibit a HD correlation. While our results
are derived for the case of an isotropic bath of gravitons, we also discuss the case in which the graviton distribution function is anisotropic. 

\end{abstract}
\maketitle
\tableofcontents

\section{Introduction}

Pulsar Timing Arrays (PTAs) were used to provide the first evidence of a stochastic gravitational wave (GW) background in the nHz band~\cite{NANOGrav:2023gor, Reardon:2023gzh, EPTA:2023sfo, Xu:2023wog}, with evidence of spatial correlation among different pulsar redshifts following the Hellings-Downs (HD) function. Various detection methods are reviewed in \cite{Romano:2016dpx}, and the physical interpretation of the HD correlation is discussed in \cite{Jenet:2014bea, Romano:2023zhb}. Methods for mapping the background have also been developed~\cite{Anholm:2008wy, Mingarelli:2013dsa, Depta:2024ykq, Semenzato:2024mtn} (see also \cite{Bernardo:2023jhs, Cusin:2018avf} for a discussion of polarisation with Stokes parameters), and it has recently been proposed that a loud source may account for the evidence of anisotropy reported in~\cite{Grunthal:2024sor}. Due to the all-sky nature of the signal, a harmonic formulation of the HD analysis has been introduced in \cite{Gair:2014rwa,Roebber:2016jzl,Qin:2018yhy,Hotinli:2019tpc,Nay:2023pwu,Bernardo:2022xzl,Bernardo:2022rif,Allen:2024bnk,Bernardo:2024bdc}, which forms the basis for our harmonics analysis. The most recent analysis of the HD signature by NANOGrav~\cite{Agazie:2024qnx} is expressed in multipole space, with a clear detection of the quadrupole and only marginal evidence for the octupole. The significance of the background detection is expected to improve in the future, as the signal-to-noise ratio (SNR) of PTA observables increases with the observation time~\cite{Siemens:2013zla,Nay:2023pwu,Pol:2022sjn}. The spectral function of the GW background from binaries is known to be approximately a power law~\cite{2008MNRAS.390..192S}, and its exponent can also be constrained. However, to eliminate sources of contamination, time residuals for each pulsar must be fit with a linear drift and a spin-down, while the modulation due to Earth's motion must be accounted for, which alters the low-frequency part of the spectrum~\cite{1976ApJ...205..580B, vanHaasteren:2008yh, Hazboun:2019vhv}.

To evaluate the error and construct estimators for the pulsar correlation function, it is crucial to understand the covariance of the quadratic estimators. Recent efforts have focused on obtaining the covariance in the most general scenarios~\cite{Allen:2022ksj}, with particular attention to source discreteness~\cite{Allen:2024mtn, Allen:2022bjz, Allen:2022dzg, Allen:2022ksj, Romano:2023zhb, Allen:2024rqk} and the anisotropic nature of the underlying large-scale structure~\cite{Grimm:2024hgi, Grimm:2024lfj, Allen:2024mtn}. It has recently been argued~\cite{Allen:2024mtn, Allen:2022bjz, Allen:2022dzg, Allen:2022ksj, Romano:2023zhb, Allen:2024rqk} that, since PTAs operate in a regime where time averaging over observation periods limits the sampling of many realizations of the Universe, any attempt to measure the HD correlation is affected by cosmic variance. This cosmic variance is often considered an irreducible limitation on our ability to constrain the HD correlation with any PTA experiment, even in the absence of instrumental noise~\cite{Allen:2022dzg, Allen:2022ksj, Romano:2023zhb, Agarwal:2024hlj}. This is in contradiction with the intuition proposed a few years ago in~\cite{Roebber:2016jzl} that since PTAs are sensitive to a range of frequencies, the effect of cosmic variance should be reduced by properly combining different frequency bins. Indeed, estimates of the SNR scaling based on simulated PTA data suggest that sensitivity continues to improve with increasing observation time~\cite{Nay:2023pwu}, with no indication of reaching a fundamental limitation. This intuition was further developed theoretically in the recent article \cite{Allen:2024uqs} where the authors build an optimal estimator for HD correlation as function of pulsar separation, exploiting the frequency information.  However, confusion persists in the literature, with cosmic variance still being referred to as a fundamental limitation in very recent studies. 

In this article, we present a pedagogical derivation of how the cosmic variance issue is mitigated by optimally combining different frequency bins. Unlike \cite{Allen:2024uqs}, when building the optimal estimator, we work in multipole space rather than in real space. The advantage of working in multipole space is that, in the strong signal limit, the covariance is fully diagonal, and a quadratic optimal estimator can be built, with a more transparent structure than the one derived in \cite{Allen:2024uqs}  in real space. 
We also consider the effect of adding instrumental noise and we present in detail how the optimal estimator is modified when accounting for the fact that in a real experiment one has to fit time residuals with a deterministic template, to subtract the effect of a linear drift and a quadratic term. We show that the final claim about the SNR is not altered by this filtering procedure.

Since PTAs measure timing residuals, which are closely related to pulsar redshifts, the HD correlation in PTAs has an analogous counterpart in CMB observations, when considering the correlation between redshift perturbations (or temperature anisotropies) induced by GWs in the early Universe. We identify the CMB analogue of the HD correlation, which, although conceptually similar, exhibits a  more complex structure compared to the PTA HD correlation. This comparison allows us to evaluate the impact of cosmic variance in both types of experiments. We find that while, for the CMB, cosmic variance is an irreducible source of uncertainty on our ability to reconstruct the angular power spectrum, it does not pose a fundamental limitation for PTA experiments, provided that the correct estimator is used. When using an optimal estimator, we find indeed that the SNR grows with the observation time and with the cadence of pulsar monitoring, hence there is no fundamental limitation in our ability  to measure the HD curve. For the CMB, the redshift is indirectly inferred from the spectrum or the energy of the photons received, which is already averaged over the duration of the experiment. Additionally, the intrinsic timescale of variations in the underlying signal is of the order of millions of years. As a result, the situation is akin to performing a single measurement of the CMB at a specific moment in time. The duration of the experiment mainly serves to reduce the experimental error in this single measurement, while it plays no role in a perfect experiment. Exploiting this parallel, we also show that, if the primordial power spectrum of tensor fluctuations was very blue with $n_T>4$, the CMB angular correlation due to tensor modes would also exhibit a HD correlation.

In our derivation of an estimator for the HD correlation, we first work in real space of time and directions. The dependence on directions can be reformulated in spherical harmonic components, and the time dependence can be expressed in frequency space. Hence we will  consider four different but equivalent formulations of the correlation function and the corresponding four (equivalent) estimators. Multipole space turns out to be a particularly suitable represention for the HD correlation as the angular power spectrum of pulsar redshfits is diagonal.  The advantage of this approach is that the covariance of any estimator is diagonal in both frequency and multipole space, taking the standard form that is familiar from CMB studies (see also \cite{Gair:2014rwa} for CMB mapping methods applied to PTA mapping).

Our article is structured as follows: in section \ref{definitions}, we present general concepts and definitions and in section \ref{PulsarRedshift} we discuss pulsar redshift in multipole space, followed by the computation of the two-point correlation in both real and multipole space in section \ref{PulsarCorrelation}. Section \ref{Quadratic} introduces different representations (in time and direction space, in frequency and direction space, and in multipole space) of a quadratic estimator for the pulsar redshift correlation function. In section~\ref{SecDiscreteFourier} we discuss how to consistently go from the ideal case of a continuous frequency spectrum, to discrete frequency space, and we compute how the pulsar two-point function gets modified when accounting for the finite observation time and filtering out spurious contributions with a proper time modeling. In section \ref{Weight}, we show that cosmic variance arises when different time (or frequency) measurements are naively combined, and we demonstrate that an optimal estimator that properly weights low frequencies mitigates the cosmic variance problem. We build formally the quadratic estimator for the HD correlation in multipolar space, for both a perfect experiment, and for a realistic pulsar noise model. For a perfect experiment, we show that the signal to noise ratio increases with the observation time and the cadence of measurements. For an imperfect experiment, the signal to noise ratio depends on an effective multipole-dependent number of frequency, which increases with the observation time, but which also depends on the frequency dependent level of noise.  We predict that with $\sim 200$ pulsars monitored for $25$ years, multipoles of the HD correlation up to $\ell=4$ can be measured. Our findings show that variance is not an intrinsic limitation, as it can be mitigated by accumulating
more observation time or improving the cadence of pulsar monitoring. Finally in section \ref{CMB} we present the terms of the analogy with the CMB and we discuss our findings in section \ref{discussions}. Our results are derived for the case of an isotropic bath of gravitons. In section \ref{Anis} we discuss the case in which the graviton distribution function is anisotropic: we build an optimal estimator for anisotropies and we compute its variance, recovering results of \cite{Hotinli:2019tpc}. The fact that a given time-residual map has a finite angular resolution fixed by the number of pulsars in a network reflects on our ability to constrain the angular power spectrum of the background energy density, even for a perfect experiment.  Technical details and mathematical derivations are presented in a series of appendices.

\section{GW background: general concepts and definitions}\label{definitions}

In this section we detail how gravitational waves are expanded on tensor valued plane waves, or harmonic functions, and we review the statistical properties of the modes of this expansion. Note that we are eventually interested in writing the redshfit of a pulsar in terms of GW from the pulsar to the observer. Since the pulsars monitored are galactic objects, we can assume spacetime to be flat (neglecting cosmological expansion). This implies that GWs  satisfy $\Box h_{ij}=0$ where the d'Alembertian is the flat space one. On the other hand, when working out the CMB analogy, we will have to extend the treatment of this section to a cosmological background, accounting for the Universe expansion.

\subsection{GW mode expansion}\label{GWExp}

Tensor fluctuations, describing GW, can be expanded on a suitable basis of tensor valued harmonic functions. In flat space, and for a wave propagating along the $z$ axis, the harmonic functions are given by\footnote{The normalization is different from the one used in \cite{Hu:1997hp} or \cite{Pitrou:2019ifq} in the context of CMB fluctuations (tensorial contributions), but it is consistent with the usual notation for the treatment of a stochastic background of GW.}
\be\label{defQ}
Q_{ij}^{(\pm 2)}({\bm k} = k \gr{e}_z,{\bm x}) =e^{(\pm 2)}_{ij}(\gr{e}_z) {\rm
  e}^{\ii {\bm k} \cdot {\bm x}}\,,
\ee
where the helicity tensor basis is 
\be\label{DefPolarization}
e^{(\pm 2)}_{ij}(\gr{e}_z) \equiv (e^x_i \pm \ii e^y_i) (e^x_j \pm \ii e^y_j)\,.
\ee

For a wave propagating in a general direction, we simply rotate the previous construction. If the mode direction is $\hat{\gr{k}}=(\theta_k,\phi_k)$, then we use the rotation $R_{\hat{\gr{k}}}=R(\phi_k,\theta_k,0)$, with the Euler angles defined in~\eqref{DefEulerAngles}, so as to build another harmonic function 
\be\label{DefQijGeneral}
Q_{ij}^{(\pm 2)}({\bm k},\gr{x}) = R_{\hat{\gr{k}}}[Q_{ij}^{(\pm 2)}(k \bm{e}_z,\gr{x})]\,.
\ee
Hence a general harmonic function can be expressed in terms of a suitable helicity basis as
\begin{subequations}
\begin{align}
Q_{ij}^{(\pm 2)}({\bm k},{\bm x}) &\equiv e^{(\pm 2)}_{ij}(\hat{\gr{k}}) {\rm
  e}^{\ii {\bm k} \cdot {\bm x}}\,,\label{DefGeneralQij}\\
e^{(\pm 2)}_{ij}(\hat{\gr{k}}) &\equiv (R_{\hat{\gr{k}}})_i^{\,\,k} (R_{\hat{\gr{k}}})_j^{\,\,l} e^{(\pm 2)}_{kl}(\gr{e}_z)  \,.
\end{align}
\end{subequations}
The essential properties of this construction are
\begin{subequations}
\bea\label{rotQ}
e^{(\pm 2)}_{ij}(-\hat{\gr{k}}) &=&e^{(\mp 2)}_{ij}(\hat{\gr{k}}) \,,\\
Q_{ij}^{(\pm 2)}(-{\bm k},\gr{x}) &=& Q_{ij}^{(\pm 2) \star}({\bm k},\gr{x})\,.\label{ConjugateQij}
\eea
\end{subequations}

With this basis, there are several, but equivalent, ways to decompose a GW background as a superposition of plane waves coming from different directions, depending on how we treat the time dependence. The first method consists in using progressive waves. The GW reads in the local Minkowski metric with time coordinate $\eta$ 
\be\label{hijExpansion}
h_{ij}(\eta,\gr{x}) = {\rm Re} \left(\int \frac{\dd^3 \gr{k}}{(2\pi)^3} \sum_{m=\pm 2}
h^{(m)}(\gr{k}) {\rm e}^{\ii k \eta}Q_{ij}^{(m)}(\gr{k},\gr{x}) \right)\,,
\ee
where we used the definition~\eqref{DefGeneralQij} and  $h^{(m)}(\gr{k})$ is the coefficient in front of a plane wave propagating in the direction $-\hat{\gr{k}}$, that is a wave which we see coming from the direction $\hat{\gr{k}}$.\footnote{This decomposition is close to but different from the (7.1) of Allen
\cite{Allen:2024bnk}: we do not use negative frequencies and we introduce both left and right polarizations. The normalization we use is also 
different.}
For future use we introduce the following modes 
\be\label{Defhcos}
h^{(m)}_{\rm cos}(\gr{k},\eta) \equiv \frac{1}{2}\left(h^{(m)}(\gr{k}) {\rm e}^{\ii k
  \eta } + h^{(m)\star}(-\gr{k}) {\rm e}^{-\ii k \eta}\right)\,,
\ee
such that 
\be\label{Propertyhcos}
h^{(m)\star}_{\rm cos}(\gr{k},\eta) = h^{(m)}_{\rm cos}(-\gr{k},\eta)\,,
\ee
and using~\eqref{ConjugateQij} the GW expansion reads
\be\label{hijhcos}
h_{ij} = \int \frac{\dd^3 \gr{k}}{(2 \pi)^3} \sum_{m=\pm 2}
h^{(m)}_{\rm cos}(\gr{k},\eta) Q_{ij}^{(m)}(\gr{k,\gr{x}})\,,
\ee
without need to take the real part as in (\ref{hijExpansion}). 
Finally, we introduce the combination
\be\label{DefH}
H^{(m)}(\gr{k},\eta) \equiv \frac{1}{2}\left(h^{(m)}(\gr{k}){\rm
    e}^{\ii k \eta}+h^{(-m)\star}(\gr{k}){\rm  e}^{-\ii k \eta}\right)\,,
\ee
which satisfies 
\be\label{PropertyH}
H^{(m)\star}(\gr{k},\eta)  = H^{(-m)}(\gr{k},\eta)\,.
\ee

\subsection{Statistics of the GW background}\label{stat}

For an unpolarized and stationary background, the two-point function of the wave modes satisfy the following relations 
\bea\label{Defstath}
\langle h^{(m)}(\gr{k}) h^{(m')\star}(\gr{k}')\rangle &=&
(2\pi)^3 \delta_{m m'} \delta(\gr{k}-\gr{k}')2 \AmpGW(\gr{k})\,,\nonumber\\
\langle h^{(m)}(\gr{k}) h^{(m')}(\gr{k}')\rangle&=&0\,,
\eea
which are independent correlation rules. The average $\langle \dots \rangle$ denotes an ensemble average  over the many possible realization of the GW background.

The  $2$-point function of the modes $h^{(m)}_{\rm cos}$  
can be derived from their definition~\eqref{Defhcos} 
\begin{align}\label{Stathcos}
\langle h^{(m)}_{\rm cos}(\gr{k},\eta) h^{(m')\star}_{\rm
  cos}(\gr{k}',\eta')\rangle =&\delta_{m m'} (2\pi)^3 \delta(\gr{k}-\gr{k}') \nonumber\\
  &\times {\AmpGW}(\gr{k}) \cos[k (\eta-\eta')]\,,\\
\langle h^{(m)}_{\rm cos}(\gr{k},\eta) h^{(m')}_{\rm cos}(\gr{k}',\eta')\rangle =&\delta_{m m'}(2\pi)^3 \delta(\gr{k}+\gr{k}') \nonumber\\
&\times {\AmpGW}(\gr{k})  \cos[k (\eta-\eta')]\,.\nonumber
\end{align}
where the latter is not independent from the former because of the relation~\eqref{Propertyhcos}.
The fact that the time dependence of the two-point correlation is only a function of $\eta-\eta'$ is a consequence of the assumed stationarity. Similarly, using  the definition~\eqref{DefH} we get for for $H^{(m)}$
\bea\label{EqStatH}
\langle H^{(m)}(\gr{k},\eta) H^{(m')\star}(\gr{k}',\eta')\rangle &=\delta_{m m'} (2\pi)^3 \delta(\gr{k}-\gr{k}')\nonumber\\
&\times {\AmpGW}(\gr{k}) \cos[k (\eta-\eta')]\,.
\eea
Again the result for $\langle H^{(m)} H^{(m')}\rangle $ is not independent and can
be found from property~\eqref{PropertyH}.

\subsection{GW background spectrum}\label{Omega}

The GW spectrum $\AmpGW(\gr{k})$, introduced in 
section \ref{stat}, can be interpreted as a distribution function of a gas of gravitons up to some factors and powers of $\gr{k}$. In the context of PTA observations, this interpretation implicitly assumes that the function ${\AmpGW}(\gr{k})$ has been promoted to a local function ${\AmpGW}(\gr{k},\gr{x}=0)$ at the observer position, in turn identified with the pulsar network position. In other words, we consider that the set of pulsars is very localized and that we have a distribution function describing the gas of gravitons at
the pulsar network position. This approach is valid only if the sources
are very far compared to the distance between pulsars (local detector
assumption). Note that when defining a distribution function from quantum field theory we do exactly the same thing: we define an occupation number
average which is in full generality global, and then we localize it (e.g. with a Wigner transform), see  \cite{Fidler:2017pkg,Pitrou:2019hqg}.

Let us parametrize the angular dependence of the distribution function as
\be\label{AnisoStat}
\AmpGW(\gr{k}) = \AmpGW(k) \hat{\AmpGW}(\hat{\gr{k}})\,,\quad \hat{\AmpGW}(\hat{\gr{k}}) = \sum_{LM} \hat{\AmpGW}_{LM} Y_L^M(\hat{\gr{k}})\,,
\ee
where $k=|\gr{k}|$ and $\hat{\AmpGW}_{L\,-M}=(-1)^M\hat{\AmpGW}_{LM}^\star$ since the GW spectrum is real valued. This decomposition assumes the existence of a factorization between a spectral shape $\AmpGW(k)$, and an angular-dependent function  $\hat{\AmpGW}(\hat{\gr{k}})$. Unless stated otherwise we shall assume an isotropic function,  that is $\hat{\AmpGW}(\hat{\gr{k}})=1$. 

The isotropic part $\AmpGW(k)$ is related to the background energy density, which can be quantified in units of the critical energy density with 
\be\label{EqOmegaGW}
\Omega_{\rm GW} = \int_0^\infty \Omega_{\rm GW}(f) \dd \ln f\,,\quad \Omega_{\rm GW}(f) = \frac{8 \pi^2}{3 H_0^2} f^3  \hat{\cal P}(f)\,,
\ee
where $k = 2 \pi f$, and the (two-sided, meaning that needs to be integrated on both positive and negative frequencies) spectral function is
\be\label{DefPf}
\hat{\cal P}(f) \equiv 4\pi f^2 {\AmpGW}(2 \pi f)\,.
\ee
The spectral function in turn is directly related to the variance of the norm of the GW fluctuations since with~\eqref{hijhcos} and \eqref{Stathcos} we find
\be\label{hijhij}
\langle h_{ij}(\eta_0) h^{ij}(\eta_0)\rangle = 4 h^2\,,
\ee
where following the notation of~\cite{Allen:2024bnk} we introduced
\be\label{hAllen}
h^2 \equiv  2 \int_0^\infty \frac{4\pi k^2 \dd k}{(2\pi)^3} {\AmpGW}(k)=\int_{-\infty}^\infty \hat{\cal P}(f)\dd f\,,
\ee
with the convention $\hat{\cal P}(-f)=\hat{\cal P}(f)$. The factor $4$ in \eqref{hijhij} comes from the normalization of the polarization basis~\eqref{DefPolarization}, and the factor $2$ in \eqref{hAllen} is the contribution of the two polarizations. The quantity $2 h^2$ is interpreted as the variance due to one polarization, and we therefore define the characteristic strain in a (one-sided) log-frequency band as $h_c^2(f) = 4 f \hat{{\cal P}}(f)$ so that $2 h^2  = \int_0^\infty h_c^2(f) \dd \ln f$.

It is common in the literature to parametrize the spectral density as \cite{Maggiore:2018sht}
\be\label{gamma}
\hat{\cal P}(f) = \AmpGWstar \left(\frac{f}{f_\star}\right)^\gamma\,,
\ee
with a spectral index $\gamma=-7/3$, such that $\Omega_{\rm GW}(f)\propto f^{2/3}$, for a background composed by the superposition of GW emitted by a population of compact binaries in the inspiralling phase (see e.g. chapter 23 of \cite{Maggiore:2018sht}).

\section{Pulsar redshift}\label{PulsarRedshift}

We detail how gravitational waves induce a redshift $z$ of the pulsar signal, distinguishing an observer and a pulsar contribution, and discussing their multipolar decomposition.

\subsection{General expression}\label{SecGeneralExpression}

For a pulsar in a given direction $\gr{n}$ and at distance $\chi \equiv |\gr{x}| = \eta_0-\eta$, the redshift to the pulsar is defined as (see e.g. chapter 23 of \cite{Maggiore:2018sht})
\be\label{Generalz}
z(\chi,\gr{n}) = \frac{1}{2}\int_0^\chi \dd \chi' n^i n^j \partial_\eta h_{ij}(\eta_0-\chi',\chi'
\gr{n})\,,
\ee
where to simplify the notation the dependence on $\eta_0$ is omitted here on the left-hand side of equations. 
In the following, we split the redshift into an observer term and a pulsar term as 
\be\label{DefPulsarObserver}
z(\chi,\gr{n}) = z^{\rm o}(\gr{n}) - z^{\rm p}(\chi,\gr{n})\,,
\ee
with 
\begin{subequations}\label{Defzozp}
\begin{align}
z^{\rm o}(\gr{n}) &= \frac{1}{2}\int_0^\infty \dd \chi' n^i n^j \partial_\eta h_{ij}(\eta_0-\chi',\chi'
\gr{n}) \,,\label{Defzo}\\
z^{\rm p}(\chi,\gr{n}) &= \frac{1}{2}\int_\chi^\infty \dd \chi' n^i n^j \partial_\eta h_{ij}(\eta_0-\chi',\chi'
\gr{n})  \,.
\end{align}
\end{subequations}
The first contribution corresponds to the redshift one would measure for an infinitely distant pulsar, the second term is the same but computed from the pulsar position (as if the observer was located on the pulsar).

We now seek to expand the observer and pulsar terms as the product of a wave term and a geometrical factor expanded in spherical harmonics, in the form
\be\label{MultDec}
z^{\rm o/p}(\gr{n}) = \int \frac{\dd^3 \gr{k}}{(2\pi)^3}
\sum_{m=\pm 2} \sum_\ell z^{{\rm o/p}}_{\ell m}(\gr{k},\eta_0) R_{\hat{\gr{k}}}[Y_\ell^m](\gr{n})\,,
\ee
where from now on we always consider $\ell \geq 2$, unless specified otherwise. 
To that purpose, we use the GW plane wave expansion~\eqref{hijExpansion}, hence we need an explicit expression for the tensor $Q^{(m)}_{ij}({\bf{k}}, {\bf{x}})$ contracted with $n^i n^j$. For a GW propagating along the $e_z$ axis one has (just like for tensor modes in the CMB context~\cite{Hu:1997hp,Pitrou:2019ifq})
\be
Q^{(m)}_{ij}(k \gr{e}_z, \gr{x}) n^i n^j = -\sum_{\ell \geq 2} c_\ell \sqrt{\frac{(\ell+2)!}{(\ell-2)!}}\frac{j_\ell(k \chi)}{(k \chi)^2} Y_\ell^m(\gr{n})\,,
\ee
and
\be
c_\ell \equiv  \ii^\ell\sqrt{4\pi(2\ell+1)}\,.
\ee
If the plane wave is in a general direction as defined in~\eqref{DefQijGeneral} we only need to rotate the result using that
\begin{align}
 R_i^{\,k} R_j^{\,l} Q^{(m)}_{kl}(k \gr{e}_z, \gr{x}) n^i n^j &=Q^{(m)}_{kl}(k
\gr{e}_z, \gr{x}) R_i^{\,k} R_j^{\,l} n^i n^j \\
&= Q^{(m)}_{ij}(k\gr{e}_z, \gr{x}) (R^{-1}\cdot n)^i (R^{-1}\cdot n)^j\,,\nonumber
\end{align}
and one finally gets
\begin{equation}\label{EqQijninj}
Q^{(m)}_{ij}(\gr{k}, \gr{x}) n^i n^j = -\sum_{\ell \geq 2} c_\ell \sqrt{\frac{(\ell+2)!}{(\ell-2)!}}\frac{j_\ell(k \chi)}{(k \chi)^2} R_{\hat{\gr{k}}}[Y_\ell^m(\gr{n})]\,,
\end{equation}
where, using~\eqref{Ylmrotation}, a rotated spherical harmonics is
\be\label{DefYlmrotation}
R_{\hat{\gr{k}}}[Y_\ell^m(\gr{n})] \equiv \sum_M D^\ell_{Mm}(R_{\hat{\gr{k}}}) Y_\ell^M(\gr{n})\,,
\ee
where $D^\ell_{Mm}$ are Wigner rotation coefficients defined in appendix~\ref{AppRotations}.

\subsection{Observer term}

We now focus on the observer term. From eqs.~\eqref{hijExpansion},~\eqref{Defzo} and~\eqref{EqQijninj} we first get the expansion
\be\label{zolm1}
z^{\rm o}(\gr{n}) = {\rm Re}\left(\int \frac{\dd^3 \gr{k}}{(2\pi)^3}
  \sum_{m=\pm 2}
  h^{(m)}(\gr{k}){\rm e}^{\ii k\eta_0} F^{(m)}_{\rm o}(\hat{\gr{k}},\gr{n})\right)\,,
\ee
where $F^{(m)}_{\rm o}$ is the antenna pattern function in the helicity basis, explicitly given by 
\be\label{antennao}
F^{(m)}_{\rm o}(\hat{\gr{k}},\gr{n}) = \sum_{\ell \geq 2} F^{\rm o}_\ell R_{\hat{\gr{k}}}[Y_\ell^m(\gr{n})] \,,
\ee
with multipolar coefficients
\be\label{DefFlo}
F_\ell^{\rm o} \equiv-\frac{\ii}{2}\sqrt{4\pi(2\ell+1)}\int_0^{\infty} \ii^\ell \sqrt{\frac{(\ell+2)!}{(\ell-2)!}}\frac{j_\ell(x)}{x^2}{\rm e}^{-\ii x}\dd x\,.
\ee
Using the relation~\cite{Bernardo:2022rif}\footnote{It is a special case of Eq.~6.621.1 in \cite{gradshteyn2007} using $j_\ell(x) = \sqrt{\pi/(2x)} J_{\ell+1/2}(x)$, $\alpha = \pm i$, $\beta=1$, $\nu=\ell+1/2$ and $\mu=-5/2$ along with ${}_2F_1(\ell/2-1/2,\ell/2;\ell+3/2;1)=2^{2+\ell}\Gamma(\ell+3/2)/\Gamma(\ell+3)/\sqrt{\pi}$.}
\be
\int_0^\infty  \frac{j_\ell(x)}{x^2}
{\rm e}^{\mp \ii x} \dd x = 2(\pm\ii)^{1-\ell} \frac{(\ell-2)!}{(\ell+2)!}\,,
\ee
one gets the compact expression 
\be\label{Fellcompact}
F_\ell^{\rm o} \equiv \sqrt{4\pi (2\ell+1)}\sqrt{\frac{(\ell-2)!}{(\ell+2)!}}\,,
\ee
where we recall that $\ell \geq 2$. Given the relation~\eqref{DlmmYslm} between Wigner rotation coefficients and spin-weighted spherical harmonics, the expansion~\eqref{antennao} with~\eqref{DefYlmrotation} is equivalent to (2.10) in \cite{Allen:2024bnk}.

Let us examine the properties of the antenna pattern function (\ref{antennao}). It is first clear that
\be
F^{(m)}_{\rm o}(R\cdot\hat{\gr{k}},R \cdot \gr{n}) = F^{(m)}_{\rm o}(\hat{\gr{k}},\gr{n})\,,
\ee
which implies that changing  the direction of a GW wave or the direction
of observation in the opposite direction have the same effect. From the conjugation properties~\eqref{ConjugateYlm} and \eqref{UsefulDproperty1} we deduce
\be
\left(R_{\hat{\gr{k}}}[Y_\ell^m]\right)^\star = (-1)^m R_{\hat{\gr{k}}}[Y_\ell^{-m}] \,,
\ee
hence using that for antenna patterns $m=\pm2$, we deduce
\be
F^{(m) \star}_{\rm o}(\hat{\gr{k}},\gr{n})  = F^{(-m)}_{\rm o}(\hat{\gr{k}},\gr{n})\,. 
\ee
With this property, the real part in~\eqref{zolm1} can be removed if we use the $H^{(m)}$ combination~(\ref{DefH}), as 
\be\label{zofHm}
z^{\rm o}(\gr{n}) = \int \frac{\dd^3 \gr{k}}{(2\pi)^3} \sum_{m=\pm 2}
  H^{(m)}(\gr{k},\eta_0)F^{(m)}_{\rm o}(\hat{\gr{k}},\gr{n})\,,
\ee
hence we deduce that the multipole expansion~\eqref{MultDec} is satisfied with
\be\label{zofH}
z_{\ell m}^{\rm o}(\gr{k},\eta_0)=H^{(m)}(\gr{k},\eta_0) F_\ell^{\rm o}\,.
\ee
However, we also get from~\eqref{Dofoppositedirection}
\be\label{RotminuskYlm}
R_{-\hat{\gr{k}}}[Y_\ell^m(\gr{n})]=(-1)^\ell R_{\hat{\gr{k}}}[Y_\ell^{-m}(\gr{n})]=R_{\hat{\gr{k}}}[Y_\ell^{-m}(-\gr{n})]\,,
\ee
hence we deduce 
\be
F^{(m)}_{\rm o}(-\hat{\gr{k}},-\gr{n}) = F^{(-m)}_{\rm o}({\hat{\gr{k}}},\gr{n})=F^{(m) \star}_{\rm o}(\hat{\gr{k}},\gr{n})\,.
\ee
Therefore, the expansion~\eqref{MultDec} is equally satisfied with
\be
z^{\rm o}_{\ell m}(\gr{k},\eta_0) = \frac{1}{2}\left[h^{(m)}(\gr{k}){\rm e}^{\ii k \eta_0}+(-1)^\ell h^{(m)\star}(-\gr{k}){\rm e}^{-\ii k \eta_0}\right] F_\ell^{\rm o}. \label{zofH2}
\ee

\subsection{Pulsar term}

We now consider the pulsar term.  We get 
\be\label{zplm1}
z^{\rm p}(\gr{n}) = {\rm Re}\left(\int \frac{\dd^3 \gr{k}}{(2\pi)^3}
  \sum_{m=\pm 2}
  h^{(m)}(\gr{k}){\rm e}^{\ii k\eta_0} F^{(m)}_{\rm p}(k \chi, \hat{\gr{k}},\gr{n})\right)\,,
\ee
where the antenna pattern function is now defined as 
\be\label{antennap}
F^{(m)}_{\rm p}(k \chi, \hat{\gr{k}},\gr{n}) = \sum_{\ell \geq 2} F^{\rm p}_\ell(k\chi)R_{\hat{\gr{k}}}[Y_\ell^m(\gr{n})] \,,
\ee
with 
\be\label{Fellpulsar}
F^{\rm p}_\ell(k\chi) =-\frac{\ii}{2}\sqrt{4\pi(2\ell+1)}\int_{k\chi}^{\infty} \ii^\ell \sqrt{\frac{(\ell+2)!}{(\ell-2)!}}\frac{j_\ell(x)}{x^2}{\rm e}^{-\ii x}\dd x\,,
\ee
and it explicitly depends on the distance $\chi$ and on the frequency $k$ (no compact expression of the type of (\ref{Fellcompact}) exists in this context even though the integral can be expressed in a quite involved manner with hypergeometrical functions). 

\begin{figure}
\centering
\includegraphics[width=\columnwidth]{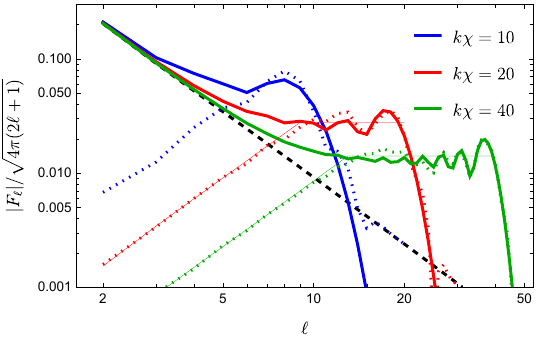}
\caption{Multipolar coefficients of the antenna pattern function: observer term ($F_{\ell}^{\rm o}$, black dashed line),  pulsar term ($F_{\ell}^{\rm p}$, dotted lines) and total term ($F_{\ell}$, continuous lines). The thin solid straight lines correspond to the approximation~\eqref{parametrisation}. Different colors correspond to different values of $k\chi$ used for evaluating the pulsar term.}
\label{Fig1}
\end{figure}

One can derive an approximation for the pulsar angular pattern modes $F^{\rm p}_\ell$. Recalling that for large arguments, spherical Bessel functions behave as
\be
j_\ell(x) \sim \frac{1}{x}\sin\left(x- \ell \frac{\pi}{2}\right)\,,
\ee
and using that
\be
\int_X^\infty \frac{\cos^2(x)}{x^3} \dd x \simeq \int_X^\infty \frac{\sin^2(x)}{x^3} \dd x \simeq \frac{1}{4 X^2} \,,
\ee
we can evaluate approximately~\eqref{Fellpulsar} for low $\ell$ (meaning $\ell \ll \ell_{\rm max} \equiv k \chi$) 
\be
\frac{-F^{{\rm p},{\rm low}}_\ell}{\sqrt{4\pi(2\ell+1)}} \simeq \frac{1}{8 \ell^2_{\rm max}} \sqrt{\frac{(\ell+2)!}{(\ell-2)!}}\,.
\ee
For larger $\ell$, one can approximate the norm of the left-hand side by a constant. Choosing 
\be\label{Flmpulsapproxi}
\frac{-F^{{\rm p},{\rm high}}_\ell(k\chi)}{\sqrt{4\pi(2\ell+1)}} 
\simeq \frac{1}{\sqrt{3} \ell_{\rm max}}\,,
\ee
allows to recover the expected form of the correlation function at zero separation, as it will become clear in section \ref{MuPulsar}. 
A reasonable description consists in choosing
\be\label{parametrisation}
-F_\ell^{\rm p} = {\rm min}\left(-F_\ell^{{\rm p},{\rm low}}, -F_\ell^{{\rm p},{\rm high}}\right)\,,
\ee
for $2 \leq \ell \leq \ell_{\rm max} =  k \chi$.
In Fig.~\ref{Fig1} we plot $F_\ell$ for the observer term and for the pulsar contribution for various $k \chi$.

\section{Pulsar redshift correlation}\label{PulsarCorrelation}

We now want to compute the expectation of the pulsar redshift correlation $z^{\rm o}(\gr{n}_1,\eta_1) z^{\rm o}(\gr{n}_2, \eta_2)$ under the statistics of the graviton gas introduced in section \ref{stat}. We will from now on denote this average as $\langle \dots \rangle_{\AmpGW}$ to avoid confusion with other averaging procedures that will be introduced later in the text.  We will focus on the observer term only, and add the contribution of the pulsar term to the pulsar redshift in a second step. The dependence on directions can be reformulated in spherical harmonic components, and the time dependence can be expressed in frequency space. Hence we will  consider four different but equivalent formulations of the correlation function.

\subsection{Direction and time basis}\label{SecDirTime}

The distributional average of the pulsar redshift correlation is  
\begin{align}\label{Startzz}
&\langle z^{\rm o}(\gr{n}_1,\eta_1) z^{\rm o}(\gr{n}_2,\eta_2) \rangle_{\AmpGW} = \int \frac{\dd^3
  \gr{k}_1}{(2\pi)^3} \frac{\dd^3 \gr{k}_2}{(2\pi)^3}\sum_{m_1 m_2}\\
&\langle H^{(m_1)}(\gr{k}_1,\eta_1)
H^{(m_2)\star}(\gr{k}_2,\eta_2) \rangle_{\AmpGW} F^{(m_1)}_{\rm o}(\hat{\gr{k}}_1,\gr{n}_1) F^{(m_2) \star}_{\rm o}(\hat{\gr{k}}_2,\gr{n}_2)\,.\nonumber
\end{align}
With the statistics~\eqref{EqStatH}, this selects $p_1=p_2$, and assuming an isotropic distribution ${\AmpGW}(k)$ for the graviton gas, this reduces to a single integral on momenta
\begin{align}
&\langle z^{\rm o}(\gr{n}_1,\eta_1) z^{\rm o}(\gr{n}_2,\eta_2) \rangle_{\AmpGW} = \int \frac{\dd^3
  \gr{k}}{(2\pi)^3} \\
  &\sum_{m=\pm 2} {\AmpGW}(k) \cos[k(\eta_1-\eta_2)] F^{(m)}_{\rm o}(\hat{\gr{k}},\gr{n}_1) F^{(m) \star}_{\rm o}(\hat{\gr{k}},\gr{n}_2)\,.\nonumber
\end{align}
Expressing the antenna pattern functions with~\eqref{antennao} and using the definition~\eqref{DefYlmrotation} we need to take the angular average over the mode direction $\hat{\gr{k}}$ since $\dd^3 \gr{k} = k^2 \dd k \dd^2 \hat{\gr{k}}$. From~\eqref{DlmmYslm} and the orthogonality relation~\eqref{OrthoYslm}, this is 
\be\label{EqIntDD}
\int \dd^2 \hat{\gr{k}} D^{\ell_1}_{M_1 m}(R_{\hat{\gr{k}}}) D^{\ell_2 \star}_{M_2 m}(R_{\hat{\gr{k}}}) = \delta^{\ell_1}_{\ell_2} \delta^{M_1}_{M_2}\frac{4\pi}{2\ell_1+1}\,.
\ee
Finally, using the addition theorem~\eqref{AdditionYlm} we get
\be\label{AAverage}
\langle z^{\rm o}(\gr{n}_1,\eta_1) z^{\rm o}(\gr{n}_2,\eta_2) \rangle_{\AmpGW}
= {\cal P}(\eta_1-\eta_2)\mu(\beta_{12})\,,
\ee
where the unequal time power is
\be
{\cal P}(\eta_1-\eta_2) \equiv \int \frac{4 \pi k^2 \dd k}{(2\pi)^3} 2 {\AmpGW}(k) \cos[k(\eta_1-\eta_2)]\,,
\ee
while the function $\mu$, expressed in terms of the angle $\beta_{12}$ between $\gr{n}_1$ and $\gr{n}_2$ ($\cos \beta_{12} \equiv \gr{n}_1 \cdot \gr{n}_2$), is the HD curve  (without pulsar term)
\be\label{Defmul}
\mu(\beta_{12})=\sum_{\ell\geq 2} \mu_\ell(\beta_{12})\,,\quad  \mu_\ell(\beta_{12})\equiv\frac{|F_\ell^{\rm o}|^2}{4
\pi} P_\ell(\cos \beta_{12})\,.
\ee
With the results of appendix~\ref{SecHDangular}, this takes the famous resummed form
\be\label{DefHD}
\mu(\beta_{12}) = \frac{1}{3} -\frac{1}{6} x + x \ln x\,,\quad x\equiv \frac{1-\cos \beta_{12}}{2}\,.
\ee
For equal times, we recover
\be
\langle z^{\rm o}(\gr{n}_1,\eta) z^{\rm o}(\gr{n}_2,\eta) \rangle_{\AmpGW}
=h^2\mu(\beta_{12})\,,
\ee
where $h^2$ is defined in~\eqref{hAllen}. The shape~\eqref{DefHD} of the HD correlation is plotted in Fig.~\ref{FigSW}. 

\subsection{Direction and frequency basis}

Defining the Fourier transform of the pulsar redshift as
\be\label{FourierDefinition}
z(\gr{n},\eta) = \int_{-\infty}^\infty \hat{z}(\gr{n},f) {\rm e}^{\ii 2 \pi f \eta} \dd f 
\ee
where 
\be\label{complexzf}
\hat{z}^\star(\gr{n},f) = \hat{z}(\gr{n},-f)\,,
\ee
and using
\be
{\cal P}(\eta_1-\eta_2) = \int_{-\infty}^{\infty} \hat{\cal P}(f) {\rm e}^{\ii 2 \pi f (\eta_1-\eta_2)}\dd f\,,
\ee
where the spectral function is defined in~\eqref{DefPf}, the correlation in frequency basis becomes
\be\label{Correlationdirf}
\langle \hat{z}^{\rm o}(\gr{n}_1,f) \hat{z}^{{\rm o},\star}(\gr{n}_2,f') \rangle_{\AmpGW}=\delta(f-f')\hat{\cal P}(f)\mu(\beta_{12})\,.
\ee

\subsection{Multipolar and time basis}

The spherical harmonics expansion of the pulsar redshift is defined as
\be\label{ExpansionzlmYlm}
z^{\rm o}(\gr{n},\eta) = \sum_{\ell m} z^{\rm o}_{\ell m}(\eta) Y_\ell^m(\gr{n})\,.
\ee
Since $z$ is real valued, $\CP{(-1)^m} z^{\rm o}_{\ell -m}=z^{{\rm o}\star}_{\ell m}$. 
We want to find the two-point function of multipolar coefficients (the angular power spectrum). To this aim, one can repeat the steps of the computation leading to to~\eqref{AAverage} until the point in which the summation on spherical harmonics is used. Then using  the orthogonality of spherical harmonics~\eqref{OrthoYslm}, the statistics of the $z^{\rm o}_{\ell m}$ is inferred immediately and one gets
\be\label{Correlationzlmtime}
\langle z^{\rm o}_{\ell m}(\eta) z^{{\rm o}\star}_{\ell'
  m'}(\eta')\rangle_{\AmpGW} = \delta_{\ell \ell'} \delta_{m m'}
C^{\rm o}_\ell(\eta,\eta')\,,
\ee
with
\be\label{Cetaetaprime}
C^{\rm o}_\ell(\eta,\eta') = {\cal P}(\eta-\eta')C_\ell^{\rm HD}\,,
\ee
and the HD multipoles are
\be\label{DefClHD}
\quad C_\ell^{\rm HD}\equiv\frac{|F_\ell^{\rm o}|^2}{2\ell+1}=4\pi \frac{(\ell-2)!}{(\ell+2)!}\,.
\ee 
For equal times we get simply $C^{\rm o}_\ell(\eta) \equiv C^{\rm
  o}_\ell(\eta,\eta)$ which is
\be\label{Ceta}
C^{\rm o}_\ell(\eta) = h^2 C_\ell^{\rm HD}\,,
\ee
and does not depend on $\eta$. 

\subsection{Multipolar and frequency basis}

Finally, we consider the Fourier transform of (\ref{ExpansionzlmYlm}). The statistics of the multipoles $\hat{z}^{\rm o}_{\ell m}(f)$, which are the Fourier transform of $z^{\rm o}_{\ell m}(\eta)$,  is given by 
\be\label{Correlationlmf}
\langle \hat{z}^{\rm o}_{\ell m}(f) \hat{z}^{{\rm o}\star}_{\ell'
  m'}(f')\rangle_{\AmpGW} = \delta_{\ell \ell'} \delta_{m m'}\delta(f-f')
C^{\rm o}_\ell(f)\,,
\ee
where
\be
C^{\rm o}_\ell(f) = \hat{\cal P}(f) C_\ell^{\rm HD}\,,
\ee
is the Fourier transform of (\ref{Cetaetaprime}). \CP{From the reality of $z$ the multipoles in Fourier space $\hat{z}^{{\rm o},\star}_{\ell m}(f) = (-1)^m \hat{z}^{\rm o}_{\ell\, -m}(-f)$.} In the multipolar and frequency basis, the correlation is fully diagonal. For the frequency part, this is due to the assumed stationarity of the underlying GW background. For the multipolar components, this is due to the fact that we assumed an isotropic function ${\AmpGW}(k)$. 

Let us briefly discuss the effects of an anisotropic spectrum: if $\hat{\AmpGW}(\hat{\gr{k}}) \neq 1$, the correlation in equation~\eqref{Correlationlmf} acquires off-diagonal contributions. This can be demonstrated by following the steps outlined in section~\ref{SecDirTime} up to equation~\eqref{EqIntDD}. However, this latter expression cannot be directly applied due to the angular dependence of $\hat{\AmpGW}(\hat{\gr{k}})$. Therefore, one must rely on the properties of spherical harmonics, as detailed in appendix~\ref{AppRotations}. The Wigner coefficients are written in terms of spin-weighted spherical harmonics, as shown in equation~\eqref{DlmmYslm}, and the integral involving three spin-weighted spherical harmonics is given by equation~\eqref{Gaunt}. Ultimately, we obtain
\begin{align}\label{GeneralCorrelation}
&\langle \hat{z}^{\rm o}_{\ell_1 m_1}(f_1) \hat{z}^{{\rm o}}_{\ell_2 m_2}(f_2) \rangle_{\AmpGW} = \delta^{\rm even}_{\ell_1+\ell_2+L}\delta(f_1\CP{+}f_2)\hat{\cal P}(f_1)F^{\rm o}_{\ell_1} F^{\rm o}_{\ell_2}\nonumber\\
&\times\sum_{LM} \frac{\hat{\AmpGW}^\star_{LM}}{\sqrt{4\pi}}  \sqrt{2L+1}\troisj{L}{\ell_1}{\ell_2}{M}{m_1}{m_2}\troisj{L}{\ell_1}{\ell_2}{0}{-2}{2}\,,
\end{align}
where $\delta^{\rm even}_{\ell_1+\ell_2+L}=1$ if $\ell_1+\ell_2+L$ is even and vanishes if it is odd. The isotropic result~\eqref{Correlationlmf} is easily recovered when using $\hat{\AmpGW}_{LM}=\delta_L^0 \delta_M^0 \sqrt{4 \pi}$ and \eqref{Troisj0ll}. The structure of the correlations~\eqref{GeneralCorrelation} is similar to the one induced in the CMB by an anisotropic power spectrum~\cite{Ackerman:2007nb,Kothari:2015tqa}. Note that the anisotropic part of $\hat{\AmpGW}(\hat{\gr{k}})$ is responsible for the off-diagonal contributions, but it also modifies the diagonal correlations.

\subsection{Adding the pulsar term}\label{MuPulsar}

\begin{figure*}
\centering
\includegraphics[width=\columnwidth]{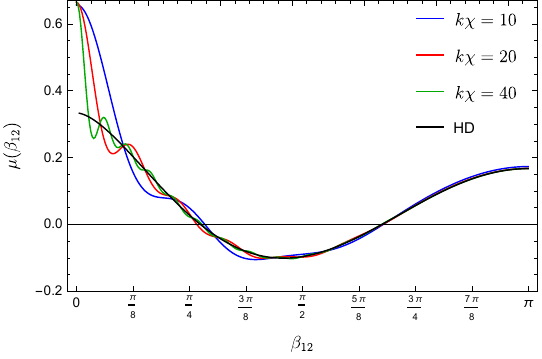}
\includegraphics[width=\columnwidth]{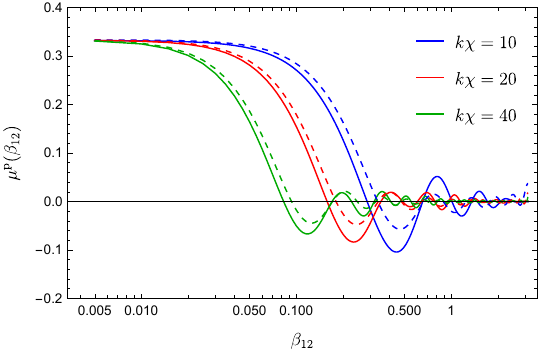}
\caption{ {\it Left:} Total correlation function, including the pulsar term contribution. Different lines correspond to different values of $k \chi$ while the black line is the HD curve. We observe that as we increase $k\chi$ the pulsar term rapidly converges to give a contribution equal to $1/3$ at zero separation, and vanishing contributions for finite separation. {\it Right:} Pulsar term contribution to the correlation function. The
  continuous line is the exact result, and the dashed line is the
  approximation~\eqref{parametrisation}.}
\label{Fig2}
\end{figure*}

Let us now look at the contribution of the pulsar term to the correlation defined in (\ref{DefHD}).  We define the total correlation function including both observer and pulsar terms  as
\be
\mu^{\rm tot}(k\chi,\beta_{12}) = \sum_{\ell\geq 2} \frac{|F_\ell^{\rm o}-F_\ell^{\rm p}(k \chi)|^2}{4\pi} P_\ell(\cos\beta_{12})\,,
\ee
and we plot it in Fig.~\ref{Fig2}, using for the mode functions $F_{\ell}$ eqs.\,(\ref{Fellcompact}) and (\ref{Fellpulsar}).  Interference
between pulsar and observer terms can be neglected and the pulsar contribution converges to the usual $(1/3)\delta_{n_1
  n_2}$ variance in the limit $k\chi  \gg1$.  Indeed, using for the pulsar term the parametrization in (\ref{parametrisation}), we observe that for $k \chi \rightarrow \infty$ 
\be
\sum_{\ell=2}^{\infty}\frac{|F^{{\rm p}}_\ell(k\chi)|^2}{4\pi} \simeq \sum_{\ell=2}^{\ell_{\rm max}}\frac{|F^{{\rm p},{\rm high}}_\ell(k\chi)|^2}{4\pi}  \to \frac{1}{3}\,.
\ee
In other words, at zero separation the pulsar contribution to the correlation function  tends to  $1/3$ for $k\chi
\to \infty$ (this is not exact for finite $k \chi$). Explicitly, for $k \chi\gg1$
\be
\mu^{\rm tot}(\beta_{12})\simeq \mu(\beta_{12})+\frac{1}{3}\delta^c(\beta_{12})\,,
\ee
where $\delta^c(\beta_{12})$ is  a kind of generalized Kronecker function $\delta$, on the set of real numbers, i.e. a function which is $1$ when $\beta_{12}=0$, and vanishes otherwise. In other words, we have  $\mu^{\rm tot}(0) \to 2/3$ as both the observer and pulsar parts converge to $1/3$, but $\mu^{\rm tot}(\beta_{12})=\mu(\beta_{12})$ for $\beta_{12} >0$. The total correlation is plotted in the left panel of Fig.~\ref{Fig2}, and the pulsar term alone is plotted in its right panel. 

Notice that in a realistic PTA situation the condition $k \chi\gg 1$ is largely satisfied. Indeed even for a distance $\chi$ as small as $156$ pc (distance to the closest pulsar) and $k$ of order of the minimum value in the PTA frequency range, $k=2\pi f\simeq 2\pi\times 3\times 10^{-9}$ 
Hz, $k \chi$ is of order $300$,  
which becomes about 
$1900$ for a millisecond pulsar at a more typical distance of $1$ kpc. 

From now on in our analysis we discard the contribution of the pulsar term, since   its support is zero (and in observations often only pairs of different pulsars  are used to get rid of noise in individual pulsars). 

\subsection{Timing residuals}

In PTA experiments we observe timing residuals which are related to the pulsar redshift evolution by 
\be\label{rz}
r(\eta) = \int_{\eta_0}^\eta z(\eta') \dd \eta'\,.
\ee
This implies that timing residuals are easily obtained from the redshift expressions by integration. As detailed below in section~\ref{CorrelationFourierSeries}, we can ignore the constant term, related to the time of reference when expressing residuals, as it is removed when fitting the residuals~\cite{1976ApJ...205..580B,Hazboun:2019vhv}. Therefore, the expressions for timing residuals are obtained from the redshift ones just by the replacements
\be\label{ReplacementTimeResiduals}
z \to r\,,\qquad \hat{\cal P}(f) \to \frac{\hat{\cal P}(f)}{(2\pi f)^2}\,,\qquad {\AmpGW}(k) \to \frac{{\AmpGW}(k)}{k^2}\,,
\ee
meaning that in practice for residuals one has to use $\gamma=-13/3$ in \eqref{gamma} when considering a background from massive black hole binaries.

\section{Quadratic estimators}\label{Quadratic}

We now introduce a quadratic estimator for the pulsar 2-point function. Similar to how we defined four types of quadratic averages based on different combinations of direction or multipoles, and time or frequency domain, we will have four corresponding flavors of quadratic estimators. In section~\ref{Anis}, we will also comment on how, in the presence of background anisotropies, an estimator can be built to extract multipoles of the spectral function.

\subsection{Directional  space and time estimator}

Observationally, in order to build the correlation function, one measures pulsar redshifts and take a spatial average over all pulsar pairs separated by a given angle:  a pulsar average. We define the following estimator  
\be\label{estimator}
\Gamma^e(\beta_{12};\eta_1,\eta_2) = \langle z(\gr{n}_1,\eta_1) z(\gr{n}_2,\eta_2) \rangle_{R} \,,
\ee
where $\beta_{12}$ is the angle between $\gr{n}_1$ and
$\gr{n}_2$. The average over pairs is defined as
\be\label{PulsarAverageBody}
\langle z(\gr{n}_1, \eta_1) z(\gr{n}_2, \eta_2) \rangle_{R} \equiv \int \frac{\dd^3 R}{8
  \pi^2} z(R \cdot \gr{n}_1,\eta_1) z(R\cdot \gr{n}_2,\eta_2)\,,
\ee
where the differential rotation element $\dd^3 R$ is defined in appendix \ref{AppRotations}. Using \eqref{antennao} along with orthogonality of Wigner rotation matrices \eqref{EqOrthogonalityDlmm}, we get 
\be\label{EqAverageFF1}
\langle F^{(m_1)}(\hat{\gr{k}},\gr{n}_1) F^{(m_2) \star}(\hat{\gr{k}},\gr{n}_2)
\rangle_{R}  = \delta_{m_1 m_2}\sum_\ell \frac{|F_\ell|^2}{4 \pi} P_\ell(\gr{n}_1 \cdot \gr{n}_2)\,,
\ee
hence using~\eqref{zofHm},  the estimator is expressed as
\begin{align}
&\Gamma^e(\beta_{12};\eta_1,\eta_2)=\int \frac{\dd^3 \gr{k}_1}{(2\pi)^3} \frac{\dd^3 \gr{k}_2}{(2\pi)^3}\times\nonumber\\
&\sum_{m=\pm 2} H^{(m)}(\gr{k}_1,\eta_1)H^{(m)\star}(\gr{k}_2,\eta_2)\sum_\ell \frac{|F_\ell|^2}{4 \pi} P_\ell(\cos\beta_{12})\,.
\end{align}

We want to compute the mean and variance of this estimator, averaging over graviton realizations (distributional average). With the statistics~\eqref{EqStatH}, the mean is
\be
\langle \Gamma^e (\beta_{12};\eta_1,\eta_2)\rangle_{\AmpGW} = {\cal P}(\eta_1-\eta_2)
\mu(\beta_{12})\,,
\ee
hence (\ref{estimator}) is a good estimator which averages to what we are looking for. 
The assumption of statistical isotropy implies that the mean of the estimator depends only on the difference of directions, while stationarity implies it depends only on the difference of times. In both cases the reciprocal space is much better adapted as we shall detail below.

To simplify the notation, to compute the covariance we consider equal-time  correlators and we define
\be
\Gamma^e (\beta_{12};\eta)  \equiv \Gamma^e (\beta_{12};\eta,\eta)\,.
\ee
The covariance is given by 
\begin{align}
&\text{Cov}(\beta_{12}, \eta; \beta_{12}', \eta')\equiv\nonumber\\
&\langle \hat\Gamma(\beta_{12},\eta) \hat\Gamma(\beta'_{12},\eta') \rangle_{\AmpGW} -
\langle \hat\Gamma(\beta_{12},\eta) \rangle \langle\hat\Gamma(\beta'_{12},\eta') \rangle_{\AmpGW}\,.
\end{align}
Using the definition of the estimator (\ref{estimator}) and replacing (\ref{zofHm}), we see that the distributional average acts on products of $4$ modes $H^{(m)}(\gr{k})$: Wick's theorem gives three pair of terms. The first is canceled by the product of averages, hence we are left with
two contributions, the first pairing $13$ and $24$, and the second
pairing $14$ and $23$, which are equal. We have to perform the integral
on the relative direction between $\gr{k}_1$ and $\gr{k}_2$ for which we use
\bea
&&\langle F^{(m_1)}(\hat{\gr{k}}_1,\gr{n}_1) F^{(m_2) \star}(\hat{\gr{k}}_2,\gr{n}_2)
\rangle_{R}  = \nonumber\\
&&\sum_\ell \frac{|F_\ell|^2}{4 \pi} D^\ell_{m_2
  m_1}(R^{-1}_{\hat{\gr{k}}_2} \cdot R_{\hat{\gr{k}}_1})
P_\ell(\gr{n}_1 \cdot \gr{n}_2)\,,
\eea
which is consistent with \eqref{EqAverageFF1} since $D^\ell_{m_2 m_1}(\mathbb{I})=\delta_{m_1 m_2}$. 
Eventually, we get 
\begin{align}
&\text{Cov}(\beta_{12}, \eta;\beta_{12}', \eta')\nonumber\\
&=\left(\frac{{\cal P}(\eta-\eta')}{2}\right)^2\sum_\ell \frac{|F_\ell|^4}{(4\pi)^2}
P_\ell(\cos \beta_{12}) P_\ell(\cos \beta'_{12}) \nonumber\\
&\qquad\times\sum_{m_1=\pm2} \sum_{m_2=\pm2} \int_{-1}^1 \dd \cos \theta |d^\ell_{m_1\,m_2}(\theta)|^2\,.
\end{align}
Using~\eqref{dlmmJacobi} and~\eqref{JacobiMagic}, the integrals are
\be
\int_{-1}^1 \dd \cos \theta
|d^\ell_{2\,2}(\theta)|^2 = \int_{-1}^1 \dd \cos \theta
|d^\ell_{2\,-2}(\theta)|^2=\frac{2}{2\ell+1}\,,
\ee
and the final result is
\begin{align}\label{CovarianceBeta}
 &\text{Cov}(\beta_{12}, \eta; \beta_{12}', \eta')=\nonumber\\
 &2{\cal P}^2(\eta-\eta')\sum_\ell \frac{|F_\ell|^4}{(4\pi)^2(2\ell+1)}P_\ell(\cos \beta_{12}) P_\ell(\cos \beta'_{12})\,,
\end{align}
in agreement with (8.4) of \cite{Allen:2024bnk}.

\subsection{Directional space and frequency estimator}

One can introduce the Fourier transform of the estimator (\ref{estimator}) to work in frequency space. The mean and the covariance are
\be
\langle \Gamma^e (\beta_{12};f_1,f_2)\rangle_{\AmpGW} = \delta(f_1-f_2)\hat{\cal P}(f_1) \mu(\beta_{12})
\ee
and 
\begin{align}\label{Covbetaf}
 &\text{Cov}(\beta_{12}, f; \beta_{12}',  f')=\delta(0)\delta(f-f')\nonumber\\
 &\hat{\cal P}^2(f)\sum_\ell \frac{2|F_\ell|^4}{(4\pi)^2(2\ell+1)}P_\ell(\cos \beta_{12}) P_\ell(\cos \beta'_{12})\,,
\end{align}
and  we observe that everything is diagonal in frequency space, as a consequence of stationarity. In  section~\ref{SecDiscreteNaive}, we will show that the function  $\delta(0)$ has to be interpreted as $\delta(0)=T$, where $T$ is the observation time.

\subsection{Multipole space and time estimator}

Now we can build an estimator for the angular power spectrum 
\be\label{estimatorCl}
C^e_\ell(\eta,\eta') \equiv \sum_{m=-\ell}^\ell
\frac{z_{\ell m}(\eta) z^{\star}_{\ell m}(\eta')}{2\ell+1}\,.
\ee
We observe that averaging over all $m$ for a given $\ell$, is the multipole space equivalent  of averaging over all pulsar pairs with a given
separation in direction space. The relation to the estimator (\ref{estimator}) can be easily found using the expansion~\eqref{ExpansionzlmYlm} and
\be
\langle Y_\ell^m(\gr{n}) Y_{\ell'}^{m' \star}(\gr{n}') \rangle_R = \delta_{\ell \ell'} \delta_{m m'} \frac{1}{4\pi}P_\ell(\gr{n}\cdot \gr{n}')\,.
\ee
One finds indeed the relations
\be\label{GammaClrelation}
\Gamma^e (\beta_{12};\eta_1,\eta_2) = \sum_{\ell} \frac{2\ell+1}{4\pi}C^e_\ell(\eta_1,\eta_2) P_\ell(\cos\beta_{12})\,,
\ee
hence the estimator $C^e_\ell$ is the
counterpart of the estimator $\Gamma^e (\beta_{12})$ in multipole space.

From the Gaussian statistics~\eqref{Correlationzlmtime}, the average of the estimator~\eqref{estimatorCl} gives
\be\label{CeetaA}
\langle C^e_\ell(\eta,\eta') \rangle_{\AmpGW} = {\cal P}(\eta-\eta') C_\ell^{\rm HD}\,.
\ee
with definition \eqref{DefClHD}. Also here, for simplicity we just focus on equal time for the covariance, introducing  
\be
C^e_\ell(\eta)  =C^e_\ell(\eta,\eta) \,,
\ee
whose variance, found from Wick theorem and the statistics~\eqref{Correlationzlmtime}, is given by  
\bea\label{CovarianceCl}
{\rm Cov}_{\ell \ell'}(\eta,\eta')&\equiv&\langle C^e_\ell(\eta)  C^e_{\ell'}(\eta')
\rangle_{\AmpGW} -  \langle C^e_\ell(\eta)  \rangle_{\AmpGW} \langle C^e_{\ell'}(\eta')
\rangle_{\AmpGW}\nonumber\\
&=&\delta_\ell^{\ell'}\frac{2}{2\ell+1} [C_\ell(\eta,\eta')]^2\,.
\eea
This is the usual expression of cosmic variance on an angular power spectrum estimator, where the factor $2\ell+1$ at the denominator corresponds to the independent values of $m$ at a given $\ell$. As expected, the covariance of the estimator (\ref{estimatorCl}) is related to the one of (\ref{estimator}) (eq.\,\eqref{CovarianceBeta})  via 
\begin{align}\label{Varianceletaspace}
&\text{Cov}(\beta_{12}, \eta; \beta_{12}', \eta')  \nonumber\\
&=\sum_{\ell} \frac{(2\ell+1)^2}{(4\pi)^2}{\rm Cov}_{\ell \ell}(\eta,\eta') P_\ell(\cos\beta_{12}) P_\ell(\cos\beta_{12}')\,,
\end{align}
as it can be checked directly. At equal time and for equal separations it is exactly Eq. (7) of \cite{Copi:2010na} which was obtained in the CMB context.

\subsection{Multipole space and frequency estimator}

Finally, we introduce the Fourier transform of the estimator (\ref{estimatorCl}) 
\be\label{estimatorCl2}
C^e_\ell(f,f') \equiv \sum_{m=-\ell}^\ell
\frac{\hat{z}_{\ell m}(f) \hat{z}^{\star}_{\ell m}(f')}{2\ell+1}\,.
\ee
Its average is given by 
\be\label{AverageEstimatorlf}
\langle C^e_\ell(f,f') \rangle_{\AmpGW} = \delta(f-f')\hat{\cal P}(f) C_\ell^{\rm HD}\,.
\ee
We compute the covariance only at equal frequency defining 
\be
C^e_\ell(f) = C^e_\ell(f,f)\,,
\ee
and we get 
\bea\label{CovarianceClf}
{\rm Cov}_{\ell \ell'}(f,f')&=&\langle C^e_\ell(f)  C^e_{\ell'}(f')
\rangle_{\AmpGW} -  \langle C^e_\ell(f)  \rangle_{\AmpGW} \langle C^e_{\ell'}(f')
\rangle_{\AmpGW}\nonumber\\
&=&\delta_\ell^{\ell'}\frac{2}{2\ell+1} \delta(0)\delta(f-f') [C_\ell(f)]^2\,. 
\eea
We observe that the covariance is now fully diagonal. As a result, the multipole and frequency representation of the quadratic estimator is the most convenient, since the inverse covariance is straightforward to compute due to its diagonal form.

\section{Fitting the residuals}\label{SecDiscreteFourier}

In this section, we
discuss how to consistently go from the ideal case of a continuous frequency spectrum, to discrete frequency space, and
we compute how the pulsar two-point function gets modified
when accounting for the finite observation time and filtering
out spurious contributions with a proper time modelling.

\subsection{Discrete Fourier series}\label{SecDiscreteNaive}

In practice, we do not have access to the full spectrum of frequency as the experiment is run only for a finite duration. For an observation time $T$, we can only access the discrete Fourier series coefficients for the frequency 
\be
f_p = p/T\quad \text{with}\quad 1 \leq |p| \leq T f_{\rm max}\,,
\ee
where 
\be
f^{-1}_{\rm max} \equiv 2\Delta \eta\,,
\ee
is set from the cadence of observations $\Delta \eta$ and the Nyquist criterion. The discrete Fourier series coefficients are defined as
\be\label{Defzn}
\tilde{z}(f_p) = \int_{-T/2}^{T/2} z(t) {\rm e}^{- 2 \pi \ii f_p t} \dd t\,.
\ee
Hence the relation between the Fourier discrete series coefficients and the Fourier transform is
\be\label{rawtildez}
\tilde{z}(f_p) = T \int \dd f \hat{z}(f) \sinc\left(\pi (f- f_p) T \right)\,.
\ee
It is standard to use the approximation
\be\label{sincdelta}
\sinc(\pi x) \simeq \delta(x)\,,
\ee
such that 
\be
\tilde{z}(f_p) \simeq \hat{z}(f_p)\,.
\ee
We chose not to normalize~\eqref{Defzn} with the observation time $1/T$ to get this approximate equality between the Fourier transform and the Fourier series coefficients.

The relations traditionally used when going from discrete to continuous and backward are
\begin{subequations}
\begin{align}
\frac{1}{T}\sum_n &\leftrightarrow \int_{\rm obs} \dd f\equiv  \int_{-f_{\rm max}}^{-1/T} \dd f + \int_{1/T}^{f_{\rm max}} \dd f \,,\label{Correspondence}\\
T \delta_{pq}&\leftrightarrow \delta(f_p-f_q)\,.\label{Correspondence2}
\end{align}
\end{subequations}
By using these results, the correlations~\eqref{Correlationdirf} and \eqref{Correlationlmf} for discrete frequencies  $f_p,f_q$,  are approximately replaced  by
\begin{subequations}
\begin{align}
\langle \tilde{z}(\gr{n}_1,f_p) \tilde{z}(\gr{n}_2,f_q) \rangle_\AmpGW &\simeq \mu(\beta_{12}) T \delta_{pq} \hat{\cal P}(f_p)\,,\\
\langle \tilde{z}_{\ell_1 m_1}(f_p) \tilde{z}_{\ell_2 m_2}(\gr{n}_2,f_q) \rangle_\AmpGW &\simeq \delta_{\ell_1}^{\ell_2} \delta_{m_1}^{m_2} C_{\ell_1}^{\rm HD} T \delta_{pq} \hat{\cal P}(f_p)\,\label{zzF}
\end{align}
\end{subequations}
Finally, given the correspondence~\eqref{Correspondence2}, it follows that the terms  $\delta(0)$ appearing in the expressions for the covariance found in the previous section for a continuous frequency spectrum (\eqref{Covbetaf} and~\eqref{CovarianceClf}) have to be interpreted as $\delta(0)=T$.

\subsection{Templates fitting}\label{CorrelationFourierSeries}

The approach in section~\ref{SecDiscreteNaive} to go from a continuous frequency spectrum to a discrete one and back, is too simplistic in two aspects. First for a very red spectrum the approximation~\eqref{sincdelta} in~\eqref{rawtildez} fails badly. And second we also need to take into account the fact that the observed timing residual is fitted with a deterministic template, and this has the effect of removing the low frequencies. We now deal with both complications.

As shown in \cite{Hazboun:2019vhv}, errors  in  the estimate of  timing model parameters lead
to deterministic features in the timing residuals. For
example, an error in the pulse period leads to timing
residuals that grow linearly with time,  while an
error in the period derivative leads to residuals that grow
quadratically with time. Thus, we can improve
our estimates of the timing model parameters by fitting for a constant, linear and quadratic term (in some cases also a cubic spin-down model is fitted for), and subtract these terms from the observed residuals. When working with redshift, related to time residual via the time integration  (\ref{rz}), one fits with a constant and a linear term (eventually with a quadratic term as well if a cubic spin-down effect is considered in the timing residuals). In the following part of this section, for consistency with the rest of the paper, we will work with pulsar redshifts. However, the treatment of time residuals is straightforward, recalling the relation (\ref{rz}), and when showing results we will consider both these observables.

The redshift of a given pulsar in real space is expressed as a function of the pulsar direction $\gr{n}$ and in multipole space as a series of multipoles $(\ell,m)$. The treatment we will present in the following can be applied to both real and multipole space, hence, to keep the discussion general, we will simply denote redshift as $z(t)$, omitting direction/multipole dependence. Pulsar redshift is 
related to its Fourier transform through 
\be
z(t) = \int \hat{z}(f) {\rm e}^{\ii 2\pi f t}\dd f\,,
\ee
with autocorrelation 
\be
\langle \hat{z}(f) \hat{z}^\star(f')\rangle_{\mathcal{A}} = \delta(f-f')S(f)\,.
\ee
Concretely, for autocorrelations in direction space $S(f) = \hat{\mathcal{P}}(f) \mu(0)$, and for multipole space autocorrelations $S(f) = \hat{\mathcal{P}}(f) C_\ell^{\rm HD}$. 
Let us define the scalar product (this corresponds to the ordinary least square described in \cite{Hazboun:2019vhv})
\be\label{DefFunctionScalarProduct1}
\{ f(t) , g(t) \} = \frac{1}{T}\int_{-T/2}^{T/2} f^\star(t) g(t)\dd t\,.
\ee
We notice that a set of functions which are orthogonal for this scalar product are the Legendre polynomials $P_n$ since
\be
\{ P_n(2t/T),P_{n'}(2t/T)\} = \delta_{n n'} \frac{1}{2n+1}\,.
\ee
Using
\be\label{ProjPnexp}
\{ P_n(2t/T), {\rm e}^{\ii 2\pi f t}\} = \ii^n j_n(\pi f T)\,,
\ee
the pulsar redshift is expanded on Legendre polynomials as
\be
z(t) = \sum_n z_n P_n(2t/T)\,,
\ee
with coefficients
\be\label{Defcoeffzn}
z_n \equiv \frac{\{P_n(2t/T), z(t)\}}{\{ P_n(2t/T),P_n(2t/T)\}}= \ii^n (2n+1)\int j_n(\pi f T)\hat{z}(f)\dd f\,.
\ee
Since we want to subtract spurious effects by fitting  a constant, a linear drift and a quadratic shape (corresponding to a pulsar spin-down), the signal after fitting is
\be\label{Defzpol}
z^{\rm pos}(t) = z(t) - \sum_{n=0}^2 z_n P_n(2t/T)\,.
\ee

The total initial power (before subtraction) is
\bea
\langle \{ z(t),z(t)\}\rangle_{\mathcal{A}} &=& \sum_{n=0}^\infty \frac{\langle z_n z_n^\star \rangle_{\mathcal{A}}}{2n+1}\\
&=& \sum_{n=0}^\infty \int (2n+1)j_n^2(\pi f T)S(f)\dd f\,,\nonumber
\eea
which using $\sum_n (2n+1)j_n^2(x)=1$, can be resummed to give 
\be\label{Eqchi2}
\langle \{ z(t), z(t) \}\rangle_{\mathcal{A}}=\int S(f)\dd f\,,
\ee
as expected. For the residual signal after fitting polynomials, we get simply
\begin{align}
\langle\{ z^{\rm pos}(t), z^{\rm pos}(t) \} \rangle_{\mathcal{A}}   &= \langle \{ z(t),z(t)\}\rangle_{\mathcal{A}}   - \sum_{n=0}^2 \frac{\langle z_n z_n^\star \rangle_{\mathcal{A}}  }{2n+1}\nonumber\\
&= \sum_{n=3}^\infty \int (2n+1)j_n^2(\pi f T)S(f)\dd f\,,
\end{align}
or equivalently
\be
\langle \{ z^{\rm pos}(t), z^{\rm pos}(t) \}\rangle_{\mathcal{A}}  =\int S(f){\cal T}(f)\dd f\,,
\ee
where we introduced the transmission, or damping, function 
\be\label{Transfer}
{\cal T}(f) = 1-\sum_{n=0}^2 (2n+1) j_n^2(\pi f T) \,.
\ee
It is easy to check that ${\cal T}(f) \propto f^6 $ when $fT \ll 1$ as found in \cite{Hazboun:2019vhv}.

\begin{figure}
\centering
\includegraphics[width=\columnwidth]{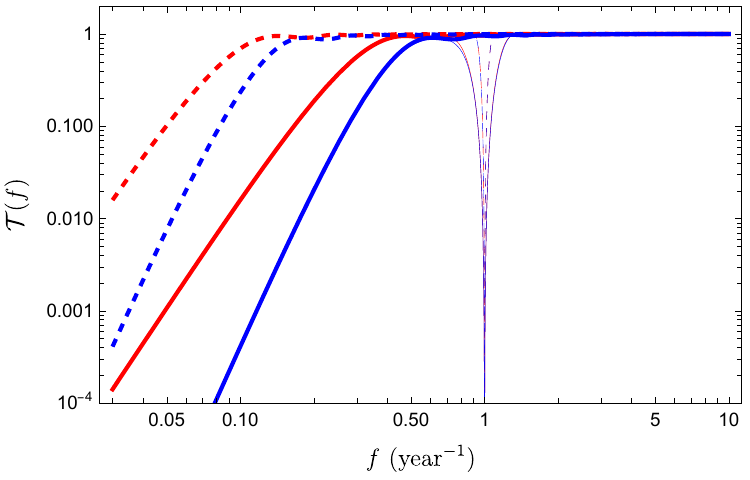}
\caption{Transmission function ${\cal T}$ (thick lines) and ${\cal T}{\cal T}^{\rm earth}$ (thin lines) for $T=3\,\,{\rm years}$ (continuous lines) and $T=10\,\,{\rm years}$ (dashed lines). In the red curves, only $P_0,P_1$ where used in the fitting, whereas for the blue curves $P_2$ was also used. Hence the first case corresponds to the transmission function for redshifts and the second to the transmission function of time residuals.  The slope at low frequency in the former case is $\propto f^4$ and it is $\propto f^6$ in the latter case. The dip at the Earth frequency has a typical width $\Delta f = 1/T$ hence gets narrower for longer observations. This figure is to be compared with Fig. 2 of \cite{Hazboun:2019vhv}, where a similar results is obtained with different filtering methods.}
\label{FigTransfer}
\end{figure}

If we also want to fit an annual modulation with frequency $f_{\rm earth}$, then we only need to fit  ${\rm e}^{\pm\ii 2\pi f_{\rm earth} t}$ along with the previous polynomials. If $f_{\rm earth} \gg 1/T$ then we have approximately $\{P_n(2t/T),{\rm e}^{\pm\ii 2\pi f_{\rm earth} t}\}\simeq 0$  for $n=0,1,2$, hence we can proceed in two steps and fit the annual modulation after the low order polynomials. This amounts to projecting in the space orthogonal to $P_0,P_1,P_2$ first and then in the space orthogonal to ${\rm e}^{\pm\ii 2\pi f_{\rm earth} t}$. After fitting we then get the residual
\be
z^{\rm pos,earth} = z^{\rm pos} - \sum_{s=\pm 1} \{{\rm e}^{\ii 2\pi s f_{\rm earth} t'}, z^{\rm pos}(t')\}{\rm e}^{\ii 2\pi s f_{\rm earth} t}\,,
\ee
where the projections are
\be
\{{\rm e}^{\pm\ii 2\pi f_{\rm earth} t'}, z(t')\} = \int \hat{z}(f) \sinc[\pi(f\mp f_{\rm earth})T]\dd f\,.
\ee
This translates into a damping of the initial power as
\be
\langle \{ z^{\rm pos,earth}(t), z^{\rm pos,earth}(t) \}\rangle_{\mathcal{A}}  \simeq\int S(f){\cal T}(f){\cal T}^{\rm earth}(f)\dd f\,,
\ee
where a product of two transmission functions appears, the transmission function defined in (\ref{Transfer}) and a new damping term  
\be
{\cal T}^{\rm earth}(f) = 1- \sinc^2[\pi(f-f_{\rm earth})T]-\sinc^2[\pi(f+f_{\rm earth})T]\,. 
\ee 
This transmission function is plotted in figure \ref{FigTransfer},  together with the transmission function $\mathcal{T}$ defined in\,(\ref{Transfer}).  Hereafter, we ignore the Earth frequency filtering and consider that we only fit the lowest order polynomials which are essential for the regularity when $f\to0$. Finally, if a given pulsar noise is not stationary, one might weight the functional scalar product~\eqref{DefFunctionScalarProduct1} by replacing $\dd t \to W(t) \dd t$, and one must revise the previous construction by using the polynomials which are orthogonal with this new measure~\cite{Hazboun:2019vhv}. For simplicity we ignore this aspect.

Up to now we assumed a continuous frequency spectrum. However, to make our model realistic,  we need to take into account that the frequency spectrum is discrete with  $f_p=p/T$, and the redshift in Fourier space has the form~\eqref{Defzn}. The Fourier modes after filtering can be found from the real space ones as a projection
\be
\tilde{z}^{\rm pos}(f) = T\{{\rm e}^{\ii 2\pi f t}, z^{\rm pos}(t)\}\,.
\ee
Hence, for $f_p=p/T$, using the decomposition~\eqref{Defzpol}, with coefficients~\eqref{Defcoeffzn} and the projections~\eqref{ProjPnexp}, we get 
\be\label{zpolfn}
\tilde{z}^{\rm pos}(f_p) = \int \dd f \hat{z}(f) \delta^T(f,f_p)\,,
\ee
where the pseudo Dirac is
\begin{align}\label{pseudoDirac}
\delta^T(f,f_p) &= T\sinc[\pi(f-f_p)T] \nonumber\\
&- T\sum_{n=0}^2 (2n+1) j_n(\pi f T) j_n(\pi f_p T)\,.
\end{align}
Since $\delta^T(f,f_p)=\delta^T(-f,-f_p)$ the reality property~\eqref{complexzf} implies $\tilde{z}^{{\rm pos},\star}(f_p) = \tilde{z}^{\rm pos}(-f_p)$.

\begin{figure*}
\centering
\includegraphics[width=\columnwidth]{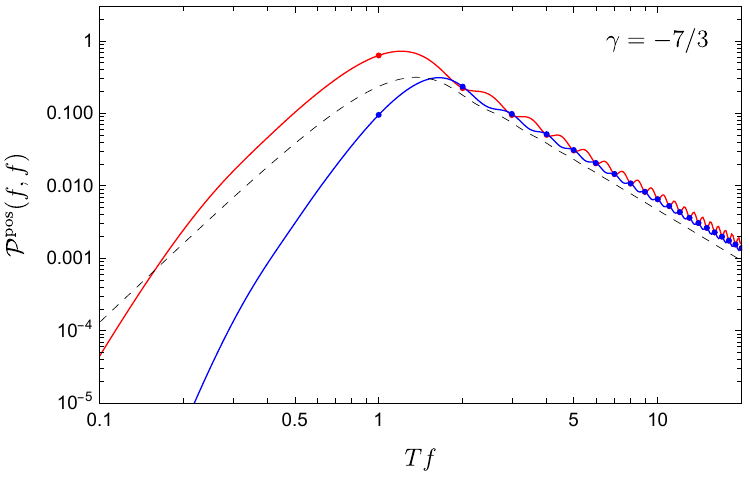}
\includegraphics[width=\columnwidth]{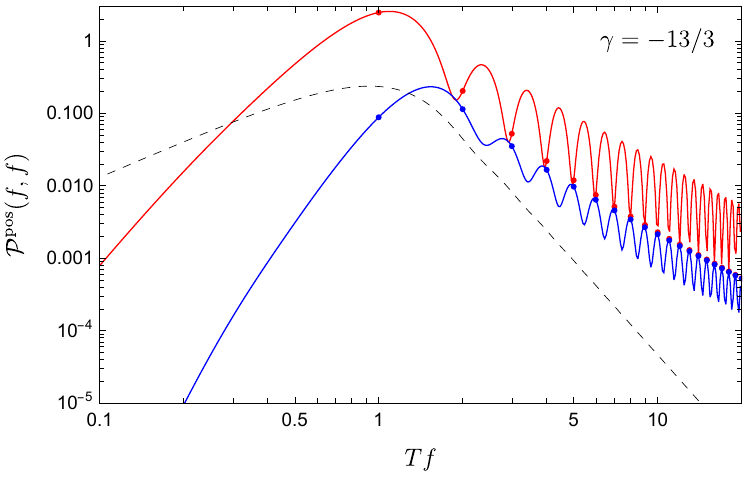}
\caption{Diagonal part of the correlation matrix defined in (\ref{KillerCorrelationlm}) and (\ref{DefSpolf1f2}). In the left panel we consider a spectral index  $\gamma=-7/3$ (corresponding to redshift measurements) whereas the right panel uses $\gamma=-13/3$ (time residual measurement). In red we subtract from redshift/time residuals only $P_0$ and $P_1$ whereas in blue we fit $P_0,P_1,P_2$. The continuous line is ${\cal P}^{\rm pos}(f,f)$ and the dots are the same function evaluated for finite frequency values, i.e. $f=f_p$ with $p=1,2,\dots$ which are the only values of frequency accessible in a realistic observation.  The dashed line is the simple estimate via  ${\cal T}(f) \hat{\cal P}(f)$ which is the initial spectrum multiplied by the transmission function~\eqref{Transfer}. We see that for timing residuals (right panel) a lot of power from the low $f$ spills to larger frequencies, i.e. ${\cal P}^{\rm pos}(f_p,f_p)$ differs substantially from  $\mathcal{T}(f)\hat{\cal P}(f_p)$ for high frequencies. The spectrum of redshift and timing residuals is so red that even a tiny spillover of low frequency over large ones completely modifies the resulting spectrum. }
\label{FigLeakage}
\end{figure*}

\begin{figure}
\centering
\includegraphics[width=\columnwidth]{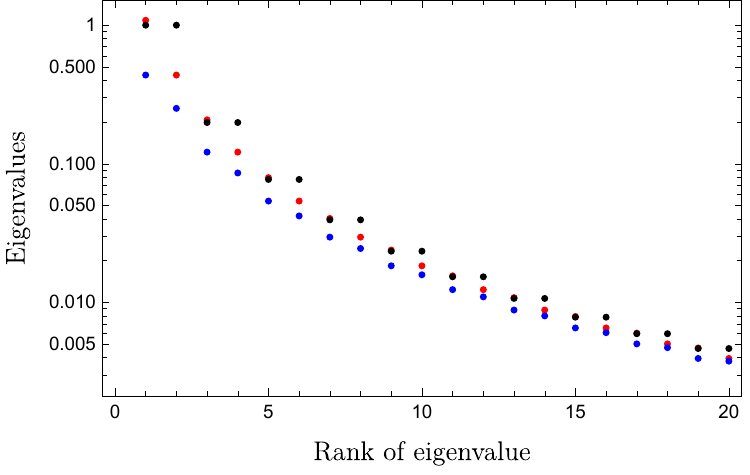}
\caption{Eigenvalues of $T^{-2} {\cal P}^{\rm pos}(f_p,f_q)$  in red (when only $P_0$ and $P_1$ are fitted) and in blue (when $P_2$ is also fitted). The black dots corresponds to the values $T^{-1}\hat{\cal P}(1/T)=T^{-1}\hat{\cal P}(-1/T)$, $T^{-1}\hat{\cal P}(2/T)=T^{-1}\hat{\cal P}(-2/T),\dots$.}
\label{FigEigenValues}
\end{figure}

Restoring the directional dependence in~\eqref{zpolfn},  let us compute the two-point function in direction space. From~\eqref{zpolfn} and \eqref{Correlationdirf}, we find
\be
\langle \tilde{z}^{\rm pos}(\gr{n}_1,f_1) \tilde{z}^{{\rm pos},\star}(\gr{n}_2,f_2)\rangle_{\AmpGW} = \mu(\beta_{12}) {\cal P}^{\rm pos}(f_1,f_2)\,,
\ee
where 
\be\label{DefSpolf1f2}
{\cal P}^{\rm pos}(f_1,f_2) \equiv \int \hat{\cal P}(f) \delta^T(f,f_1)\delta^T(f,f_2)\dd f\,.
\ee
This matrix can be defined for general values of $f_1,f_2$ but it only makes sense for the discrete frequency values and we use the compact notation ${\cal P}^{\rm pos}_{pq} \equiv {\cal P}^{\rm pos}(f_p,f_q)$.  If we correlate multipoles, from~\eqref{Correlationlmf}, we now get 
\be\label{KillerCorrelationlm}
\langle \tilde{z}_{\ell_1 m_1}^{\rm pos}(f_p) \tilde{z}_{\ell_2 m_2}^{{\rm pos},\star}(f_q)\rangle_{\AmpGW} = \delta^{\ell_1}_{\ell_2} \delta^{m_1}_{m_2} C_\ell^{\rm HD} {\cal P}^{\rm pos}_{pq}\,.
\ee
The diagonal part of this correlation matrix ${\cal P}^{\rm pos}_{pp}$ is plotted in Fig.~\ref{FigLeakage}.
We observe that there is a huge leakage from the low frequencies to the high frequencies since at high frequencies ${\cal P}^{\rm pos}(f_p,f_p)$ differs substantially from $\mathcal{T}(f)\hat{\cal P}(f_p)$ that is what one obtains assuming the spectrum of frequencies is continuous (see eq.\,(\ref{Correlationdirf})) and applying a transmission function that accounts for the effect of having filtered out linear drift and constant term. The effect becomes more apparent the more the spectrum is red, i.e. $\gamma \ll -1$. Let us try to develop an intuition of why this is the case, by inspecting the definition of the correlation ${\cal P}^{\rm pos}(f_p,f_p)$ (\ref{DefSpolf1f2}). We consider first only the term $\sinc[\pi(f-f_p)T]$ in the definition of the pseudo-Dirac function $\delta^T(f_1,f_2)$ (see eq.\,(\ref{pseudoDirac})). With the approximation $\sinc(\pi(f-f_p)T)\sim fT/p$ valid at small $f$, and using the parametrization (\ref{gamma}) of the spectral function, the integral (\ref{DefSpolf1f2}) 
defining ${\cal P}^{\rm pos}$ is dominated by a term of the type $\sim T^2 \int \dd f f^{\gamma+2}/(pq)$. For timing residuals this strongly diverges in $f\to 0$ since $\gamma=-13/3$. However with the extra terms of~\eqref{pseudoDirac} coming from the fitting, $\delta^T(f-f_p)\sim (fT)^3/p$ and the integral is regular. However, it is still the low frequency part of the integral~\eqref{DefSpolf1f2}, that is the range in which $|f|T$ is of order unity, which dominates, even for large values of $f_p$ and $f_q$. 
It follows that 
\be
{\cal P}^{\rm pos}_{pq}\propto \frac{T^2}{pq}\,.
\ee
This dominant contribution has very strong off-diagonal correlations.\footnote{Note that we assumed that the biggest contribution to the integral comes from $|f|T$ of order unity, but part of the integral comes from the usual Dirac type contribution of $\sinc[\pi(f-f_p)T]$ and $\sinc[\pi(f-f_q)T]$ which is around $f=f_p$ and $f=f_q$. However, the redder the spectrum is, the less important these contributions are compared to the low $f$ part. } However, in a realistic data analysis, one will maximize the likelihood with respect to the injected parameter, e.g. as in Fig.\,2a of \cite{2013PhRvD..87j4021L}, instead of measuring the diagonal values of ${\cal P}^{\rm pos}_{pq}$. Because of the off-diagonal structure, it is expected that the eigenvalues of the correlation matrix are the relevant quantities to estimate the effect of power removal at low frequencies, and we find that they differ substantially from the diagonal value as the eigenmodes mix substantially the various functions $f_p$. This is illustrated in Fig~\ref{FigEigenValues}: while for large frequencies the eigenvalues of ${\cal P}^{\rm pos}_{pq}$ tend to follow the values of $\hat{\cal P}(f_p)$, the first eigenvalues are lower since power has been removed at low frequency via the fitting process.

To summarize, when working with the postfitting residuals $\tilde{z}^{\rm pos}_{\ell m}(f_p)$ or $\tilde{z}^{\rm pos}(\gr{n},f_p)$,
the correspondence~\eqref{Correspondence2} between Fourier transform (evaluated on the discrete set of frequencies) and Fourier series coefficients is modified to 
\be\label{SposttoP}
{\cal P}^{\rm pos}(f_p,f_q)={\cal P}^{\rm pos}(-f_p,-f_q)\leftrightarrow \delta(f_p-f_q) \hat{\cal P}(f_p)\,.
\ee
In other words, once postfitting residuals are used,  when going from continuous to discrete frequency space the correlation function acquires off-diagonal terms.

\section{Combining different estimators}\label{Weight}

In section~\ref{Quadratic}, we considered quadratic estimators of pulsar redshifts for a given pair of times or frequencies. Exploiting statistical isotropy, directions were combined with a rotational average (pulsar average), which in multipole space corresponds to a summation over $m$. We now need to combine the various times or frequencies to construct estimators that account for the entire dataset, making use of  the underlying stationarity of the GW background. There are two key features of the signal we wish to measure. 
\begin{itemize}
\item First, we aim to verify that the multipole correlations follow the shape of the $ C_\ell^{\rm HD}$ given in equation~\eqref{DefClHD}, which involves measuring these multipoles individually~\cite{NANOGrav:2023gor}. We thus use the parametrization
\be\label{Defclparameters}
C_\ell^{\rm HD} \to c_\ell C_\ell^{\rm HD}
\ee
and wish to constrain the $c_\ell$. In the CMB context, this corresponds to revealing the Sachs-Wolfe plateau~\cite{PeterUzanBook} dependence $ C_\ell^{\rm CMB} \propto 1/[\ell(\ell+1)] $ for a scale-invariant primordial scalar spectrum (see e.g. Fig.~2.4 of \cite{DurrerBook}). 
\item Second, we seek to constrain the global amplitude $ \AmpGWstar $ and, possibly, the spectral index $ \gamma $, by gathering all available information. In the CMB context it is equivalent of constraining the global amplitude $A_s^2$ of scalar fluctuations and the primordial spectrum spectral index $n_S$~\cite{PeterUzanBook,Planck:2018vyg}.
\end{itemize}

PTA work in the regime where the average over the observation time does not allow to sample many realisations of the Universe, contrary to the case of Earth-based and space-based instruments. For a given observation time $T$, the frequency resolution of a given instrument is $\Delta f \simeq 1/T$. Assuming $T$ of the order of three year, one has $\Delta f \simeq 10^{-8}$ Hz, which is about two orders of magnitude below the maximum frequency set by the cadence of pulsar observations. 
Note that  for Earth- and space-based detectors 
the frequency resolution is similar to the PTA one, being set by the observation time, however the frequency we have access to is order of magnitude higher (mHz for space-based and Hz for ground-based instruments). 
Recently, some studies have claimed that cosmic variance imposes an irreducible limitation on our ability to reconstruct the Hellings-Downs correlation \cite{ Allen:2022dzg,Allen:2024bnk, Romano:2023zhb}. In this section, we demonstrate in multipole space that this limitation arises only when different time measurements are combined in a naive manner (see~\cite{Allen:2024uqs} for an analogue result derived in configuration space). However, in section~\ref{SecOptimal} we detail that when appropriate weighting is applied, the impact of the so-called cosmic variance is mitigated. 

\subsection{Nonweighted time average}\label{brute}

Let us introduce the time-averaged version of the equal-time estimator (\ref{estimatorCl}) (observationally taking such an average helps in reducing instrumental noise disturbance) 
\be\label{BruteForce}
C^{Te}_\ell = \frac{1}{T} \int_{-T/2}^{T/2} \dd \eta
C^e_\ell(\eta)\,.
\ee
From~\eqref{CeetaA} and using ${\cal P}(0)=h^2$ where $h^2$ is defined in (\ref{hAllen}), its distributional average is given by 
\be
\langle C^{Te}_\ell \rangle_\AmpGW = h^2 C_\ell^{\rm HD}\,.
\ee
From~\eqref{CovarianceCl}, the variance of the estimator reads 
\bea
&&\langle C^{Te}_\ell C^{Te}_{\ell'} \rangle_\AmpGW -
\langle C^{Te}_\ell \rangle_\AmpGW \langle C^{Te}_{\ell'} \rangle_\AmpGW \nonumber\\
&& =\delta_{\ell \ell'}\frac{2}{2\ell+1}\int_{-T/2}^{T/2} \frac{\dd \eta}{T} \int_{-T/2}^{T/2} \frac{\dd \eta'}{T}
[C_\ell(\eta,\eta')]^2\,.
\eea
To simplify our notation, and following~\cite{Allen:2024bnk}, we introduce 
\bea\label{mathcalh}
 {\calh}^4&\equiv&\int_{-T/2}^{T/2} \frac{\dd \eta}{T} \int_{-T/2}^{T/2} \frac{\dd \eta'}{T} {\cal
  P}^2(\eta-\eta')\\
&=& 4\int \frac{4\pi k_1^2 \dd k_1}{(2\pi)^3} \frac{4\pi k_2^2 \dd
  k_2}{(2\pi)^3}{\AmpGW}(k_1){\AmpGW}(k_2)f_T(k_1,k_2)\,, \nonumber
\eea
where
\begin{align}
f_T(k_1,k_2)&\equiv\iint _{-T/2}^{T/2}  \frac{\dd \eta \dd \eta'}{T^2} \cos[k_1(\eta-\eta')] \cos[k_2(\eta-\eta')]\nonumber\\
&=\frac{1}{2}\left[{\rm sinc}^2((k_1+k_2)T/2) + {\rm sinc}^2((k_1-k_2)T/2)\right]\,,\nonumber
\end{align}
as in (7.14) of \cite{Allen:2024bnk}. 
Then the variance can be written as 
\be\label{VarianceToMean}
\langle C^{Te}_\ell C^{Te}_{\ell'} \rangle_\AmpGW -
\langle C^{Te}_\ell \rangle_\AmpGW \langle C^{Te}_{\ell'} \rangle_\AmpGW =\frac{2 \delta_{\ell \ell'}}{2\ell+1} \frac{\calh^4}{(h^2)^2} (\langle C^{Te}_\ell \rangle_\AmpGW)^2\,.
\ee
The relative size variance to mean is set by the ratio $\calh^4/(h^2)^2$. Indeed for an ideal experiment, one gets for a given $\ell$
\be\label{SNRBrute}
{\rm SNR}^2_\ell=\frac{2 \ell+1}{2}N_f^{\rm eff}\,,\qquad N_f^{\rm eff}\equiv\left(\frac{h^4}{\calh^4}\right)\,.
\ee
Approximating $\sinc(\pi x)\simeq \sinc^2(\pi x) \simeq \delta(x)$ we have roughly
\be\label{hh}
\calh^4 \simeq \frac{1}{T}\int_{\rm obs} [\hat{\cal P}(f)]^2\dd f\,,\quad
h^2 =  \int_{\rm obs} \hat{\cal P}(f)\dd f\,.
\ee

If the frequency integrals in (\ref{hh}) are not sensitive to the low frequency boundary $1/T$, that is if the spectrum is not red, then it is clear $\sqrt{N_f^{\rm eff}}=h^2/\calh^2 \propto \sqrt{T}$ and the SNR improves with time. However for a red spectrum (expected in PTA experiment, see eq.\,(\ref{gamma})) with $\gamma < -1$ there is always more power at low frequency, hence $\calh^4 \propto T^{-2-2\gamma}$ and $h^2 \propto T^{-1-\gamma}$, such that 
\be\label{ratioH}
N_f^{\rm eff} \simeq -\frac{(2+4\gamma)}{(1+\gamma)^2} \,,
\ee
implying that the $\sqrt{T}$ scaling is lost. For the time residual spectrum with $\gamma=-13/3$ as expected from an astrophysical population, $N_f^{\rm eff}$ is of order (slightly larger than) unity and one gets ${\rm SNR}^2_\ell\simeq (2\ell+1)/2$. This fact is at the origin of the cosmic variance problem discussed in \cite{Allen:2022dzg,Allen:2024bnk, Romano:2023zhb}.  

In order to understand the issue more clearly, let us insert the Fourier transform~\eqref{FourierDefinition} in the time average estimator (\ref{BruteForce}), to cast it as
\be
C^{Te}_\ell = \int \dd f \int \dd f' \sinc[\pi(f-f')T] C_\ell^e(f,f')\,.
\ee
Using approximation~\eqref{sincdelta}, this is approximately
\be\label{BruteForceF}
C^{Te}_\ell \simeq \frac{1}{T}\int C^e_\ell(f)  \dd f= \int \sum_m \frac{\hat{z}_{\ell m}(f) \hat{z}^\star_{\ell m}(f)}{(2\ell+1)T}\dd f\,,
\ee
where it is understood that the integral over frequencies is $\int_{\rm obs}$ defined in~\eqref{Correspondence}. 
The integral~\eqref{BruteForceF} is not an optimal way to combine the various frequencies because all frequencies are weighted in the same way. As a consequence, the SNR, which is given by (\ref{SNRBrute}), is not optimal either. Combining frequencies as in (\ref{BruteForceF}) effectively amounts to consider as contributing to the signal only the lowest frequency $1/T$ and a few larger ones, and as explained in appendix~\ref{AppQuadratic} this is the worst choice if one aims to have a high SNR estimator. The effective number of frequencies used with this estimator is precisely $N_f^{\rm eff}$. If we instead aim to constrain the global amplitude $\AmpGWstar$, in the infinite pulsar number we will combine the information from all $\ell$, and even though the frequencies are not optimally combined, this brings and infinite number of angular modes as PTA are sensitive to all $\ell$ in perfect experiments, hence the cosmic variance in that case vanishes (see e.g. section VII.A of \cite{Allen:2022ksj}).

We conclude this section with a remark. As detailed in section~\ref{SecDiscreteFourier}, the integral on frequencies in~\eqref{BruteForceF} is in fact a sum on the discrete functions $f_p=p/T$, and the $\hat{z}_{\ell m}(f)$ are to be replaced by the $\tilde{z}_{\ell m}^{\rm pos}(f_p)$ obtained as in~\eqref{Defzpol} from $\hat{z}_{\ell m}(f)$ by fitting low order polynomials and extracting the Fourier series components. Since the power in low frequencies is reduced by this procedure, one might wonder whether this alleviates the variance problem, giving rise to a higher SNR compared to the case in which the SNR size is set by $N_f^{\rm eff}$ given by \eqref{ratioH}. The estimator~\eqref{BruteForce} for a discrete frequency spectrum and after having filtered low-order polynomial,  becomes
\begin{align}
C_\ell^{Te} &\equiv \frac{1}{T} \int_{-T/2}^{T/2} \sum_{m=-\ell}^\ell \frac{\tilde{z}^{\rm pos}_{\ell m}(\eta)\tilde{z}^{{\rm pos}\star}_{\ell m}(\eta)}{2\ell+1} \dd \eta\nonumber\\
&= \frac{1}{T^2} \sum_{|p|\geq 1} \sum_{m=-\ell}^\ell \frac{\tilde{z}^{\rm pos}_{\ell m}(f_p)\tilde{z}^{{\rm pos}\star}_{\ell m}(f_p)}{2\ell+1}\,,
\end{align}
where the Parseval relation has been used to obtain the second line. Repeating the computation of the signal-to-noise ratio and using~\eqref{KillerCorrelationlm} we find that we must replace in~\eqref{SNRBrute}
\be\label{ratioHbetter}
N_f^{\rm eff} \to \frac{\left(\sum_{|p|\geq 1} {\cal P}^{\rm pos}_{pp}\right)^2}{\sum_{|p|,|q|\geq 1} |{\cal P}^{\rm pos}_{pq} |^2 }\,.
\ee
\begin{figure}
\centering
\includegraphics[width=\columnwidth]{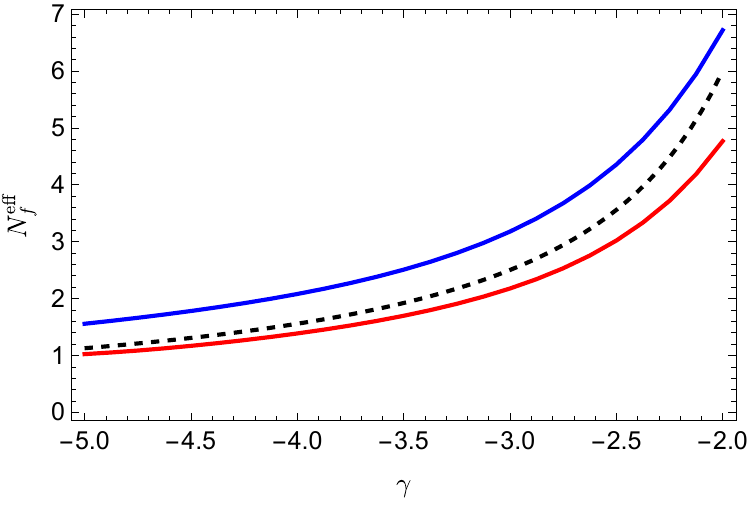}
\caption{The effective number of frequency modes $N_f^{\rm eff}$ as a function of the spectral index $\gamma$. The dashed line is the approximation~\eqref{ratioH}. The colored lines are the estimation from the Fourier series coefficients correlations~\eqref{ratioHbetter}. The red line corresponds to the residual after fitting only $P_0,P_1$ whereas the blue line corresponds to also fitting $P_2$ from the raw signal.}
\label{h2g2}
\end{figure}
A comparison between these two quantities is presented in  Fig.~\ref{h2g2}, which shows that the correct estimation with Fourier series coefficients (\ref{ratioHbetter}) leads to a slightly better SNR ($N_f^{\rm eff}\simeq1.9$ for $\gamma=-13/3$), but overall the approximation~\eqref{ratioH} is not far off. For very negative $\gamma$ we checked that it converges toward the minimum value which is unity.

After reviewing briefly in the next section the various noise contributions that affect the measurements of the angular power spectrum of the pulsar redshift two-point function, we explain in section~\ref{SecOptimal} how cosmic variance can be mitigated by properly weighting the different frequencies so that the effective number of frequencies reaches the maximum. 

\subsection{Noise contributions}

The first noise contribution arises from the fact that, in our theoretical derivation of the correlation function, we assumed an infinite number of pulsar pairs. However, in a realistic experiment, the number of pairs is finite, which introduces a Poisson noise, or shot noise, contribution to the angular power spectrum. We emphasize that this shot noise is due to the discrete distribution of pulsars, not the discreteness of the emitting sources. This effect has been discussed in \cite{Allen:2022dzg, Allen:2024bnk}, where the authors examine the redshift correlation function in real space.\footnote{Although the authors in these references do not explicitly mention pulsar shot noise, they introduce the concept of the total variance of the HD correlation, which accounts for the fact that a realistic experiment involves a finite number of pulsars.} The second noise contribution comes from the noise of individual pulsars, that add up and give a flat in $\ell$-space (shot-noise-like) offset to the angular power spectrum of the pulsar redshift correlation function, usually modeled as a sum of white and red noises~\cite{2015PhRvD..91d4048C}.

We find it more convenient to work in multipole space, where the structure of the estimator is diagonal, as discussed in Section \ref{Quadratic}.

\subsubsection{Noise due to Poissonian pulsar statistics}\label{SectionDefzlmN}

Let us estimate the effect of extracting the $z_{\ell m}$ coefficients from a finite number of pixels, which may not necessarily be optimally placed.  If we have a finite number $N$ of pulsars we get the naive estimation of the multipoles (either functions of time or frequency) as
\be\label{DefzlmN}
z_{\ell m}^N = \frac{4 \pi}{N} \sum_{i=1}^N z(\gr{n}_i) Y_\ell^{m \star}(\gr{n}_i)\,.
\ee
If we perform a double average on realization of the GW background, and on pulsar positions, we get 
\be\label{AverageAP}
\langle z_{\ell m}^N z_{\ell' m'}^{N\star}\rangle_{\AmpGW,{\rm P}} = \delta_\ell^{\ell'} \delta_m^{m'} (C_\ell^{\rm GW} + N_\ell^{\rm P})\,,
\ee
where we have split the total angular power spectrum on the right hand side into a contribution that is always present and a contribution going to zero in the limit $N \rightarrow \infty$, that we identify with a shot noise contribution. It is given by 
\be\label{Cp}
N_\ell^{\rm P} = \frac{1}{N}\left( -C_\ell^{\rm GW} + \sum_L (2L+1)C^{\rm GW}_L\right)\,.
\ee
The computation is quite technical and can be found in appendix \ref{PoissonPulsarAverage}. Note that the second term in the parenthesis is $4\pi\mu^{\rm tot}(\beta_{12}=0)$, that is it is directly proportional to the total variance at zero separation and this contributes equally for all multipoles $N_\ell$. We observe that this  noise contribution (\ref{Cp}) scales as $1/N$ with nearly no $\ell$ dependence as the second term is the dominant one: this is precisely the reason why we decided to call it shot noise, even if it is not $\ell$-independent. To summarize, even in a perfect experiment we will have shot noise from the finite number of pulsars used to evaluate the multipoles.

We observe that the shot noise of the  estimator~\eqref{DefzlmN} can be reduced by introducing proper weights in the sum over pulsars (using a quadrature scheme). Concretely, the estimator can be replaced by
\be\label{BetterEstimate}
z_{\ell m}^N = 4 \pi \sum_{i=1}^N w_i \,z(\gr{n}_i) Y_\ell^{m \star}(\gr{n}_i)\,,
\ee
where the $w_i$ are appropriate weights (the naive estimate~\eqref{DefzlmN} corresponds to $w_i=1/N$). A simple choice consists in extending the trapezoidal Riemann sum to a sphere. For a set of points randomly placed this is equivalent to weighting each point by the fraction of the area encompassed. The fractional area $w_i$ around each pulsar direction can be obtained by generating many random directions and evaluating the fraction which are the closest to each individual pulsar.
Another choice consists in finding the best quadrature coefficients, see e.g.~\cite{KeinerPotts,Drake:2019dys}, such that~\eqref{DefzlmN} is exact when $z(\gr{n})$ can be expressed as a finite sum of spherical harmonics. 

\begin{figure}
\centering
\includegraphics[width=\columnwidth]{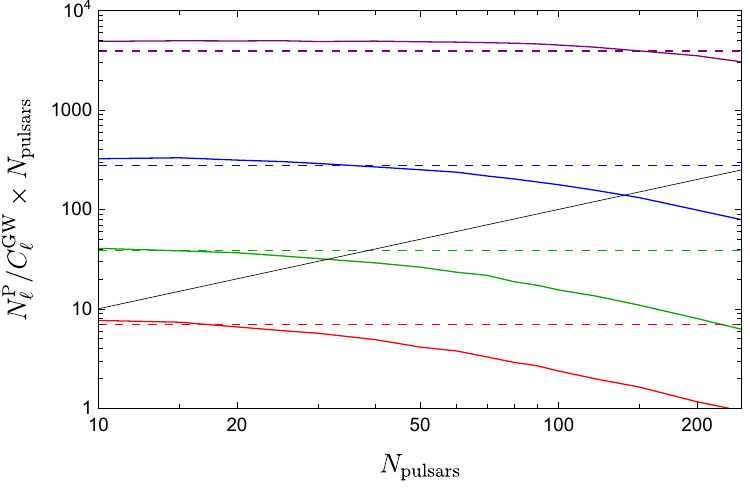}
\caption{Ratio of the shot noise contribution and the correlation function (rescaled by the pulsar number), as a function of the pulsar number. The dashed lines are the pulsar shot noise~\eqref{Cp} of the naive estimate~\eqref{DefzlmN} whereas the continuous lines are obtained with the quadrature weighting~\eqref{BetterEstimate} where the weights are the area of each pulsar direction. The curves are from bottom to top for $\ell=2,3,5,10$ respectively. The thin black line corresponds to $C_\ell^{\rm GW} = N_\ell^{\rm P}$ where the pulsar noise is as large as the HD correlation, hence points located above this line are dominated by pulsar shot noise.  We check that $N_\ell^{\rm P} \propto 1/N$ with the naive estimate whereas it scales as $1/N^2$ when weighting by the area. However we see that the prefactor of this latter scaling increases with $\ell$, i.e. the weighting suppression is more efficient at small multipoles.}
\label{FigPoissonPulsar}
\end{figure}

The pulsar shot noise with the naive method~\eqref{DefzlmN} and with the quadrature scheme~\eqref{BetterEstimate} (using area weights) is represented in Fig.\,\ref{FigPoissonPulsar} and we check that with the quadrature the pulsar shot noise decreases as $1/N^2$.\footnote{This does not mean that we can beat the optimal estimator~\eqref{PerfectTotalSNR} whose ${\rm SNR}^2$ scales as $1/N$, because when estimating the total variance we must add the contributions of all multipoles (all $\ell$ values) and the prefactor in front of the $1/N^2$ scaling increases with $\ell$, as shown in Fig.~\ref{FigPoissonPulsar}.}

\subsubsection{Noise of individual pulsars}

Finally, we need to account for the noise in individual pulsars, either coming from intrinsic irregularity or experimental errors. Each pulsar signal is supplemented with an independent noise $n_i$, and if stationary, the statistics in Fourier space is of the form
\be \label{Ncorsivo}
\langle \hat{n}_i(f) \hat{n}^\star_j(f')\rangle_{{\cal N}} = (4\pi)^{-1}\delta_{ij} \delta(f-f'){\cal N}_i(f)\,.
\ee
where the average on the left-hand side has to be understood as an average of different noise realisations. Different  
 noise contributions will add up to give a shot noise contribution to the angular power spectrum, flat in multipole space. Indeed, replacing $\hat{z}(\gr{n}_i,f) \to \hat{z}(\gr{n}_i,f) + \hat{n}_i(f)$ in~\eqref{DefzlmN}, and repeating the analysis which led to~\eqref{AverageAP}, the average over the GW ensemble, the pulsar directions and the pulsar noises becomes (accounting for weights as in eq.\,(\ref{BetterEstimate}))
\begin{align}\label{AllCovariances}
\langle \hat{z}^N_{\ell m}(f) \hat{z}_{\ell' m'}^{N\star}(f')\rangle_{\AmpGW,{\rm P},{\cal N}} &= \delta_\ell^{\ell'} \delta_m^{m'}\delta(f-f')\nonumber\\
&\quad\left[C_\ell^{\rm GW}(f) + N_\ell^{\rm P}(f) + {\cal N}(f)\right]\,,
\end{align}
where 
\be\label{DefNoisef}
{\cal N}(f) \equiv\sum_{i=1}^N w_i^2 {\cal N}_i(f)\,,
\ee
and where from~\eqref{Correlationlmf} $C_\ell^{\rm GW}(f) = \hat{\cal P}(f)C_\ell^{\rm HD}$. It is understood that the pulsar shot noise now takes the same form as in~\eqref{Cp} if the naive estimate~\eqref{DefzlmN} is used, but with $C_\ell^{\rm}(f) \to C_\ell^{\rm}(f) + {\cal N}(f)$. For a family of pulsars whose noises are of the same order of magnitude, increasing their number reduces the noise since the weights are of order $1/N$ hence ${\cal N}(f) \propto 1/N$.

In the case where the noise~\eqref{DefNoisef} is larger than the pulsar shot noise, we could decide to use weights that minimize this contribution instead of weights that minimize the pulsar shot noise. Under the constraint $\sum_i w_i = 1$, this leads to the choice
\be
w_i = \frac{{\cal N}_i^{-1}}{\sum_{j=1}^N {\cal N}_j^{-1}},
\ee
such that
\be\label{EqCombinationNoise}
{\cal N}^{-1}(f) = \sum_{i=1}^{N} {\cal N}_i^{-1}(f).
\ee
However, with this choice while ${\cal N}(f)$ is minimized, 
the second term of the pulsar shot noise~\eqref{Cp} is worsened by a multiplicative factor
\be
\frac{N \left( \sum_{i=1}^N {\cal N}_i^{-2} \right)}{\left( \sum_{i=1}^N {\cal N}_i^{-1} \right)^2}.
\ee 
This factor is always larger than unity unless all the noises are equal.\footnote{This follows from Cauchy-Schwarz inequality applied to the $N$ dimensional vectors $\gr{A}$ and $\gr{B}$ with components $A_i = 1/{\cal N}_i$ and $B_i=1$. It saturates when $\gr{A}$ is aligned with $\gr{B}$ hence when all noises are equal.} 
In that case, this corresponds to directly using the naive estimate~\eqref{DefzlmN}. Ideally, a good set of weights would take into account both the area of each pixel and its intrinsic noise to minimize the total noise.

\subsection{Optimal weighting}\label{SecOptimal}

In this section, we explain  how cosmic variance can be mitigated by properly weighting the angular power spectrum of pulsars at different frequencies, to build an optimal estimator. We work in multipole space as this makes derivations simpler. We first consider the case of a perfect experiment, for which an optimal quadratic estimator for the correlation amplitude can be built using the procedure detailed in appendix~\ref{AppQuadratic}. When adding instrumental noise, optimal estimators are formally more involved and use the Fisher information matrix. We will show that in the limit of strong signal, the SNR scales as the product of the observation time and cadence of observation, hence there is no fundamental limitation that prevents one from improving it in future observations.   

\subsubsection{Perfect experiment}

Let us first discuss how an optimal estimator can be built in a perfect experiment, that is one in which there is no noise (no intrinsic pulsar noise nor pulsar shot noise). The $\hat{z}^N_{\ell m}(f) = \hat{z}_{\ell m}(f)$  can be viewed as independent measurements where it is understood that we must consider the discrete set of frequencies (one should use the $\tilde{z}_{\ell m}^{\rm pos}(f_p)$). 
Let us use a generalized index $a=(\ell,m,f)$ to make the notation compact. If we know the shape of the correlation of these measurements, and want to measure its global amplitude, as detailed in appendix~\ref{AppQuadratic}, we need to form an optimal estimator as
\be\label{DefQ}
Q \propto \sum_{ab}  z^\star_b C^{-1}_{ba} z_a\,.
\ee
This estimator, applicable to perfect experiments, is formally simpler than the estimator for imperfect experiments reviewed in appendix~\ref{AppQuadraticImperfect}. As a result, it offers a simpler explanation of how the optimal combination of frequencies consistently mitigates cosmic noise.

When detecting the HD correlation, we are first interested in measuring the correlation for each $\ell$ separately~\cite{Agazie:2024qnx}, that is we aim at constraining the $c_\ell$ defined in~\eqref{Defclparameters}, hence we fix $\ell$ and perform the previous sums on $a,b=(m,f)$ only. Using  that the correlation matrix~\eqref{AllCovariances} is diagonal (and ignoring the noise), the optimal estimator (with the continuous frequency approximation) is
\be
Q_\ell = \sum_{m=-\ell}^\ell\int_{\rm obs} \frac{\dd f}{N_f} \frac{\hat{z}_{\ell m}(f)\hat{z}^\star_{\ell m}(f)}{(2\ell+1) \overline{\hat{\cal P}}(f) C_\ell^{\rm HD}}= \int_{\rm obs} \frac{\dd f}{N_f}§\frac{C^e_\ell(f)}{\overline{\hat{\cal P}}(f) C_\ell^{\rm HD}} 
\ee
where $\overline{\hat{\cal P}}(f) \equiv \hat{\cal P}(f)/\AmpGWstar$ is the normalized spectral function and $N_f = 2 T f_{\rm max}$ the number of frequency modes. The average and variance of this optimal estimator are 
\begin{subequations}
\begin{align}
\langle Q_\ell \rangle_{\AmpGW} &=  \AmpGWstar \,,\\
\langle (Q_\ell)^2 \rangle_{\AmpGW}-(\langle Q_\ell \rangle_{\AmpGW})^2 &= \frac{2}{2\ell+1} (2T f_{\rm max})^{-1} \AmpGWstar^2\,,
\end{align}
\end{subequations}
hence the estimator can be used to measure the $c_\ell$. The associated signal-to-noise ratio (SNR) is
\be\label{IdealSNRl}
{\rm SNR}_\ell^2 = \frac{2\ell+1}{2} (2 T f_{\rm max})\,.
\ee

In practice, when considering the Fourier series coefficients $\tilde{z}^{\rm pos}_{\ell m}(f_p)$ instead of the Fourier transform $\hat{z}_{\ell m}(f)$, the correlation is not diagonal anymore, as detailed in section~\ref{CorrelationFourierSeries}. But the argument is the same and one needs only to retain the correct correlation matrix in~\eqref{DefQ} to combine the discrete set of measured frequencies, that is we must replace
\be\label{TozposruleOptimal}
\int_{\rm obs} \frac{\dd f}{N_f} \frac{\hat{z}_{\ell m}(f)\hat{z}^\star_{\ell m}(f)}{\overline{\hat{\cal P}}(f) } \to \frac{1}{N_f}\sum_{pq}  \tilde{z}^{\rm pos}_{\ell m}(f_p)\tilde{z}^{{\rm pos}\star}_{\ell m}(f_q) \bar{S}^{{\rm pos},-1}_{qp}\,,
\ee
where the sums on $p,q$ run on the values $1\leq|p|\leq T f_{\rm max}$ and $1\leq|q|\leq T f_{\rm max}$ and where the normalized covariance matrix $\bar{S}^{{\rm pos}}_{pq}$ is defined from~\eqref{DefSpolf1f2} with $\overline{\hat{\cal P}}(f)$ instead of ${\hat{\cal P}}(f)$. In all cases for a perfect experiment we have $2\ell+1$ modes at a given $\ell$ from the isotropy assumption (all the possible values of $m$), and we have $2 T f_{\rm max}$ frequency modes. The product of these quantities gives the number of modes available and the (square of) SNR is $1/2$ for each of them. 

Now if we aim at constraining the global amplitude $\AmpGWstar$, that is the second goal mentioned at the beginning of section~\ref{Weight},  several $\ell$ can also be optimally combined as 
\be
Q_{\rm opt} \equiv \Sigma_{\rm max}^{-1}\sum_{\ell\geq 2}^{\ell_{\rm max}} (2\ell+1) Q_\ell\,,
\ee
where $\Sigma_{\rm max}\equiv \sum_{\ell\geq 2}^{\ell_{\rm max}} (2\ell+1)=(\ell_{\rm max}+3)(\ell_{\rm max}-1)$ is the number of multipoles available until $\ell_{\rm max}$. This estimator is also obtained directly from~\eqref{DefQ} with $a,b=(\ell,m,f)$. The corresponding SNR with which $\AmpGWstar$ can be measured is
\be
{\rm SNR}^2 = \frac{\Sigma_{\rm max}}{2} (2 T f_{\rm max})\,.
\ee
Since we can only combine $N_{\rm P}$ pulsars, it is expected that $\Sigma_{\rm max} \simeq N_{\rm P}$. 

Let us show that this is indeed the case by building the optimal estimator in real space directly from the pulsar pairs so that we do not have to account for the pulsar shot noise inherently contained in the multipole estimators~\eqref{DefzlmN}. With the results of appendix~\ref{AppQuadratic}, that is using~\eqref{DefQ} with $a,b=(\gr{n},f)$ and the covariance~\eqref{Correlationdirf}, an optimal estimator for $\AmpGWstar$ is
\be\label{Qopt}
Q_{\rm opt}^{\rm real}=\frac{1}{N_{\rm P}} \sum_{i=1}^{N_{\rm P}}\sum_{j=1}^{N_{\rm P}} \int_{\rm obs} \frac{\dd f}{N_f} \frac{\hat{z}(\gr{n}_i,f)\hat{z}^\star(\gr{n}_j,f)}{\overline{\hat{\cal P}}(f)} [\mu(\beta_{ji}]^{-1}\,.
\ee
It is clear using~\eqref{Correlationdirf} that $\langle Q_{\rm opt}^{\rm real} \rangle_{\mathcal{A}} =  \AmpGWstar$ and $\langle (Q_{\rm opt}^{\rm real})^2 \rangle-(\langle Q_{\rm opt}^{\rm real}\rangle)^2 = (2 T f_{\rm max})^{-1} \AmpGWstar^2/N_{\rm P}$. Again, in a rigorous treatment, one should not use the continuous frequency approximation, and one should consider a modification of the type~\eqref{TozposruleOptimal} for pulsar directions, that is
\be\label{Qopt2}
Q_{\rm opt}^{\rm real}=\frac{1}{N_{\rm P} N_f} \sum_{i,j=1}^{N_{\rm P}}\sum_{pq} \tilde{z}^{\rm pos}(\gr{n}_i,f_p)\tilde{z}^{{\rm pos}\star}(\gr{n}_j,f_q) \overline{S}^{{\rm pos},-1}_{qp} [\mu(\beta_{ji}]^{-1}\,,
\ee
but this does not alter these results. Hence, when measuring $\AmpGWstar$ with this optimal estimator 
\be\label{PerfectTotalSNR}
{\rm SNR}^2 = \frac{N_{\rm P}}{2} (2T f_{\rm max}) \,.
\ee
This result is to be compared with (7.12) of \cite{Allen:2022ksj} in which $N_f=2 T f_{\rm max}$ is replaced by $N_f^{\rm eff}=h^4/\calh^4 \ll 2 T f_{\rm max}$. Indeed in this reference while the angular information is optimally combined, frequencies are not. It corresponds to the perfect experiment version of (30) in~\cite{Allen:2024uqs}.

\subsubsection{Imperfect experiment}\label{SecImperfect}

When considering an imperfect experiment with noise, a similar procedure has to be followed to weight different frequencies in an optimal manner. If the noise is exactly known, then a quadratic estimator which takes noise into account can be built~\cite{Allen:2024uqs}. Since the noise influences the signal, the signal in turn affects the estimation of the noise, meaning that the noise can never be fully separated from the signal. Therefore, to accurately measure the correlations induced by the gravitational wave (GW) background, the estimation should involve maximizing the likelihood over the GW parameters, the template-fitting parameters, and the noise parameters. The final step is to analyze the marginalized distribution. With the compact index notation, the Gaussian likelihood which is maximized is of the form
\be\label{LikelihoodGauss}
{\cal L} = {\rm G}(0, C_{ab}; z_a)\,,
\ee
where $G(M_a,C_{ab};z_a)$ is a multidimensional Gaussian of average $M_a$ and total covariance $C_{ab}$ evaluated in the set of values $z_a$. 

The set of variables which is directly related to what is observed is the set of $\hat{z}(\gr{n}_i,f)$, or more precisely the $\tilde{z}^{\rm pos}(\gr{n}_i,f_p)$ (after filtering). As detailed in appendix~\ref{AppQuadraticImperfect}, if we can estimate the noise statistics independently, it becomes possible to build quadratic estimators for imperfect experiments, either to measure the $c_\ell$ of~\eqref{Defclparameters} or to constrain the global amplitude $\AmpGWstar$. However estimating the SNR is very difficult analytically as it requires the inversion of the covariance matrix, and this is not feasible analytically for noisy experiments with finite number of pulsars.

Due to the impossibility of assessing analytically the SNR in imperfect experiments with the optimal quadratic estimators built in appendix~\ref{AppQuadraticImperfect}, let us work directly on the multipole estimators~\eqref{DefzlmN} and let us assume that they follow a Gaussian distribution of the type~\eqref{LikelihoodGauss},  whose covariance~\eqref{AllCovariances} is due to three sources of randomness: the GW signal, the intrinsic noises, and the pulsar directions. Instead of considering the fundamental likelihood of the $\tilde{z}^{\rm pos}(\gr{n}_i,f_p)$ observables, we thus consider a likelihood in the intermediate quantities~\eqref{DefzlmN} built on sums of the former observables. In addition, even though the directions of the pulsars are well measured, we will estimate the pulsar noise induced by the finite number of pulsars, that is the pulsar shot noise, by averaging over the realizations of the pulsar directions. This means that we replace the pulsar noise of a given realization by the average pulsar noise on all possible direction realizations. Hence an estimator which maximizes this hybrid likelihood will not be optimal. Nonetheless its associated SNR will shed light on the SNR of the optimal estimator. Indeed, under these three averages (GW signal, intrinsic noises and pulsar directions) the covariance is diagonal in the $a=(\ell,m,f)$ indices, hence its inversion, which is required to estimate the SNR, is straightforward. In fact, as we work with the Fourier series coefficients after fitting, one should use~\eqref{KillerCorrelationlm} which is not diagonal in frequency space. For simplicity and to make our derivation more transparent, we use the continuous frequency approximation hereafter.
The Fisher matrix provides the SNR with which different parameters can be measured. Focusing on the space of the parameters $c_\ell$, the corresponding Fisher matrix is  diagonal and given by
\be
F_{\ell \ell} \equiv \frac{1}{2} {\rm Tr}\left[ C^{-1}\cdot \partial_{c_\ell} C \cdot C^{-1} \cdot \partial_{c_\ell} C \right]\,,
\ee
where the matrix products and trace are taken in the space $a=(m,f)$ at fixed $\ell$. We find (using the continuous frequency approximation for simplicity)
\be
F_{\ell \ell} = \frac{2\ell+1}{2} T \int_{\rm obs} \left[\frac{C^{\rm GW}_\ell(f)}{c_\ell C^{\rm GW}_\ell(f) + N^{\rm P}_\ell(f) + {\cal N}(f)} \right]^2\dd f \,.
\ee
Then the SNR for the $c_\ell$ is
\bea\label{SNRFisher}
{\rm SNR}^2_\ell &=& \left. c_\ell^2 /F^{-1}_{\ell \ell} \right|_{c_\ell =1}=\left. c_\ell^2 F_{\ell \ell} \right|_{c_\ell =1}\nonumber\\
&=&\frac{2\ell+1}{2} N_f^{\rm eff}(\ell)\,,
\eea
where the $\ell$-dependent effective number of frequencies\footnote{Note that in section~\ref{brute}, we used the term "effective number of frequencies" $N_f^{\text{eff}}$ when an optimal estimator is not used. Here \emph{effective} rather means that the computation accounts for instrumental noise. We always have $N_f^{\rm eff}(\ell) \leq N_f$ with the total number of available frequencies $ N_f= 2 T f_{\rm max}$, and the equality is only reached for perfect experiments with an infinite number of pulsars.} is defined by
\be\label{DefNeffl}
N_f^{\rm eff}(\ell) \equiv \int_{\rm obs} T \left[\frac{C^{\rm GW}_\ell(f)}{C^{\rm GW}_\ell(f) + N^{\rm P}_\ell(f) + {\cal N}(f)} \right]^2\dd f \,.
\ee
We deduce immediately that when increasing observation time or the number of pulsars (the latter improvement decreasing both the pulsar shot noise and the total intrinsic noise) the SNR increases, as already predicted with mock PTA data (see e.g. Fig.~4 in \cite{Nay:2023pwu}). When translated into discrete frequencies with~\eqref{Correspondence}, and retaining only the dominant frequency mode $1/T$ (and its mirror negative frequency $-1/T$), this reduces to the estimation (26) of \cite{Roebber:2016jzl}. When ignoring pulsar shot noise $N_f^{\rm eff}(\ell) \simeq 2 T f_{\rm max}^{\rm eff}(\ell)$ where the effective maximum frequency is the frequency for which $C_\ell^{\rm GW}(f_{\rm max}^{\rm eff}(\ell)) = {\cal N}(f^{\rm eff}_{\rm max}(\ell))$ since for larger frequencies the integrand becomes negligible. If the noise has a spectral index $\gamma_{n}>\gamma$ defined by ${\cal N}(f) \propto (f/f_\star)^{\gamma_n}$ then from the shape~\eqref{DefClHD} we deduce the scaling $f_{\rm max}^{\rm eff} \propto \ell^{4/(\gamma-\gamma_{n})}$. Therefore the effective number of frequencies depends on $\ell$ contrary to what is assumed in the equivalence (7) of~\cite{Allen:2024uqs}. This explains why, even when using many pulsars such that the pulsar shot noise becomes subdominant, the larger multipoles remain the most challenging to constrain~\cite{Agazie:2024qnx}.

We now seek to get a rough estimate of $N_f^{\rm eff}(\ell)$ with the pulsar noise characteristics of~\cite{NANOGrav:2023hde}. For any pulsar in a set, the covariance matrix of timing residuals is written as 
\be
\langle r(\eta_a) r(\eta_b)\rangle=C^{\text{GW}}_{ab}+C^{\cal N}_{ab}\,,
\ee
where we have distinguished a signal and a noise component, with noise covariance 
\be
C^{\cal N}_{ab}=\int^{f_{\text{max}}}_{-f_{\text{max}}} \dd f\,\cos\left(2\pi f (\eta_a-\eta_b)\right) \frac{\mathcal{N}(f)}{4\pi}\,.
\ee
$\mathcal{N}(f)$ is defined in (\ref{Ncorsivo}) for a given pulsar  and  $P_{\cal N}(f)\equiv \mathcal{N}(f)/(4\pi)$ is the two-sided noise spectral density (for timing residuals it has dimensions of inverse cubic frequency).
This quantity is usually parametrized for each pulsar separating a contribution of white noise and a red noise one
\be
P_{\cal N}(f)=\frac{\mathcal{N}(f)}{4\pi}=\sigma^2 \Delta \eta + \frac{1}{2}A^2_{\text{RN}}\left(\frac{f}{1 \text{yr}^{-1}}\right)^{\gamma_{\text{RN}}}\,,
\ee
where $\sigma$ is the standard deviation due to white noise and $\Delta \eta$ is the cadence of observation.
We consider the 68 pulsars in tables 5 and 6 of~\cite{NANOGrav:2023hde}, and take into account red noise only for the 23 pulsars for which it has been measured. The cadence of observation is estimated by dividing the time of observation by the number of epochs for each pulsar, which is a crude estimation of the cadence since there are a few gaps in the observations. The noises, divided by the transmission factor~\eqref{Transfer} computed with their respective observing time, are combined with~\eqref{EqCombinationNoise}, that is using weights which are optimized to reduce the total noise. As a consequence, when estimating the shot noise due to the finite number of pulsars which we also take into account, we must replace $1/N$ by $\sum_{i=1}^{N} w_i^2$ in \eqref{Cp} and this factor is approximately $0.12$ for $f = 1\,{\rm yr}^{-1}$, hence larger than $1/N=1/68$.
Finally, the signal spectrum amplitude at $f_\star = 1 {\rm yr}^{-1}$ is set to ${\cal P}_\star = \tfrac{1}{4}(h_c^{1\,{\rm yr}})^2 \, {\rm yr}$ with $h_c^{1\,{\rm yr}} = 3\times10^{-15}$, taking into account~\eqref{ReplacementTimeResiduals} for time residuals.

Eventually, we find that $N^{\text{eff}}_f(\ell=2)\simeq 4.2$, $N^{\text{eff}}_f(\ell=3) \simeq 0.43$ and $N^{\text{eff}}_f(\ell=4) \simeq 0.06$. This corresponds to ${\rm SNR}_2 \simeq 3.2$ and  ${\rm SNR}_3 \simeq 1.2$ (and a combined ${\rm SNR} \simeq 3.5$) which is in line with the constraints on the HD quadrupole and octupole in Fig.~2 of~\cite{Agazie:2024qnx} and the amplitude detection at $(3.5-4)\sigma$ reported in~\cite{NANOGrav:2023gor}. If we had three times more pulsars (204) with the same observational characteristics we would find $N^{\text{eff}}_f(\ell=2)\simeq 10$ (${\rm SNR}_2 \simeq 5$) and $N^{\text{eff}}_f(\ell=3)\simeq 2.2$ (${\rm SNR}_3 \simeq 2.7$). If the 68 pulsars were all observed during $25$ years, we would get $N^{\text{eff}}_f(\ell=2)\simeq 8.6$ (${\rm SNR}_2 \simeq 4.6$) and $N^{\text{eff}}_f(\ell=3)\simeq 1$ (${\rm SNR}_3 \simeq 1.9$). If we combine both improvements, that is 204 pulsars all monitored during 25 years, we would get $N^{\text{eff}}_f(\ell=2)\simeq 18$ (${\rm SNR}_2 \simeq 6.8$), $N^{\text{eff}}_f(\ell=3)\simeq 4.7$ (${\rm SNR}_3 \simeq 4.0$), and $N^{\text{eff}}_f(\ell=4)\simeq 1$ (${\rm SNR}_4 \simeq 2.1$) and a combined ${\rm SNR}\simeq 8.3$ for the amplitude of the background. Since both the observing time and the number of pulsars monitored is set to increase, we can state that we have entered the era of effective number of frequencies much larger than unity. From the above values we can hope that the octupole of the HD correlation should soon be clearly detected, and in the long run one should be able to detect the hexadecapole ($\ell=4$). The multipole $\ell=5$ would however require around 450 pulsars monitored during 25 years.

We also recover that in a regime where the sources of noise are negligible (ideal experiment), the SNR scales as in~\eqref{IdealSNRl}. Again, this has to be compared to (\ref{SNRBrute}), which is what we obtain when combining naively different frequencies. Taking one residual per month, that is $2 T f_{\text{max}}=1/12\,{\rm yr}^{-1}$, and an observation time of $15$ years, one gets (independent of the spectrum) ${\rm SNR}^2_\ell\sim 180(2\ell+1)/2$, i.e. a factor about $120$ larger than  what one gets by doing (\ref{SNRBrute}) for a spectrum with $\gamma=-13/3$. Notice that both the cadence of observation, which determines $f_{\text{max}}$, and the observation period will improve in the future, further increasing the SNR. 

Summarizing: if we correctly combine the contributions of different frequencies (using optimal weights), in the context of a perfect experiment, we improve the SNR over a single frequency by a factor corresponding to the number of frequency modes. In contrast, by performing a brute-force computation as in Eq.\,(\ref{SNRBrute}), where no weights are applied, we only gain a factor which is an effective number of frequency modes $N_f^{\rm eff}=h^4/\calh^4$, and it can be very small for red spectra (since we effectively give too much weight to a few low frequencies). For an imperfect experiment (i.e. when instrumental noise is accounted for) the counting of the frequency bins is weighted by the noise, and given by the multipole-dependent integral~\eqref{DefNeffl}. Still, it grows with the observation time.
Our simple derivation in multipole space makes it clear that by optimally combining different frequency bins, we can construct an estimator for the HD correlation multipoles whose SNR grows with observation time, even for red spectra. Hence, cosmic variance is actually not an irreducible noise component in this context, contrary to what happens for the CMB where we effectively have access to one single measurement for each direction and there is nothing to combine to improve the SNR of the correlation estimator.

\section{Background anisotropy estimator}\label{Anis}

Up to now we have considered an isotropic bath of gravitons, see  section \ref{Omega}: in this context, the HD correlation is a signature of an isotropic spectral function $\AmpGW(k)$. The anisotropic part $\hat{\AmpGW}(\hat{\gr{k}})$ of the GW background is reflected instead in the off-diagonal correlations in the $(\ell,m)$ space, see (\ref{GeneralCorrelation}). For a finite number of pulsars, the constraints on anisotropy are better placed using a set of adapted functions~\cite{Ali-Haimoud:2020ozu,Ali-Haimoud:2020iyz}. However for a large number of pulsars one can estimate (see section~\ref{SectionDefzlmN}) $z_{\ell m}(f)$, and estimators of anisotropy can be built out of the two-point function of $z_{\ell_1 m_1}(f_1) z_{\ell_2 m_2}(f_2)$, looking at the off-diagonal structure of the correlation. As an illustration, let us restore the anisotropic part of the GW spectrum in this section only. Our starting point is~\eqref{GeneralCorrelation} for the two-point function of pulsar redshift in the presence of anisotropy. Our goal is to build an estimator for the anisotropic coefficient ${\AmpGW}_{LM}$ in~\eqref{GeneralCorrelation}, evaluate its variance (along the lines of \cite{Hotinli:2019tpc}) due to the limited angular resolution of a time-residual map, in the context of a perfect experiment with zero instrumental noise. We will then discuss how this noise propagates into a variance on the angular power spectrum of the background energy density, which is the observable usually used to quantify the deviation from isotropy.   We will show that, even for a perfect experiment and in the absence of other noise components, the fact that the time-residual map has a limited angular resolution, limits our ability to measure the angular power spectrum from the background intensity (energy density) map.

\subsection{Anisotropy in the time residual map: estimator and variance}

We start by isolating the ${\AmpGW}_{LM}$ modes in~\eqref{GeneralCorrelation}  by exploiting the orthonormality of Wigner 3j. Hence we introduce the combinations (contractions of the 2-point function with Wigner 3j symbol) 
\begin{align}\label{DefElllLM}
{\cal E}^{LM}_{\ell_1 \ell_2}(f_1,f_2) &\equiv \sqrt{2L+1}\sum_{m_1 m_2}\troisj{L}{\ell_1}{\ell_2}{M}{m_1}{m_2} \nonumber\\
&\qquad\times \hat{z}_{\ell_1 m_1}(f_1) \hat{z}_{\ell_2 m_2}(f_2)\,,
\end{align}
which satisfies ${\cal E}^{{LM}\,\star}_{\ell_1 \ell_2}(f_1,f_2) = (-1)^M {\cal E}^{L\,-M}_{\ell_1 \ell_2}(-f_1,-f_2)$ and obviously
\be\label{EqSymmetryELM}
{\cal E}^{{LM}}_{\ell_1 \ell_2}(f_1,f_2) = {\cal E}^{{LM}}_{\ell_2 \ell_1}(f_2,f_1)\,.
\ee
The average of~\eqref{DefElllLM} can be computed using~\eqref{GeneralCorrelation} and \eqref{Ortho3j} and is given by 
\begin{align}\label{EqAverageELMl1l2}
\langle {\cal E}^{LM}_{\ell_1 \ell_2}(f_1,f_2) \rangle_\AmpGW &= \delta_{\ell_1+\ell_2+L}^{\rm even}\delta(f_1+f_2) \hat{{\cal P}}(f_1)\nonumber\\
&\quad\frac{\hat{\AmpGW}^\star_{LM}}{\sqrt{4\pi}} F_{\ell_1} F_{\ell_2} \troisj{L}{\ell_1}{\ell_2}{0}{-2}{2}\,. 
\end{align}
Hereafter we only consider parameters for which $\ell_1 + \ell_2 +L $ is even. In the case where the anisotropy is mild, we can assume the covariance is only due to the isotropic part. In that case, using~\eqref{Correlationlmf} and \eqref{Ortho3j}, the covariance takes the simple form  
\begin{align}\label{VarE}
&\langle{\cal E}^{LM}_{\ell_1 \ell_2}(f_1,f_2) {\cal E}^{L'M' \star}_{\ell'_1 \ell'_2}(f'_1,f'_2)\rangle_\AmpGW\\
&-\langle{\cal E}^{LM}_{\ell_1 \ell_2}(f_1,f_2)\rangle_\AmpGW \langle {\cal E}^{L'M' \star}_{\ell'_1 \ell'_2}(f'_1,f'_2)\rangle_\AmpGW\nonumber\\
&\simeq\delta_L^{L'}\delta_M^{M'}\left[\delta_{\ell_1}^{\ell'_1}\delta_{\ell_2}^{\ell'_2} \delta(f_1-f_1')C_{\ell_1}(f_1) \delta(f_2-f_2')C_{\ell_2}(f_2)\right. \nonumber\\
&\qquad\qquad+\left.\delta_{\ell_1}^{\ell'_2}\delta_{\ell_2}^{\ell'_1}\delta(f_1-f_2')C_{\ell_1}(f_1)\delta(f_2-f_1')C_{\ell_2}(f_2)\right]\,.\nonumber
\end{align}

If we want to combine the expressions~\eqref{DefElllLM} to form an optimal estimator for the $\hat{\AmpGW}_{LM}$, we must normalize them and weight them by their inverse variance~\cite{Hotinli:2019tpc}. Explicitly, after having discretized the frequency space (with the $f=0$ mode removed), one builds the following estimator
\begin{equation}
\hat{\AmpGW}^{\ell_1 \ell_2,e}_{LM, f}\equiv\frac{\sqrt{4\pi}}{T F_{\ell_1}F_{\ell_2}}\frac{1}{\hat{\mathcal{P}}(f)}\troisj{L}{\ell_1}{\ell_2}{0}{-2}{2}^{-1}{\cal E}^{LM}_{\ell_1 \ell_2}(f,-f)^{*}\,,
\end{equation}
where on the left-hand side we have explicitly indicated that one can build an estimator for each given value of $\ell_1$ and $\ell_2$. 
From~\eqref{EqAverageELMl1l2} the mean of this object is trivially given by $\hat{\AmpGW}_{LM}$ while, assuming that anisotropies are mild, and using (\ref{VarE}), the variance of this object (we define $\text{Var}(X) \equiv \langle X X^\star \rangle - \langle X \rangle \langle X^\star \rangle $ hence $\sqrt{\text{Var}}$ is the associated error) is given by  
\be\label{VarALM}
\text{Var}\left(\hat{\AmpGW}^{\ell_1 \ell_2,e}_{LM, f}\right)=\frac{4\pi } {(2\ell_1+1)(2\ell_2+1)}\troisj{L}{\ell_1}{\ell_2}{0}{-2}{2}^{-2}\,.
\ee
We recall that all averages here are taken over GW realisations, i.e. we make use of the average $\langle \dots \rangle_{\mathcal{A}}$ defined in section \ref{PulsarCorrelation}.  
We then combine all the estimators $\hat{\AmpGW}^{\ell_1 \ell_2,e}_{LM, f}$ with inverse 
variance weighting to obtain the minimum-variance estimator. However, from~\eqref{EqSymmetryELM} we deduce the symmetry property
\be\label{EqSymmetryALM}
\hat{\AmpGW}^{\ell_1 \ell_2,e}_{LM, f} = \hat{\AmpGW}^{\ell_2 \ell_1,e}_{LM, -f}\,,
\ee
hence in order to avoid double counting when combining estimators we can either restrict to $f>0$ and use all $\ell_1,\ell_2$ combinations, or we can use both $f>0$ and $f<0$ and restrict to $\ell_1 \leq \ell_2$. We make the former choice which is simpler in our case;\footnote{The latter choice is however more adapted when we consider an equal time correlator, as is implicitly the case in~\cite{Hotinli:2019tpc}. The associated estimator is $\hat{\AmpGW}^{\ell_1 \ell_2,e}_{LM} \equiv \sum_f \hat{\AmpGW}^{\ell_1 \ell_2,e}_{LM, f} $ where the sum is on both positive and negative frequencies. When considering the symmetry property~\eqref{EqSymmetryALM} for $\ell_1 = \ell_2$, we deduce that $\text{Var}\left(\hat{\AmpGW}^{\ell \ell,e}_{LM}\right)$ is twice its counterpart with $\ell_1 \neq \ell_2$ due to covariance between positive and negative frequencies, that is $\text{Var}\left(\hat{\AmpGW}^{\ell_1 \ell_2,e}_{LM}\right)$ is given by the r.h.s of \eqref{VarALM} multiplied by $1+\delta_{\ell_1 \ell_2}$ as in Eq.~(26) of \cite{Hotinli:2019tpc}. When combining the estimators we must also restrict to $\ell_1 \leq \ell_2$ so as to avoid double counting. Then, when the variance of the optimal estimator is recast as a sum on all $\ell_1,\ell_2$ combinations, the $1/(1+\delta_{\ell_1 \ell_2})$ is exactly what is needed to recover~\eqref{varA} with $N_f=1$, as is expected when only a single equal time measurement is performed.} and obtain (with an implied restriction of the sums below to $\ell_1+\ell_2+L$ being even)
\be
\hat{\AmpGW}_{LM}^e \equiv \frac{\sum_{\ell_1 \ell_2} \sum_{f>0}\hat{\AmpGW}^{\ell_1 \ell_2,e}_{LM, f} \,\text{Var}^{-1}\left(\hat{\AmpGW}^{\ell_1 \ell_2,e}_{LM, f}\right)}{\sum_{\ell_1 \ell_2}\sum_{f>0} \text{Var}^{-1}\left(\hat{\AmpGW}^{\ell_1 \ell_2,e}_{LM, f}\right)}\,.
\ee
The variance of this estimator is by construction
\be
\text{Var}^{-1}\left(\hat{\AmpGW}^e_{LM}\right)=\sum_{\ell_1 \ell_2}\sum_{f>0} \text{Var}^{-1}\left(\hat{\AmpGW}^{\ell_1 \ell_2,e}_{LM, f}\right)\,,
\ee
or explicitly 
\begin{align}\label{varA}
&\text{Var}^{-1}\left(\hat{\AmpGW}^e_{LM}\right)\\
&=\frac{N_f}{2}\frac{1}{(4\pi)}\sum_{\ell_1 \ell_2 }\delta_{\ell_1+\ell_2+L}^{\text{even}}(2\ell_1+1)(2\ell_2+1) \troisj{L}{\ell_1}{\ell_2}{0}{-2}{2}^2\nonumber\,.
\end{align}
In this equation we assumed that in the absence of instrumental  noise all frequencies equally contribute, and $N_f$ is equal to the number of frequency modes, i.e. $N_f = 2 T f_{\rm max}$. In the presence of noise, a proper weighting should be introduced. This result agrees with what obtained in \cite{Hotinli:2019tpc} (where it is implicitly assumed that $N_f=1$). 
 Notice that all these sums run over $\ell_1, \ell_2 \leq \ell_{\text{max}}$ with $\ell_{\text{max}}\sim \sqrt{N_p}$, where $N_p$ is the number of pulsars in the network, and the multipole measured is constrained by $0\leq L \leq 2 \ell_{\text{max}}$. 

\subsection{Angular power spectrum of energy density: estimator and variance}

We conclude this section by relating this anisotropy estimator to the estimator of the energy density angular power spectrum, which is the observable usually used to quantify anisotropies \cite{Cusin:2017fwz}. 
Let us define $\left(\Omega_{\text{GW}}\right)_{LM} = 8 \pi^2/(3 H_0^2) f^5 \mathcal{A}(2\pi f) \hat{\mathcal{A}}_{LM}$ whose estimator is
 \begin{align}
\left(\Omega_{\text{GW}}\right)^e_{LM}&\equiv\frac{8 \pi^2}{3 H_0^2} f^5 \mathcal{A}(2\pi f) \hat{\mathcal{A}}_{LM}^e\,.
 \end{align}
The mean and variance of this estimator can be trivially computed from the ones of $\hat{\mathcal{A}}_{LM}^e$. 
 An estimator for the angular power spectrum of the anisotropies is given by
 \be
\mathcal{C}_L^{\Omega_{\text{GW}},e} \equiv\frac{1}{2L+1}\sum_M \left({\Omega}_{\text{GW}}\right)^e_{LM}\left( {\Omega}_{\text{GW}}\right)^{e*}_{LM}\,,
\ee
 with variance given by 
\begin{align}
&\text{Var}\left(\mathcal{C}_L^{\Omega_{\text{GW}},e}\right)=\frac{2}{(2 L+1)^2}\\
&\times \sum_M \left[\text{Var}^2\left(\left(\Omega_{\text{GW}}\right)_{LM}^e \right)+ 2 \text{Var}\left(\left(\Omega_{\text{GW}}\right)_{LM}^e \right)|(\Omega_{\text GW})_{LM}|^2\right]\,.\nonumber
 \end{align}
We stress again that the average is an average over GW realizations, and there is a residual stochasticity due to cosmological initial conditions  (see also \cite{Grimm:2024hgi} for a detailed discussion of this double average procedure). In other words, the result above holds for any cosmological  realization. To get predictions, one has to take an additional average over cosmological initial conditions, and using~\eqref{EqOmegaGW} and~\eqref{DefPf} the cosmological average of the variance is
 \begin{align}\label{VarCl}
&\langle \text{Var}\left(\mathcal{C}_L^{\Omega_{\text{GW}},e}\right)\rangle_{\text{cos}}=\frac{2}{2 L+1}\times\\
&\left[\left(\frac{\bar{\Omega}_{\text{GW}}}{4\pi} \right)^4 \text{Var}^2 \left(\hat{\mathcal{A}}_{LM}^e\right) + 2 \left(\frac{\bar{\Omega}_{\text{GW}}}{4\pi} \right)^2 \text{Var}\left(\hat{\mathcal{A}}_{LM}^e\right) \mathcal{C}_L^{\Omega_{\text{GW}}}\right]\,,\nonumber
 \end{align}
 which can be directly computed from (\ref{varA}). It follows that for a given pulsar number $N_p$ (determining the angular resolution of a time-residual map, as $\ell_{\text{max}}\sim \sqrt{N_p}$), (\ref{VarCl}) provides us with the precision with which, in a perfect experiment, the angular power spectrum of the background energy density can be computed from a time-residual map.

\section{The Hellings-Downs curve in the CMB}\label{CMB}

There is a formal analogy between the redshift to a pulsar and the tensor modes contribution to the CMB temperature fluctuations. We want to address the following questions: is there a HD-like correlation in the context of the CMB temperature fluctuations? And what is cosmic variance in this case?
These questions have been addressed in~\cite{Namikawa:2019tax,Ng:2021waj} in the case of a narrow GW background generated within the horizon. We study  them for the GW background generated by amplification of the vacuum fluctuations in the primordial Universe, highlighting the main differences when identifying the HD-like correlation. In this section,  $\eta$ will denote the conformal time associated with a cosmological scale factor $a(\eta)$.

\subsection{GW in cosmology}

The background of gravitational waves generated in the early Universe~\cite{PeterUzanBook} during an inflationary phase, differs from the background considered so far for PTA by three important details.
\begin{itemize}
\item First, the modes satisfy a damped wave equation instead of a simple wave equation, because of Hubble friction. Namely the GW, which are tensor cosmological perturbations, satisfy 
\be
h''_{ij}+ 2 {\cal H}h'_{ij} + k^2 h_{ij} =0\,.
\ee
where a prime indicates a derivative with respect to conformal time $\eta$. In a matter dominated era, the conformal Hubble rate scales as ${\cal H}=a'/a=2/\eta$ and the independent solutions of this equations are 
\be\label{SolCosmo}
\frac{j_1(k \eta)}{(k \eta)}\,,\qquad \frac{n_1(k \eta)}{(k \eta)}\,.\qquad 
\ee
When generalizing (\ref{Defhcos}) to a matter dominated cosmological context,  we thus need to use the
replacement
\be\label{ChangeTimeEvolution}
{\rm e}^{\ii k \eta} \to s(k \eta)\,,\qquad {\rm e}^{-\ii k \eta} \to s^\star(k \eta)\,,
\ee
with the cosmological solution $s$ defined in terms of Hankel functions
\be
s(x) = -\frac{j_1(x) + \ii n_1(x)}{x}\,.
\ee
The solutions~\eqref{SolCosmo} are only valid for a matter dominated Universe. In radiation domination, $\HH = 1/\eta$ and the solutions are simply $j_0(k \eta)$ and $n_0(k \eta)$, and $s(x)= [j_0(k \eta) + \ii n_0(k\eta)]$. 

\item The solution must be regular when outside the Hubble horizon at early times ($k\eta \ll 1$). Hence since only $j_1(x)/x$ is regular for the matter era solution (resp. $j_0(x)$ for the radiation era solution), the initial conditions for a mode $\gr{k}$ and $-\gr{k}$ must be related to eliminate $n_1(k \eta)/(k\eta)$ (resp. $n_0(k\eta)$). It follows that the initial conditions are not set with random phases in cosmology. This translates into a modification of the statistics of the modes as we must have
\be\label{link}
h^{(m)\star}(\gr{k}) = h^{(m)}(-\gr{k})\,,
\ee
such that
\be\label{Defhcos3}
h^{(m)}_{\rm cos}(\gr{k},\eta) =h^{(m)}(\gr{k}) {\rm Re}[s(k \eta)]\,.
\ee
The link~\eqref{link} implies that the statistics of the mode is modified, and instead of~\eqref{Defstath}, they must obey
\begin{align}\label{Defstathcosmo}
\quad\,\,\langle h^{(m)}(\gr{k}) h^{(m')\star}(\gr{k}')\rangle_{\mathcal{A}} =&
(2\pi)^3 \delta_{m m'} \delta(\gr{k}-\gr{k}'){\AmpGW}(k)\,,\nonumber\\
\quad\,\,\langle h^{(m)}(\gr{k}) h^{(m')}(\gr{k}')\rangle_\AmpGW=&(2\pi)^3 \delta_{m
  m'} \delta(\gr{k}+\gr{k}'){\AmpGW}(k)\,.
\end{align}
The statistics of $h_{\rm cos}^{(m)}$ is also altered compared to~\eqref{Stathcos}, and it becomes
\begin{align}
&\quad\,\langle h^{(m)}_{\rm cos}(\gr{k},\eta) h^{(m')\star}_{\rm cos}(\gr{k}',\eta')\rangle_{\mathcal{A}} =\delta_{m m'}(2\pi)^3 \delta(\gr{k}-\gr{k}') \\
&\qquad\qquad\qquad\qquad\qquad\times{\AmpGW}(k) {\rm
  Re}(s(k \eta)) {\rm  Re}(s^\star(k \eta'))\,. \nonumber
\end{align}

\item The general expression for the redshift~\eqref{Generalz} and its splitting between an observer and a pulsar term remains valid, if we replace the pulsar distance $\chi$ by the last-scattering surface (LSS) distance $\chi_{\rm LSS}$ (and the integrals~\eqref{Defzozp} do not extend to infinity but go up to the big-bang distance). However a major difference, detailed hereafter, is that it is generally not correct to neglect the pulsar term by assuming $k \chi_{\rm lSS} \gg 1$ since in cosmology the spectrum is nearly scale invariant, and large modes which evade this condition, contribute substantially. 

\end{itemize}

The difference in statistics, that is namely the replacement of~\eqref{Defstath} with~\eqref{Defstathcosmo} when the time evolution is modified according to~\eqref{ChangeTimeEvolution}, has an important consequence for the equivalent of the HD correlation in the CMB correlations. Apart from the argument of regularity on super-Hubble scales, the origin of the difference is rooted in the quantum origin of fluctuations. A correlation of the type~\eqref{Defstath} might be of classical origin, due to independence of a large number of binary sources. But equivalently it could be of quantum origin as we could promote, as in (2) of~\cite{Ng:2021waj}, the tensor perturbations to a quantum field by promoting the mode functions to operators (note that in this discussion a hat refers to an operator and not to a Fourier transform)
\be
h^{(m)}(\gr{k}) \to \hat{h}^{(m)}(\gr{k}) = \hat{a}_m(\gr{k}) h^{(m)}(\gr{k}) \,,
\ee
where the creation and annihilation operators satisfy the usual commutation rules
\be
[\hat{a}_m(\gr{k}),\hat{a}^\dagger_{m'}(\gr{k}')]=(2\pi)^3 \delta_{m m'} \delta^3(\gr{k}-\gr{k}')\,.
\ee
The statistics~\eqref{Defstath} is then interpreted as an average in the quantum vacuum using $\hat{a}_m(\gr{k})|0\rangle=0$, and defining $2 \AmpGW(\gr{k}) = |h^{(2)}(\gr{k})|^2=|h^{(-2)}(\gr{k})|^2$ for an unpolarized background.
However tensor fluctuations generated during inflation are assumed to be generated in the Bunch-Davies vacuum when asymptotically inside the Hubble radius, and then this leads to frozen solutions when their scale exits the Hubble radius. It is this frozen value which serves as an initial state for the subsequent standard cosmological model, but the statistical properties are then inherited from the inflationary phase. The important consequence is that in inflation this translates for the initial conditions of the standard model into the operators 
\be
\hat{h}^{(m)}(\gr{k}) =\hat{A}_m(\gr{k}) h^{(m)}(\gr{k})\,,\quad \hat{A}_m(\gr{k}) \equiv \frac{\hat{a}_m(\gr{k}) + \hat{a}^\dagger_m(-\gr{k})}{\sqrt{2}}
\ee
and the mode functions must satisfy \eqref{link}. Although of quantum origin, the tensor modes behave classically since $[\hat{A}_m(\gr{k}),\hat{A}_{m'}(\gr{k}')]=0$, and we find the modified statistics~\eqref{Defstathcosmo}. Physically, the tensor modes of inflationary origin are a superposition of counter propagating waves forming a standing wave, hence a mode with wavevector $\gr{k}$ and the one with opposite wavevector $-\gr{k}$ are not statistically independent. This is very different from a classical GW background from binary sources, where the binary sources in one directions have absolutely no reason to be correlated with the ones in the opposite direction. As we detail hereafter this triggers a small difference for the HD correlation in the CMB.

\subsection{Redshift multipole transfer function}

Similarly to the PTA case, we can decompose CMB redshifts in a basis of spherical harmonics~\cite{DurrerBook} as~\eqref{MultDec}
\be\label{MultDec2}
z^{\rm CMB}(\gr{n}) = \int \frac{\dd^3 \gr{k}}{(2\pi)^3}
\sum_{m=\pm 2} \sum_\ell z^{\rm CMB}_{\ell m}(\gr{k},\eta_0) R_{\hat{\gr{k}}}[Y_\ell^m](\gr{n})\,.
\ee
Observe that what is measured in CMB observations is the temperature: on this context $z^{\rm CMB}(\gr{n})$ stands for $-\delta T^{\rm CMB}(\gr{n})/ \bar{T}^{\rm CMB}$ as the GW affect energies, and thus temperatures, via the integrated Sachs-Wolfe effect. 
This can be split into the sum of an observer and a pulsar term, where the pulsar term is in fact a LSS term as we must use $\chi=\chi_{\rm LSS}$ in~\eqref{Generalz}. Note that  $\eta$ cannot go infinitely in the past but is restricted to $0\leq
\eta \leq \eta_0$.

The multipole transfer function is defined in Boltzmann codes such as CLASS~\cite{CLASSI,CLASSII}, as the ratio
\be\label{DefTransferlm}
{\cal T}_{\ell m}(\gr{k}) \equiv \frac{z^{\rm CMB}_{\ell m}(\gr{k},\eta_0)}{h^{(m)}(\gr{k})}\,.
\ee
A key difference with the background measured by PTA, is that we must replace the evolution of the modes with the damped solution according to~\eqref{ChangeTimeEvolution}. The general form of the transfer requires the resolution of the Boltzmann hierarchy, usually with a line of sight method. However we can estimate rather easily the observer contribution by approximating the time evolution for large modes (locally, near the observer) by
\be
s(k\eta ) \sim \frac{{\rm e}^{\ii k \eta}}{(k \eta)^2} \simeq \frac{{\rm e}^{\ii k \eta}}{(k \eta_0)^2}\,,
\ee
that is we keep the oscillatory behaviour but we consider only the final value for the damping which has occurred due to Hubble friction. Then we can just follow the same steps as for a GW background observed with PTA. The final result~\eqref{zofH2} is then only altered by this damping $1/(k \eta_0)^2$ but can be written in a simpler form due to the relation~\eqref{link}. Eventually we find that the observer contribution can be approximately written as 
\be\label{zt}
z_{\ell m}^{{\rm CMB},{\rm o}}(\gr{k},\eta_0) \simeq (-\ii)^\ell \frac{\cos\left(k \eta_0 + \ell
  \frac{\pi}{2}\right)}{(k \eta_0)^2} h^{(m)}(\gr{k}) F^{\rm o}_\ell\,,
\ee
with the same $F^{\rm o}_\ell$ as in~\eqref{Fellcompact}.
The observer part of the transfer function ${\cal T}^{\rm o}_{\ell m}$ can be read directly from this result~(\ref{zt}) recalling its definition~\eqref{DefTransferlm}, and it depends only on $k$.

\begin{figure}
\centering
\includegraphics[width=\columnwidth]{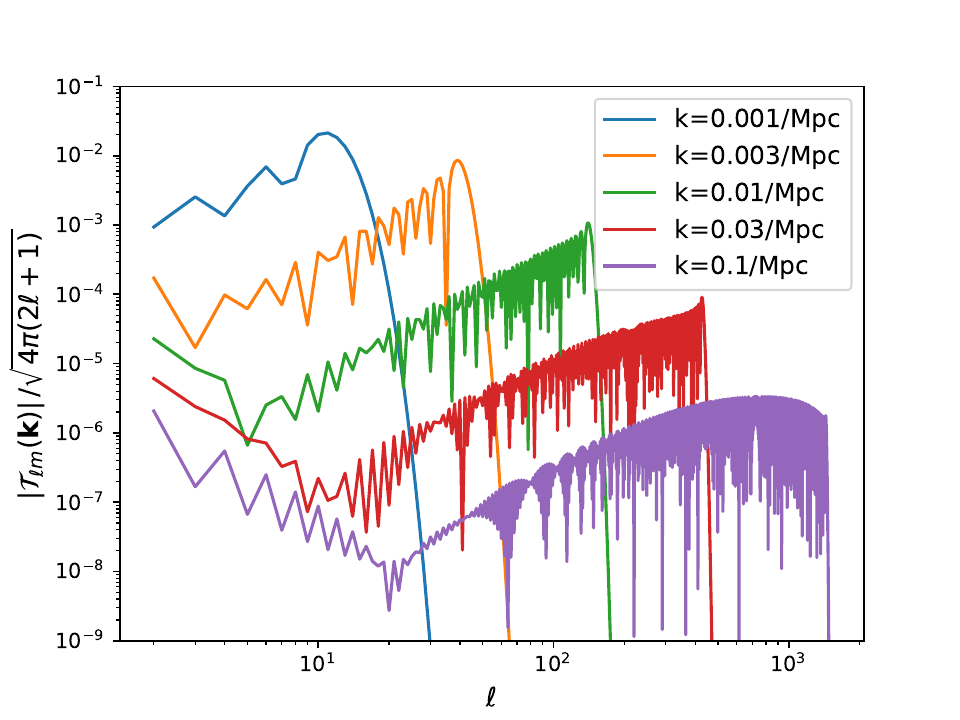}
\caption{Transfer function~\eqref{DefTransferlm} for several modes and $m=2$ obtained from a modified version of the Boltzmann solver CLASS~\cite{CLASSI,CLASSII}. For large modes ($k>0.01/{\rm Mpc}$ approximately) and low $\ell$ we see an observer term with its characteristic decay $\propto (\ell-2)!/(\ell+2)!$ with a modulation coming from the factor $\cos\left(k \eta_0 + \ell
  \frac{\pi}{2}\right)$, which is then dominated by the pulsar (also known as the LSS term) for large values of $\ell$. It is the observer part of the transfer function which is responsible for the appearance of a HD-like correlation in the CMB correlation function. The background cosmology parameters are $\Omega_\Lambda=0$, $\tau_{\rm reio}=0$ and $h_0=0.4$.}
\label{FigCosmo1}
\end{figure}

The multipole transfer function~\eqref{DefTransferlm} is plotted in Fig.~\ref{FigCosmo1}, for different values of $k$. The modulation $\cos\left(k \eta_0 + \ell\frac{\pi}{2}\right)$, is the main difference with a classical GW background measured by PTA. Since cosmological GW are standing waves, they have nodes and maxima, and depending on where these are located with respect to the observer, this affects how the integrated SW effect is split in the various multipoles. It is fundamentally due to using the statistics~\eqref{Defstathcosmo} instead of~\eqref{Defstath} as required by the origin of primordial GW modes. This feature is absent for backgrounds generated within inside the horizon in a later phase, hence the modulation is not seen in the Fig.~1 of~\cite{Ng:2021waj} even though it corresponds to sharply peaked spectra. Apart from this modulation we see that the rescaled transfer function ${\cal T}_{\ell m}/\sqrt{4\pi(2\ell+1)}$  scales like the observer term $F_\ell^{\rm o}/\sqrt{4\pi(2\ell+1)} \propto \sqrt{(\ell-2)!/(\ell+2)!}$ for small $\ell$ and large modes and then it gets dominated by a pulsar term (which is a last scattering term in cosmology).

We also notice that this pulsar term, which has
roughly a slope $\propto \ell^2$ stops when $\ell \geq k \eta_0$ as expected since the distance to the last scattering surface is approximately $\eta_0$. 
For small modes $k$, we are not in the regime where the wavelength is much shorter than the distance to the last scattering surface (playing the counterpart of pulsars in our analogy) and the observer term is always subdominant compared to the pulsar term.

\subsection{Correlation function }

In full analogy with the PTA case, we define a redshift (or CMB temperature fluctuations) correlation function or its equivalent in multipolar space when decomposing $z^{\rm CMB}(\gr{n}) = \sum_{\ell m}z^{\rm CMB}_{\ell m} Y_\ell^m(\gr{n})$ which is 
\be
\langle z_{\ell m}^{\rm CMB} z^{{\rm CMB}\star}_{\ell' m'} \rangle_{\AmpGW} = \delta_{\ell \ell'} \delta_{m m'} C^{\rm CMB}_\ell\,.
\ee
All time dependence has been eliminated compared to the PTA case because an observation of the CMB sky amounts to a single observation given the extremely slow evolution of the CMB (over a scale of millions of years) compared with the observation time. The expression of the correlation in multipole space (the angular power spectrum) is rather simple in terms of the transfer function
\be
C^{\rm CMB}_\ell = \int \frac{k^2 \dd k}{2 \pi^2} \AmpGW(k) \sum_{m=\pm 2} |{\cal T}_{\ell m}(k)|^2\,.
\ee

\begin{figure}
\centering
\includegraphics[width=\columnwidth]{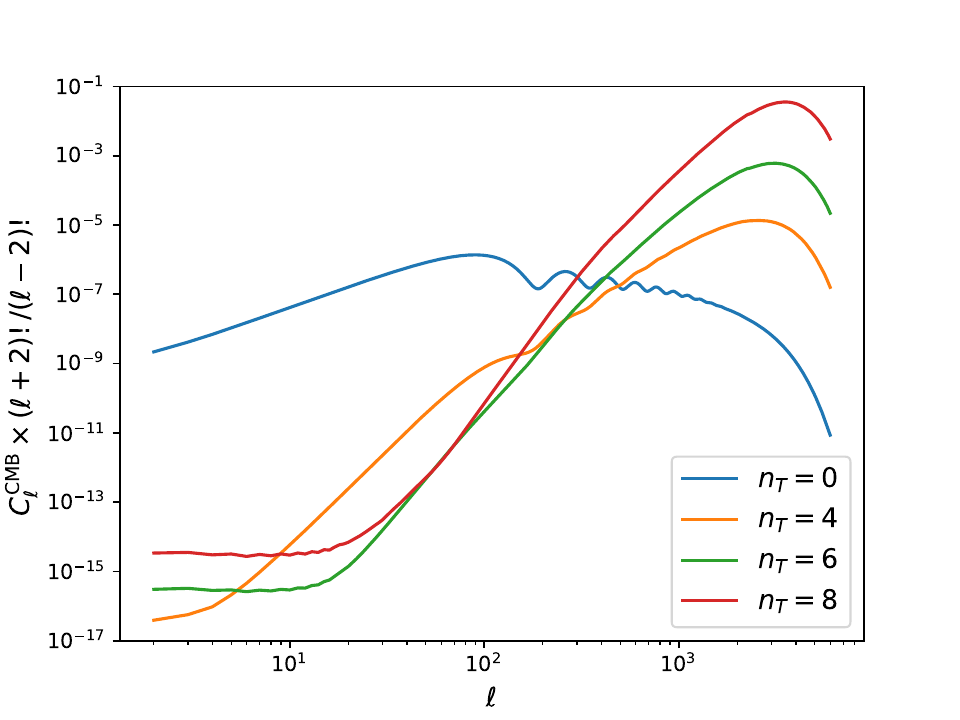}
\caption{Multipoles of the CMB fluctuations from tensor modes obtained from CLASS~\cite{CLASSI,CLASSII}. The background cosmology parameters are $\Omega_\Lambda=0$, $\tau_{\rm reio}=0$, $h_0=0.4$, $A_s=2.1\,\times10^{-9}$ and unit tensor to scalar ratio. We plot $(\ell+2)!/(\ell-2)! C_\ell$ so that a HD correlation appears as a plateau, which is visible for $n_T=6$ and $n_T=8$.}
\label{FigCosmo2}
\end{figure}

In Fig.~\ref{FigCosmo2} we plot the $C^{\rm CMB}_\ell$ for various spectral indices defined in the cosmological context as $\AmpGW(k) \propto (k/k_\star)^{n_T-3}$. We see that for the scale invariant cosmological spectrum with
  $n_T=0$, we do not see the Hellings-Downs correlation as large angular scales are dominated by small modes which exhibit only the pulsar contribution (see Fig.\,\ref{FigCosmo1}). However if we consider a very blue spectrum, then these small modes are subdominant and even the large scale angular correlations are dominated by the observer Hellings-Downs correlation of the large modes. As we plot  $(\ell+2)!/(\ell-2)! C^{\rm CMB}_\ell$, the HD correlation appears as a plateau. The modulation in $\cos\left(k \eta_0 + \ell  \frac{\pi}{2}\right)$ of the transfer is not visible because when integrating on $k$ it is averaged over since the modes which contribute to the HD plateau satisfy $k \eta_0 \gg1$. In order to have a large scale HD plateau we need at least $n_T >4$ so that the primordial conditions can compensate the subsequent damping $[1/(k\eta_0)^2]^2$ due to Hubble friction.

Finally, as done in the PTA context,  we can build an estimator for the redshift correlations as (we stress that for the sake of the comparison with PTA, we are here assuming  unrealistically that there are no scalar fluctuations in the CMB anisotropies) $C_\ell^{{\rm CMB},e} = \sum_{m}z^{\rm CMB}_{\ell m}(\eta_0)z^{{\rm CMB}\star}_{\ell m}(\eta_0)/(2\ell+1)$. Since in this context there are no different time of observations, nor frequency bins, this is the only estimator one can build. The variance is estimated in the usual way with Wick's theorem and the signal to noise reduces to ${\rm SNR}_\ell^2 = (2\ell+1)/2$ which is the usual CMB cosmic variance, irreducible, since there is no way to mitigate it, contrary to the PTA case.\footnote{When we say that nothing can be done to mitigate cosmic variance in the context of the CMB, we mean that we will never achieve the sensitivity required to directly observe the timescale of GWs that influence the CMB. If we could observe over a timescale of one million years, techniques like those used in PTA might allow us to mitigate cosmic variance. In other words, when we refer to a \emph{perfect experiment}, we are implying an experiment that can be conducted within a humanly feasible timescale.}

\section{Discussion and Conclusions}\label{discussions}

In this article, we critically examined the derivation of the cosmic variance associated with the HD correlation curve and explored its implications for PTA observations. By working in real and multipole space, we showed that an optimal estimator for the angular power spectrum of the HD correlation can be constructed. This estimator's signal-to-noise ratio improves with increased observation time and increasing the cadence of observations, even in the presence of a very red spectrum (see eq.\,(\ref{IdealSNRl})), hence it will keep on improving in the future. The key to this improvement lies in the introduction of an optimal weighting scheme, which reduces the impact of cosmic variance by mitigating the contribution of low-frequency components in the presence of a red spectrum. We also considered the effect of instrumental noise, we accounted for pulsar shot noise, and discussed a weighting procedure to minimize the impact of these noise components. Additionally, we provided a detailed analysis of how the optimal estimator is modified when accounting for the fact that, in a real experiment, time residuals must be fitted with a deterministic template to subtract the effects of a linear drift and a quadratic term. We showed that this filtering procedure does not alter the final conclusion regarding the SNR in the strong signal limit.

In this regard, the term \emph{cosmic
variance} is somewhat of an abuse of language in the PTA context, as it suggests that nothing can be done to reduce its impact, even with perfect experiments, and the authors of~\cite{Nay:2023pwu} have advocated to rename it {\it sample variance}.  This stands in contrast to the situation in CMB observations, where even in a perfect experiment, more observation time cannot reduce cosmic variance as the CMB signal is essentially static on any human time scale. In the CMB context, cosmic variance is indeed an irreducible limitation. 

To make this point clear, we developed a formal analogy between the redshift to a pulsar and the tensor modes contribution to the redshift to the large scattering surface. We showed that two contributions can be distinguished in the CMB redshift: an observer term which dominates small multipoles and a pulsar term which becomes important at higher multipoles (smaller scales). The transition scale depends on the mode $k$ under consideration (the observer term becoming visible at small $\ell$ for $k>0.01/$Mpc approximately). We computed the redshift correlation function, and we showed that its multipolar decomposition (i.e. the angular power spectrum) exhibits a HD-like behaviour at small $\ell$ for a sufficiently blue spectrum with $n_T>4$, while for redder spectra small modes, which exhibit only the pulsar contribution, dominate the spectrum hence the HD-behaviour is lost. Since the tensor modes have a subdominant effect compared to the scalar modes, and since in standard inflationary models the tensor spectrum is nearly scale-invariant, the HD correlation in the CMB will realistically never be observed. We recovered that any estimator for the angular power spectrum is cosmic-variance limited in this context since no time weighting is possible.

We conclude with two remarks. First, we stress that in PTA data analysis techniques, estimators for the correlation function are constructed using optimal weightings, as discussed, for example, in \cite{Maggiore:2018sht}. However, since several recent theoretical papers have referred to cosmic variance as a fundamental limitation for PTA observations, we believe it is important to avoid further confusion by providing a pedagogical derivation of the weighting procedure used to optimally combine frequency bins. To this end, we also draw a parallel with the CMB case, where such a weighting procedure is not possible, even though a HD correlation os present for very blue primordial spectra. 
While working in multipole space offers a direct parallel with CMB observables, it is also the simplest representation for HD analysis. Indeed, we have shown that in multipole space, the covariance matrix is diagonal, significantly simplifying computations. This suggests that the multipole space framework, rather than real space, is the most effective approach for studying HD correlations in PTA observations.

\begin{center}
\textbf{Acknowledgements}
\end{center}

 We thank Camille Bonvin, Nastassia Grimm, Guillaume Faye, Marc Kamionkowski and Martin Pijnenburg for discussions.  G.C. acknowledges support from CNRS and from the Swiss National Science Foundation (Ambizione grant, ``Gravitational wave propagation in the clustered universe").

\appendix

\section{Rotations and spherical harmonics}\label{AppRotations}

We gather all relations related to spherical harmonics and rotations which are used in this article. We use the conventions of \cite{Challinor:2005jy} to define spherical harmonics and the Wigner rotation coefficients, which are defined by
\be
Y_\ell^m = \langle {\gr{n}} | \ell m \rangle\,,\qquad
D^\ell_{m' m} = \langle \ell m' | R | \ell m \rangle
\ee
where an active rotation is parameterized with Euler angles as
\be\label{DefEulerAngles}
R(\phi,\theta,\gamma) = R_z(\phi) \cdot R_y (\theta) \cdot R_z (\gamma)\,.
\ee
The active rotation of spherical harmonics is defined by the inverse passive transformation of the direction since 
\be\label{Ylmrotation}
R[Y_\ell^m](\gr{n}) \equiv \langle {\gr{n}}| R | \ell m \rangle=  Y_\ell^m(R^{-1} \cdot \gr{n})=D^\ell_{Mm}(R) Y_\ell^M(\gr{n})\,.
\ee
The Wigner coefficients are related to the small Wigner coefficients $d^\ell_{m m'}$ assocoiated with a $R_y$ rotation only by
\be
D^\ell_{m m'}(\phi,\theta,\gamma) = {\rm e}^{-\ii m \phi} d^\ell_{m
  m'}(\theta) {\rm e}^{-\ii m' \gamma}\,.
\ee
In particular when $m=\pm2$ and $m'\pm2$ and noting $\cos \theta = \beta$, the small Wigner coefficients are expressed  in terms of Jacobi polynomials as
\be\label{dlmmJacobi}
d^\ell_{m' m}(\theta)=\left(\frac{1+\beta}{2}\right)^\frac{|m+m'|}{2}\left(\frac{1-\beta}{2}\right)^\frac{|m-m'|}{2}P_{\ell-2}^{(|m-m'|,|m+m'|)}(\beta)\,,
\ee
which satisfy from their orthonormality relation the particular relations
\bea\label{JacobiMagic}
\frac{2}{2\ell+1} &=& \int_{-1}^1 \dd x \left(\frac{1+x}{2}\right)^2 \left(P_{\ell-2}^{(0,4)}(x)\right)^2 \nonumber\\
&=&\int_{-1}^1 \dd x \left(\frac{1-x}{2}\right)^2 \left(P_{\ell-2}^{(4,0)}(x)\right)^2 \,.
\eea

The Wigner coefficients are also related to the spin-weighted spherical harmonics through
\be\label{DlmmYslm}
D^\ell_{ms} (\phi,\theta,\gamma) =  {}_s Y_\ell^{-m}(\theta,\phi)
(-1)^m \sqrt{\frac{4\pi}{2\ell+1}} {\rm e}^{-\ii s \gamma}\,.
\ee
The most useful properties are
\be\label{UsefulDproperty1}
D^\ell_{m' m}(R^{-1}) = D^{\ell \star}_{m m'}(R) = (-1)^{m-m'}D^{\ell}_{-m \,-m'}(R)\,,
\ee
\be\label{Dofoppositedirection}
D^\ell_{Mm}(\varphi+\pi,\pi-\theta,0) = (-1)^\ell D^\ell_{M\, -m}(\varphi,\theta,0)\,,
\ee
from which we deduce in particular
\be\label{ConjugateYlm}
{}_s Y_\ell^{m \star} = (-1)^{s+m} {}_{-s} Y_\ell^{-m}\,.
\ee

The orthogonality property of Wigner coefficients is expressed as
\be\label{EqOrthogonalityDlmm}
\int \frac{\dd^3 R}{8 \pi^2} D^\ell_{ms}(R) D^{\ell' \star}_{m' s'}(R) =
\frac{\delta_{\ell \ell'} \delta_{m m'} \delta_{s s'}}{2\ell+1}\,,
\ee
where an integral on all rotations is defined as
\be
\int \dd^3 R \equiv \int_0^{2\pi}\dd \gamma \int_0^\pi \sin \theta \dd
\theta \int_0^{2\pi} \dd \phi = 8\pi^2\,.
\ee
In particular this leads to the orthonormality of spin-weighted spherical harmonics
\be\label{OrthoYslm}
\int \dd^2 \gr{n} {}_s Y_\ell^m(\gr{n}) {}_{s}Y_{\ell'}^{m'\star}(\gr{n}) = \delta_{\ell \ell'} \delta_{m m'}\,,
\ee
whose counterpart for $s=0$ are the addition theorems
\bea
\sum_{m} Y_\ell^m(\gr{n}) Y_\ell^{m\star}(\gr{n}') &=& \frac{2\ell+1}{4\pi}
P_\ell(\gr{n} \cdot \gr{n}') \,,\label{AdditionYlm}\\
\sum_\ell \frac{(2\ell+1)}{4\pi}P_\ell(\gr{n}\cdot\gr{n}') &=& \delta^2_D(\gr{n},\gr{n}')\,,
\eea
where $\delta^2_D$ is the Dirac delta on the unit sphere.

The integral of three spin-weighted spherical harmonics is
\begin{align}\label{Gaunt}
&\int {}_{s_1} Y_{\ell_1}^{m_1} \,{}_{s_2} Y_{\ell_2}^{m_2}\, {}_{s_3} Y_{\ell_3}^{m_3} \dd^2 \gr{n} =\sqrt{\frac{(2 \ell_1 +1)(2 \ell_2 +1)(2 \ell_3 +1)}{4\pi}}\nonumber\\
&\qquad\times\troisj{\ell_1}{\ell_2}{\ell_3}{m_1}{m_2}{m_3}\troisj{\ell_1}{\ell_2}{\ell_3}{-s_1}{-s_2}{-s_3}\,.
\end{align}
The Wigner 3j symbols which enter the previous expression satisfy
\be\label{Ortho3j}
\sum_{m_1 m_2} \troisj{\ell_1}{\ell_2}{L}{m_1}{m_2}{M} \troisj{\ell_1}{\ell_2}{L'}{m_1}{m_2}{M'} = \frac{\delta^L_{L'} \delta^M_{M'}}{2L+1}\,,
\ee
when the triangular inequality is satisfied, and
\be\label{Ortho3jbis}
\sum_{LM} (2L+1)\troisj{L}{\ell_1}{\ell_2}{M}{m_1}{m_2}\troisj{L}{\ell_1}{\ell_2}{M}{m'_1}{m'_2} = \delta^{m_1}_{m'_1}\delta^{m_2}_{m'_2}\,.
\ee
Compatibility of~\eqref{Gaunt} with~\eqref{OrthoYslm} implies
\be\label{Troisj0ll}
\troisj{0}{\ell}{\ell}{0}{m}{-m'} = \delta_{m m'}\frac{(-1)^{\ell+m'}}{\sqrt{2\ell+1}}\,.
\ee

\section{Quadratic estimator for perfect experiments}\label{AppQuadratic}

Let us assume we have $N$ real-valued Gaussian random variable $x^i$ with corelation $\langle x_i x_j \rangle = A C_{ij}$. The $x^i$ correspond to residual or redshift measurements at the $i/T$ frequency for instance. We want to evaluate $A$ which is unknown, assuming that the overall shape of the correlations, $C_{ij}$, is known theoretically. 

Let us consider a quadratic estimator (we use implicit Einstein summation)
\be\label{QxWx}
Q = x_j W_{ji} x_i\,,
\ee
where $W_{ij}$ is symmetric. Its average and variance are
\begin{subequations}
\begin{align}
\langle Q \rangle  &= A W_{ij} C_{ji} = A{\rm Tr}(C \cdot W)\,,\\
\langle Q^2 \rangle-(\langle Q \rangle)^2 &= 2 A^2C_{j' i}  W_{ij} C_{j i'} W_{i' j'}\nonumber\\
&= 2 A^2 {\rm Tr}(C \cdot W \cdot C \cdot W)\,.
\end{align}
\end{subequations}
With a positive definite symmetric matrix ${\cal Q}$ one can always define a scalar product in the space of symmetric matrices by 
\be\label{DefScalarMat}
\{M_{ij}; N_{ij} \}_{\cal Q} \equiv {\cal Q}_{j' i} M_{ij} {\cal Q}_{j i'} N_{i' j'}={\rm Tr}({\cal Q}\cdot M\cdot {\cal Q} \cdot N)\,,
\ee
which is positive definite since the correlation matrix is also positive definite. Then we can rewrite
\begin{subequations}\label{AverageVarianceQ}
\begin{align}
\langle Q \rangle &= A \{W_{ij}; C^{-1}_{ij} \}_C\,,\\
\langle Q^2 \rangle-(\langle Q \rangle)^2 &= 2 A^2 \{ W_{ij}; W_{ij}\}_C\,.
\end{align}
\end{subequations}
The SNR ratio is
\be
{\rm SNR} = \frac{\langle Q \rangle}{\sqrt{\langle Q^2 \rangle-(\langle Q \rangle)^2}}\,,
\ee
which can be written as ${\rm SNR} = \sqrt{N^{\rm eff}/2}$ with
\be
N^{\rm eff}\equiv \frac{\{W_{ij}; C^{-1}_{ij} \}_C^2}{\{ W_{ij}; W_{ij}\}_C}\,. 
\ee
We find from Cauchy-Schwarz inequality
\be
{\rm SNR} \leq \sqrt{\frac{\{ C^{-1}_{ij}; C^{-1}_{ij}\}_C }{2}}=\sqrt{\frac{N}{2}}\,,
\ee
and this inequality saturates for
\be\label{Wijopt}
W^{\rm opt}_{ij} = C^{-1}_{ij}\,.
\ee
To summarize, the optimal weight in~\eqref{QxWx} is the inverse of the correlation shape for which the effective degrees of freedom used is $N^{\rm eff}=N$, which is the maximum available. Note that since $\{W_{ij}; C^{-1}_{ij} \}_C$ can be made as small as desired (it can be made to vanish by projection of $W_{ij}$ in the space orthogonal to $C^{-1}_{ij}$) the ${\rm SNR}$ can be extremely small, by taking  a suboptimal choice for $W$.

Clearly the minimum requirement is that $W_{ij}$ is positive definite. But even in that case the SNR can be quite damaged without optimization, as in full generality $N^{\rm eff}$ is only constrained by $N^{\rm eff} \geq 1$. To show this we can diagonalize the correlation in an orthogonal basis, hence there exists an orthogonal matrix $P$ and diagonal matrix $D$ with positive entries such that $C = P\cdot D\cdot P^{-1}$ and the square root is defined by $\sqrt{C}=P\cdot\sqrt{D}\cdot P^{-1}$. Then $C\cdot W = \sqrt{C}\cdot (\sqrt{C}\cdot W\cdot\sqrt{C}) \sqrt{C}^{-1}$ shows that $C \cdot W$ is similar to the symmetric matrix $\sqrt{C}\cdot W\cdot\sqrt{C}$ which can also be diagonalized in an orthonormal basis by using the matrix $R$, with diagonal matrix $d$ such that $\sqrt{C}\cdot W\cdot\sqrt{C} = R \cdot d \cdot R^{-1}$. The diagonal values $d_i$ are positive if $W$ is positive definite. Indeed for any vector $X$, ${}^T X \cdot \sqrt{C}\cdot W\cdot\sqrt{C} \cdot X = {}^T Y \cdot W \cdot Y $ with $Y = \sqrt{C} \cdot X$. Eventually on obtains $C \cdot W = (\sqrt{C}\cdot R)\cdot d \cdot (\sqrt{C}\cdot R)^{-1}$ and \be
N^{\rm eff} = \frac{\left({\rm Tr}(C\cdot W)\right)^2}{{\rm Tr}(C\cdot W \cdot C\cdot W)} = \frac{(\sum_i d_i)^2}{\sum_i d_i^2}\,.
\ee
Multiplying all $d_i$ by the same factor does not alter this ratio. However if we have two different eigenvalues $d_I$ and $d_J$ with $d_I > d_J$ then for $\epsilon>0$, the shifts $d_I \to d_I + \epsilon$ and $d_J \to d_J - \epsilon$ reduce the denominator. Therefore for a given fixed $\sum_i d_i$ the minimum of $N^{\rm eff}$ is reached when one eigenvalue is maximal and all other vanish and in that case $N^{\rm eff}=1$.
For instance we could choose $ W_{ij} = \delta^1_i \delta^1_j $  (plus negligible positive values for all other diagonal entries to make sure it is positive definite), such that ${\rm SNR} \simeq 1/\sqrt{2}$. It amounts to throwing all points except the first one. This is essentially what is achieved when considering the naive time-average with a very red spectrum for which $N^{\rm eff}$ is only slightly larger than unity.

Finally, redshift Fourier components are complex valued, but satisfy the condition~\eqref{complexzf}. 
Let us build an estimator for complex-valued random variables such that $x_i^\star = x_{-i}$ and $\langle x_i x_j^\star \rangle = A C_{ij}$. The hermitian correlation satisfies $C_{-j\,-i} = C_{ji}^\star=C_{ij}$. The general quadratic estimator of $A$ is of the form
\be
Q = x_j^\star W_{ji} x_i\,,
\ee
with $W_{-j\,-i} = W_{ji}^\star=W_{ij}$. Repeating the previous analysis, now~\eqref{DefScalarMat} becomes a hermitian product on hermitian matrices which is again positive definite because $C$ is. The average and variance of the estimator are then similar to~\eqref{AverageVarianceQ}, hence we find that the optimal estimator is~\eqref{Wijopt} also in this context. 

\section{Quadratic estimator for imperfect experiments}\label{AppQuadraticImperfect}

We aim at finding an estimator which maximizes the likelihood~\eqref{LikelihoodGauss} where the set of variables are the $\tilde{z}^{\rm pos}(\gr{n}_i,f_p)$. We use the multi-indices of the type $a=(\gr{n}_i,f_p)$, $b=(\gr{n}_j,f_q)$, or in compact form $a=(i,p)$, $b=(j,q)$, $c=(k,r)$, $d=(l,s)$, such that in~\eqref{LikelihoodGauss} $z_a= {\tilde{z}}^{\rm pos}(\gr{n}_i,f_p)$. In that case the correlation in~\eqref{LikelihoodGauss}, which we separate into signal and noise, is
\begin{align}\label{DefCab}
C_{ab} &\equiv \langle z_a z^\star_b \rangle_\AmpGW = C^S_{ab} + C^{\cal N}_{ab} \nonumber\\
&\equiv\mu(\beta_{ij}) {\cal{P}}^{\rm pos}(f_p,f_q) + \delta_{ij} (4\pi)^{-1} {\cal N}^{\rm pos}_i(f_p,f_q)\,,
\end{align}
where ${\cal N}^{\rm pos}_i(f_p,f_q)$ is defined from ${\cal N}_i(f)$ in the same way as the signal part was defined using~\eqref{DefSpolf1f2}. The correlation of the signal is further expanded as
\be\label{DefCslab}
C^S_{ab} = \sum_\ell c_\ell C^{S,\ell}_{ab} \,,\quad C^{S,\ell}_{ab} \equiv \mu_\ell(\beta_{ij}) {\cal{P}}^{\rm pos}(f_p,f_q)\,,
\ee
where we used the multipolar component of the HD curve $\mu_\ell(\beta)$ defined in~\eqref{Defmul}.

Following~\cite{Bond:1998zw}, a simple quadratic estimator can be devised to update the values of the $c_\ell$ of~\eqref{Defclparameters} in an iterative procedure which aims at maximizing the likelihood. Let us first define a hermitian inner product in the space of hermitian matrices using the (inverse) total covariance matrix in~\eqref{DefScalarMat}. With the current notation, this is
\be
\{M_{ab} ; N_{cd}\}_{C^{-1}} \equiv {\rm Tr}(M \cdot C^{-1} \cdot N \cdot C^{-1}) = M_{ab} C^{-1}_{ca} C^{-1}_{bd} N_{dc}\,,
\ee
with implicit summation in the last expression. The Fisher matrix for the $c_\ell$ parameters is 
\be\label{DefFisher}
F_{\ell \ell'}\equiv \frac{1}{2} {\rm Tr}\left[ C^{-1}\cdot \partial_{c_\ell} C \cdot C^{-1} \cdot \partial_{c_{\ell'}} C \right] = \frac{1}{2}\{ C^{S,\ell}_{ab} ; C^{S,\ell'}_{cd}\}_{C^{-1}}\,.
\ee
The quadratic estimator which updates $c_\ell$, requires to inverse the Fisher matrix, see e.g. (2.18) of \cite{Bond:1998zw}, and is
\be\label{Defquadratic}
c^e_\ell = F^{-1}_{\ell \ell'} X_{\ell'}\,,\qquad  X_\ell\equiv \frac{1}{2}\{ C^{S,\ell}_{ab} ; z_c z^\star_d - C^N_{cd} \}_{C^{-1}}\,.
\ee
From~\eqref{DefCab} and the expansion~\eqref{DefCslab} it is immediate to obtain $\langle X_\ell \rangle_\AmpGW = F_{\ell \ell'} c_{\ell'}$, hence the estimator~\eqref{Defquadratic} is unbiased, that is if the guess was correct, there is nothing to update. This can also be seen by rewriting $c_\ell^e -c_\ell = \frac{1}{2} F^{-1}_{\ell \ell'}\{ C^{S,\ell} ; z_c z^\star_d - C_{cd} \}_{C^{-1}}$ which vanishes if the initial guess is already unbiased, as can be seen  using~\eqref{DefCab}. This estimator is the counterpart in multipole space of the estimator in angular separation space built in~\cite{Allen:2024uqs}. While submitting the present article, a similar approach based on the Fisher matrix inversion was developed in~\cite{Allen:2024cgm}, where the sums are only performed on pairs of pulsars such that there is no need for an offset in the definition of $X_\ell$. The estimator is obtained by approximating that the quadratic observables $z_a z_b$ follow a Gaussian distribution with covariance $C_{ab;cd}=C_{ac}C_{bd}+C_{ad}C_{bc}$ and thus it uses a maximization of the associated $\chi^2$. On the other hand, the method~\cite{Bond:1998zw} that we follow is directly based on the maximization of the Gaussian distributions for the $z_a$ with correlation $C_{ab}$. Contrary to the former approach, in the latter one the estimator must be computed iteratively until convergence is reached so as to take into account that the Fisher matrix itself depends on the estimated parameters.

The SNR can be evaluated with the Fisher matrix. Indeed, since $\langle c_\ell^e c^e_{\ell'}\rangle_{\AmpGW} = F^{-1}_{\ell \ell'}$, the SNR for $c_\ell$ is ${\rm SNR}_\ell^2 = c_\ell^2/F^{-1}_{\ell\ell}$. Sadly it is in general impossible to find analytic forms for the Fisher matrix and thus for the SNR, since when there is noise \eqref{DefCab} does not factorize into a frequency part and an angular part. Hence let us first lower our ambitions and consider the case where there is no instrumental noise but shot noise due to the finite pulsar number is present, i.e. the imperfection comes from the finite number of pulsars. We assume that there are $N_f = 2 T f_{\rm max}$ frequencies which are measured perfectly. The inverse of the correlation factorizes as
\be\label{FactorCminusone}
C^{-1}_{ab} = \left[\mu(\beta_{ij})\right]^{-1} {{\cal P}}^{{\rm pos},-1}_{pq}\,,
\ee
where it is understood that $\left[\mu(\beta_{ij})\right]^{-1}$ means the inverse of the matrix $\mu(\beta_{ij})$. Hence the Fisher matrix reduces  to
\be\label{Fll2}
F_{\ell \ell'} = \frac{N_f}{2} \sum_{ijlk} \mu_\ell(\beta_{ij})[\mu(\beta_{ik})]^{-1} [\mu(\beta_{jl})]^{-1} \mu_{\ell'}(\beta_{kl})\,,
\ee
where the frequency-dependent part has disappeared into a trace of the identity of the frequency space, which yields a factor $N_f$. Again we must accept that there is no general simplification for a generic given set of pulsar directions. We would like to average this expression over the possible realizations of the pulsar directions to assess the noise introduced by the randomness of a finite number of pulsars, but there is to our knowledge no simple analytic method. At best, we could use the approximation $\left[\mu(\beta_{ij})\right]^{-1} \simeq (4\pi/N_{P})^2\mu^{-1}(\beta_{ij})$, where the latter is the inverse in the limit of infinite pulsar number, obtained by inversion in the $\ell\geq 2$ multipolar subspace (see section IV.C of \cite{Allen:2022ksj})
\be
\mu^{-1}(\beta_{ij}) = \sum_{\ell\geq2} \frac{1}{C_\ell^{\rm HD}} \frac{(2\ell+1)}{4\pi} P_\ell(\cos \beta_{ij})\,.
\ee
In that case the Fisher matrix simplifies drastically, as sums over pulsar directions are replaced by angular integrations, and with~\eqref{OrthoYslm} and~\eqref{AdditionYlm} we recover
\be\label{Fll3}
F_{\ell \ell'} = \delta_{\ell \ell'} \frac{N_f}{2}(2\ell+1)\,.
\ee
Having neglected noise, and assumed an infinite number of pulsars, our imperfect experiment is now perfect hence we are not surprised to recover~\eqref{IdealSNRl}. The difference between~\eqref{Fll2} and~\eqref{Fll3} is precisely due to the finite number of pulsars. However we know no simple method to assess the magnitude of this difference, hence in section~\ref{SecImperfect} we take an alternative approach based on the estimators~\eqref{DefzlmN} and their variance under the realizations of the pulsar directions to estimate this pulsar shot noise.

Finally, if we want to constrain the global amplitude $\AmpGWstar$ with an imperfect experiment, the Fisher matrix is just a number
\be
F = \frac{1}{2}\AmpGWstar^{-2}\{ C^S_{ab}; C^S_{cd} \}_{C^{-1}}
\ee
and the estimator is simply
\be
\AmpGWstar^e \equiv F^{-1} X \,,\qquad X\equiv \frac{1}{2}\AmpGWstar^{-1}\{ C^S_{ab}; z_c z^\star_d - C^N_{cd} \}_{C^{-1}}\,.
\ee
Up to the constant shift due to noise, this is exactly the estimator (7.2) of \cite{Allen:2022ksj}. For a perfect experiment $C^S_{ab}=C_{ab}$, $F =1/2\AmpGWstar^{-2} N_f N_P$, and this reduces to $\AmpGWstar^e = \AmpGWstar/(N_P N_f) \sum_{ab} z_a z^\star_b C^{-1}_{ba}$, hence to the estimator~\eqref{Qopt2} when using~\eqref{FactorCminusone}.

\section{Pulsar shot noise}\label{PoissonPulsarAverage}

Let us start from the estimator~\eqref{DefzlmN}. We want to compute its covariance when averaging over both the statistics of $z(\gr{n}_i)$ which is a statistics of the $z_{\ell m}$, and over the randomness of the position of the pulsars (indicated with $\langle \rangle_{\rm P}$ in that case). The underlying redshift function is decomposed as
\be
z(\gr{n}) = \sum_{L\geq 2} \sum_{M=-L}^L  z_{LM} Y_L^M(\gr{n})\,,
\ee
and the multipole ensemble statistics is of the form
\be
\langle  z_{LM} z_{L' M'}^\star\rangle_\AmpGW = \delta_L^{L'} \delta_M^{M'} C_\ell\,.
\ee
Now we shall also need
\begin{align}
\langle Y_\ell^m(\gr{n}_i) \rangle_{{\rm P}} &= \int \frac{\dd^2 \gr{n}}{4 \pi} Y_\ell^m(\gr{n}) =  \delta^\ell_0 \delta^m_0 \frac{1}{\sqrt{4\pi}}\,,\\
\langle Y_\ell^m(\gr{n}_i) Y_{\ell'}^{m' \star}(\gr{n}_i) \rangle_{{\rm P}} &= \int \frac{\dd^2 \gr{n}}{4 \pi} Y_\ell^m(\gr{n}) Y_{\ell'}^{m'\star}(\gr{n})= \frac{\delta^\ell_{\ell'} \delta^m_{m'}}{4 \pi}\,,
\end{align}
whereas for two distinct pulsars
\be
\langle Y_\ell^m(\gr{n}_i) Y_{\ell'}^{m' \star}(\gr{n}_j) \rangle_{{\rm P}} = \frac{\delta^\ell_0 \delta^m_0 \delta^{\ell'}_0 \delta^{m'}_0}{4\pi}\,.
\ee
Let us evaluate now the statistics of the estimator \eqref{DefzlmN} under the distribution of pulsar positions. The one-point function is simply
\be
\langle z_{\ell m}^N \rangle_P = z_{\ell m}\,.
\ee
In order to evaluate the covariance, we need to distinguish when we have the same pulsar or not hence this brings the two contributions 
\begin{align}
&\langle z_{\ell m}^N  z_{\ell' m'}^{N\star} \rangle_{\rm P} \\
&=\frac{N-1}{N} z_{\ell m} z_{\ell' m'}^\star + \frac{4 \pi}{N} \sum_{L L' M M'} z_{L M} z_{L' M'} \int \dd^2 \gr{n} Y_\ell^{m\star} Y_{L'}^{M'\star} Y_{\ell'}^{m'} Y_L^M\nonumber\,.
\end{align}
The first term comes from the $N(N-1)$ terms for which $\gr{n}_i \neq \gr{n}_j$ and the second term from the $N$ other terms.

Then we take the GW ensemble statistics and get
\be
\langle\langle z_{\ell m}^N \rangle_{\rm P} \rangle_\AmpGW = \langle z_{\ell m} \rangle_\AmpGW = 0\,.
\ee
For the covariance we have a product of four spherical harmonics. We can pair them using the addition of spherical harmonics which is inferred  from~\eqref{Ortho3j}, \eqref{OrthoYslm} and \eqref{ConjugateYlm}
\bea
Y_{\ell}^{m} Y_{\ell'}^{m'} &=& \sum_{LM}(-1)^{M}\sqrt{\frac{(2\ell+1)(2\ell'+1)(2L+1)}{4\pi}}\nonumber\\
&&\troisj{\ell}{\ell'}{L}{m}{m'}{-M} \troisj{\ell}{\ell'}{L}{0}{0}{0} Y_L^M\,.
\eea
We pair $Y_\ell^{m\star}$ with $Y_{L'}^{M'\star}$ to give $Y_{{\cal L}'}^{{\cal M}'\star}$, and we pair $Y_{\ell'}^{m'}$ with $Y_L^M$ to give $Y_{\cal L}^{\cal M}$. Using the GW ensemble average brings $\delta_L^{L'}\delta_M^{M'}$ and using the orthogonality property of spherical harmonics brings $\delta^{\cal L}_{{\cal L}'} \delta^{\cal M}_{{\cal M}'}$. Putting all this together we obtain
\begin{align}
&\langle\langle z_{\ell m}^N  z_{\ell' m'}^{N\star} \rangle_{\rm P}\rangle_\AmpGW =\frac{N-1}{N} \delta^\ell_{\ell'} \delta^m_{m'} C_\ell \\
&+ \frac{1}{N} \sum_{L M } \sum_{{\cal L} {\cal M}} C_L (2L+1)(2{\cal L}+1)\sqrt{(2 \ell+1)(2\ell'+1)}\nonumber\\
&\troisj{L}{\ell'}{{\cal L}}{M}{m'}{-{\cal M}}\troisj{L}{\ell'}{{\cal L}}{0}{0}{0} \troisj{L}{\ell}{{\cal L}}{M}{m}{-{\cal M}}\troisj{L}{\ell}{{\cal L}}{0}{0}{0}\,.\nonumber
\end{align}
With the orthonormality~\eqref{Ortho3j}, we get
\begin{align}
\langle\langle z_{\ell m}^N  z_{\ell' m'}^{N\star} \rangle_{\rm P}\rangle_\AmpGW &=\frac{N-1}{N} \delta^\ell_{\ell'} \delta^m_{m'} C_\ell\\
&+ \frac{1}{N} \sum_{L {\cal L}} C_L (2L+1)(2{\cal L}+1) \delta^\ell_{\ell'} \delta^m_{m'}\troisj{L}{\ell}{{\cal L}}{0}{0}{0}^2\,.\nonumber
\end{align}
Using finally
\be
\sum_L (2L+1)\troisj{\ell_1}{\ell_2}{L}{0}{0}{0}^2 =1\,,
\ee
which is a particular case of~\eqref{Ortho3jbis}, this reduces to
\be
\langle\langle z_{\ell m}^N  z_{\ell' m'}^{N\star} \rangle_{\rm P}\rangle_\AmpGW =\delta^\ell_{\ell'} \delta^m_{m'} (C_\ell + N_\ell^{\rm P})\,,
\ee
where the pulsar shot noise is
\be
N_\ell^{\rm P} = -\frac{1}{N} C_\ell + \frac{1}{N} \sum_{L} (2L+1) C_L \,.
\ee
Finally if we repeat the analysis of this section starting from~\eqref{BetterEstimate}, that is with general weights $w_i \neq 1/N$, then the factors $1/N$ in the previous expression are replaced by $\sum_{i=1}^N w_i^2 \geq 1/N$.

\section{Hellings-Downs correlation in angular space}\label{SecHDangular}

\begin{figure}
\centering
\includegraphics[width=\columnwidth]{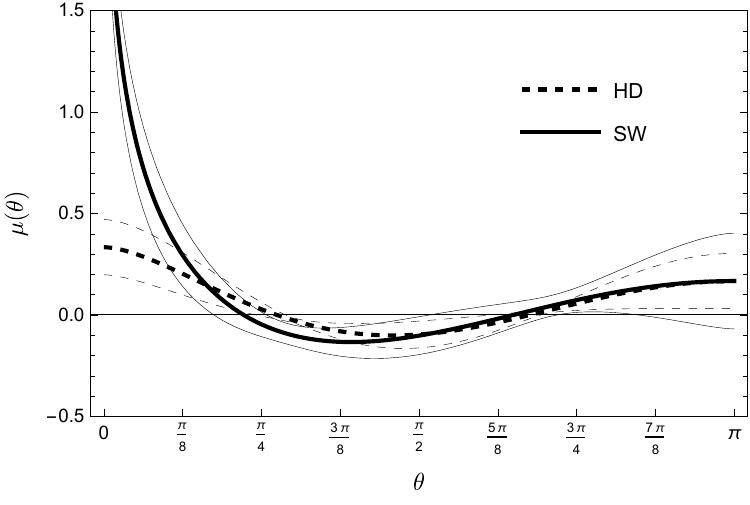}
\caption{Hellings-Downs correlation~\eqref{DefHD} in thick dashed line and Sachs-Wolfe plateau correlation from scalar perturbations~\eqref{EqSW} in thick continuous line, the latter being normalized to the same value as the former when $\theta=\pi$. The thin lines correspond to the one standard deviation contour due to cosmic variance for $N^{\rm eff}_f=1$, obtained from~\eqref{Varianceletaspace}. The contours will of course shrink when the value of $N^{\rm eff}_f$ is increased by improving observation time and pulsar monitoring cadence.}
\label{FigSW}
\end{figure}

We summarize how \eqref{DefHD} is obtained with a method adapted from Sec III.E of \cite{Gair:2014rwa}. Let us start by decomposing $ 
\ln x$, where $x=(1-\cos \beta)/2 = \sin^2(\beta/2)$, on Legendre polynomials, that is we seek $\ln x = \sum_\ell a_\ell P_\ell(\beta)$ and the coefficients are given from the orthogonality of Legendre polynomials as
\bea
a_\ell &=& \frac{2\ell+1}{2}\int_{-1}^1 \ln x P_\ell(\cos \beta)\dd \cos \beta \nonumber\\
&=& (2\ell+1)\int_0^1 \ln x \,P_\ell(1-2 x) \dd x\,.
\eea
Using 
\be
P_\ell(1-2x) = \frac{1}{\ell!} \frac{\dd^\ell}{\dd x^\ell}[x^\ell(1-x)^\ell]\,,
\ee
and 
\be
\frac{\dd^n}{\dd x^n} \ln x = \frac{(-1)^{n-1} (n-1)!}{x^{n}}\,,
\ee
the coefficients are easily evaluated with successive integration by parts and for $\ell\geq 1$we get
\be
a_\ell = -\frac{2\ell+1}{\ell(\ell+1)}\,,
\ee
whereas $a_0=-1$. Therefore we find
\be\label{lnxexpansion}
-\ln x = -\ln \sin^2(\theta/2) = 1 + \sum_{\ell \geq 1} \frac{2\ell+1}{\ell(\ell+1)} P_\ell(\cos \theta)\,.
\ee
The Sachs-Wolfe plateau in the CMB temperature fluctuations for a scale-invariant spectrum of scalar perturbations behaves as $C_\ell^{\rm SW} \propto 1/[\ell(\ell+1)]$. Hence once translated with~\eqref{lnxexpansion} as a correlation for angular separation where the dipole is conventionally removed, the Sachs-Wolfe plateau correlation function is
\bea\label{EqSW}
\mu^{\rm SW}(\theta)&\equiv& \sum_{\ell\geq2}\frac{(2\ell+1)}{4\pi}C_\ell^{\rm SW}P_\ell(\theta)\nonumber\\
&\propto& -1 - \ln \sin^2(\theta/2)-\frac{3}{2}\cos \theta\,.
\eea 
This captures the shape of the CMB correlations in angular separation space (see e.g. Fig.~2 of \cite{Copi:2010na}), and it diverges for $\theta \to 0$ contrary to the Hellings-Downs curve, see Fig.~\ref{FigSW}.

Finally, we need only to use
\be
\cos \theta P_\ell(\cos \theta) = \frac{\ell+1}{2\ell+1}P_{\ell+1}(\cos \theta)+\frac{\ell}{2\ell+1}P_{\ell-1}(\cos \theta)\,,\nonumber
\ee
in combination with~\eqref{lnxexpansion}, and rearranging the summations so as to sum only on the $P_\ell$, we obtain
\be
x \ln x = \sum_{\ell \geq 2} (2\ell+1)\frac{(\ell-2)!}{(\ell+2)!} P_\ell(\cos \theta) -\frac{1}{3} + \frac{x}{6}\,,
\ee
which can be used to find the Hellings-Downs correlation function~\eqref{DefHD}.

\newpage

\bibliography{HDbib}

\end{document}